\documentclass[preprint,12pt]{elsarticle}

\usepackage{amssymb}
\usepackage{algorithm}
\usepackage{algpseudocode}
\usepackage{booktabs}
\usepackage{multirow}
\usepackage{amsmath}
\usepackage[dvipsnames]{xcolor}
\usepackage{tablefootnote}
\usepackage{hyperref}

\newcommand{\Rey}[0]{\mathrm{Re}}
\newcommand{\figref}[1]{Figure \ref{#1}}
\newcommand{\equref}[1]{Equation \ref{#1}}
\newcommand{\tabref}[1]{Table \ref{#1}}

\newcommand{\secref}[1]{Section \ref{#1}}

\newcommand{\algoref}[1]{Algorithm \ref{#1}}

\journal{International Journal of Heat and Fluid Flow}

\begin{document}

\begin{frontmatter}



\title{FR3D: Three-dimensional Flow Reconstruction and Force Estimation for Unsteady Flows Around {Extruded} Bluff Bodies via Conformal Mapping Aided Convolutional Autoencoders}


\author[inst1]{Ali Girayhan \"Ozbay}
\author[inst1]{Sylvain Laizet}

\affiliation[inst1]{organization={Department of Aeronautics, Imperial College London},
            addressline={Exhibition Road}, 
            city={London},
            postcode={SW7 2AZ}, 
            country={United Kingdom}}

\begin{abstract}
In many practical fluid dynamics experiments, measuring variables such as velocity and pressure is possible only at a limited number of sensor locations, \textcolor{black}{for a few two-dimensional planes, or for a small 3D domain in the flow}. However, knowledge of the full fields is necessary to understand the dynamics of many flows. Deep learning reconstruction of full flow fields from sparse measurements has recently garnered significant research interest, as a way of overcoming this limitation. This task is referred to as the flow reconstruction (FR) task. In the present study, we propose a convolutional autoencoder based neural network model, dubbed FR3D, which enables FR to be carried out for three-dimensional flows around extruded 3D objects with different cross-sections. An innovative mapping approach, whereby multiple fluid domains are mapped to an annulus, enables FR3D to generalize its performance to objects not encountered during training. We conclusively demonstrate this generalization capability using a dataset composed of 80 training and 20 testing geometries, all randomly generated. We show that the FR3D model reconstructs pressure and velocity components with a few percentage points of error. Additionally, using these predictions, we accurately estimate the Q-criterion fields as well lift and drag forces on the geometries.
\end{abstract}




\end{frontmatter}


\section{Introduction and related work}
\label{sec:intro}
Flow reconstruction (FR) involves the prediction of dense fields such as velocity based on sparse measurements. Since typical experiments in fluids involve only point measurements of the flow via simple and inexpensive methods such as pitot tubes, FR techniques can provide researchers additional insight into flows when more advanced techniques such as particle image velocimetry (PIV) are not available.

Various statistical tools have been applied to FR {such as linear stochastic estimation (LSE) \cite{flow_reconstruction_history_stochastic_est}, gappy proper orthogonal decomposition (gappy POD) \cite{gappy_pod_ex2}, extended proper orthogonal decomposition (EPOD) \cite{epod}, and sparse representation \cite{flow_reconstruction_1}. Though these techniques are time-tested and have been applied in practical experiments, for instance to estimate and control the flow in a backward-facing step case via LSE \cite{lse_experimental_control}, their linear nature limit their capability to deal with complex flows.} 

{Neural networks, owing to their universal approximation capabilities \cite{nn_universal_approx1}, are capable of learning arbitrary non-linear and high-dimensional relationships in datasets. This capability makes them very attractive for FR tasks. As a result, the recent explosion of interest in neural networks (NNs) -- enabled by substantial increases in computing power, theoretical advances, and the availability of open-source deep learning software -- has coincided with a shift towards NN-based FR,} and substantial strides were made recently with the application of NNs to the field. {Notably, Erichson et al.\ \cite{flow_reconstruction_2} produced a seminal study exploring the usage of neural networks to reconstruct flows past cylinders. A number of works followed Erichson et al., a selection of which are presented: Fukami et al. \cite{flow_reconstruction_8} demonstrated that NN-based methods can outperform linear FR methods for the reconstruction of flows past cylinders and flapped airfoils \cite{flow_reconstruction_3}, and also coupled NN-based FR with Voronoi tessellations to achieve flexibility in terms of the sensor setup \cite{flow_reconstruction_8}. Sun and Wang \cite{flow_reconstruction_6} investigated the application of physics-informed Bayesian NNs in FR, demonstrating high robustness to noise when reconstructing flows in simulated vascular structures. Dubois et al \cite{flow_reconstruction_4} extended FR to variational autoencoder architectures. Kumar et al.\ \cite{flow_reconstruction_7} developed a recurrent NN architecture to carry out FR with extremely sparse sensor setups. Xu et al.\ \cite{flow_reconstruction_5} considered the usage of physics-informed loss functions to train FR models using gappy data. He et al.\ \cite{flow_reconstruction_9} explored the usage of graph attention NNs to reconstruct flows past cylinders. Carter et al.\ \cite{flow_reconstruction_nn_experimental} investigated the usage of an NN-based FR model on experimental data of flow past an airfoil.} Applications outside typical wind tunnel-like scenarios, such as in porous media \cite{flow_reconstruction_porous_media} or the reconstruction of atmospheric flows based on satellite imagery \cite{flow_reconstruction_satellite}, have also been developed. 

However, the applicability of even these recent approaches to practical scenarios is limited. The first obstacle is tackling multi-geometry FR. Even recently published works typically investigate FR for a single geometry only (often a 2D circular cylinder), and as a result the models used in such works must be re-trained for every case investigated; a model trained on e.g.\ a circular cylinder will not work well for a square cylinder. This necessitates laborious data collection and a computationally expensive training process. To overcome this limitation, a growing body of works have investigated 2D multi-geometry FR using techniques such as graph convolutional neural networks \cite{mgfr_1, mgfr_2, mgfr_3} and conformal mappings \cite{our_journal_article}.

The second challenge pertains to the reconstruction of three-dimensional flows, regarding which relatively few works exist compared to the reconstruction of two-dimensional flows. Three dimensional flows exhibit substantially more complicated dynamics than two-dimensional flows \cite{2d_vs_3d}, and require much greater computing power to process. Despite this, a growing body of works is tackling the challenge of 3D FR. Particularly, reconstruction of 3D flows past cylinders \cite{flow_reconstruction_3d_6} and of flows concerning domains without embedded objects such as channel flows \cite{flow_reconstruction_3d_1, flow_reconstruction_3d_2} have been investigated recently. A number of studies have also investigated the reconstruction of 2D slices of flows past square cylinders \cite{flow_reconstruction_3d_3, flow_reconstruction_3d_5} and vice versa \cite{flow_reconstruction_3d_4} (i.e.\ reconstruction of 3D fields from 2D slices).

In this work, we introduce a method enabling the reconstruction of unsteady three-dimensional flows around objects with arbitrarily shaped cross-sections. This is achieved via an autoencoder-based convolutional neural network architecture which incorporates conformal mappings to achieve geometry invariance \cite{our_journal_article, our_tsfp_article}. In our previous study \cite{our_journal_article, our_tsfp_article}, we have shown that it is possible to reconstruct dense contemporaneous or future vorticity fields of two-dimensional flows past various objects from current sparse sensor measurements, even for objects not encountered during training. To achieve optimal performance in these tasks, Schwarz-Christoffel mappings were used for choosing the sampling points of the dense fields. The results showed that the mapping aided approach provides a substantial boost in accuracy for all model and sensor setup configurations, enabling percentage errors under 3\%, 10\% and 30\% for reconstructions of pressure, velocity and vorticity fields, respectively.

For the present study, we reconstruct unsteady three-dimensional flows around bluff bodies with periodic spanwise boundary conditions at Re = 500 (based on the freestream velocity and the characteristic length of the bluff body). Additionally, we use these reconstructions to accurately predict the aerodynamic forces experienced by the investigated geometries as well as the Q-criterion \cite{qcriterion}, which defines vortices as areas where the vorticity magnitude is greater than the magnitude of the rate of strain. 

The paper is organized as follows: first, in \secref{sec:data}, we detail the procedure used to generate our dataset and the experiments to be carried out using the dataset. Next, in \secref{sec:model}, we expound upon the FR3D model architecture and its training procedure. Subsequently, in \secref{sec:results}, we display the performance of the FR3D model.  Finally, in \secref{sec:conclusion}, we summarize the results, and identify avenues for further research in 3D flow reconstruction.

\section{Data and experimental setup}
\label{sec:data}

\subsection{Geometries, meshing and flow simulations}

\begin{figure}
    \centering
    \includegraphics[width=0.45\textwidth]{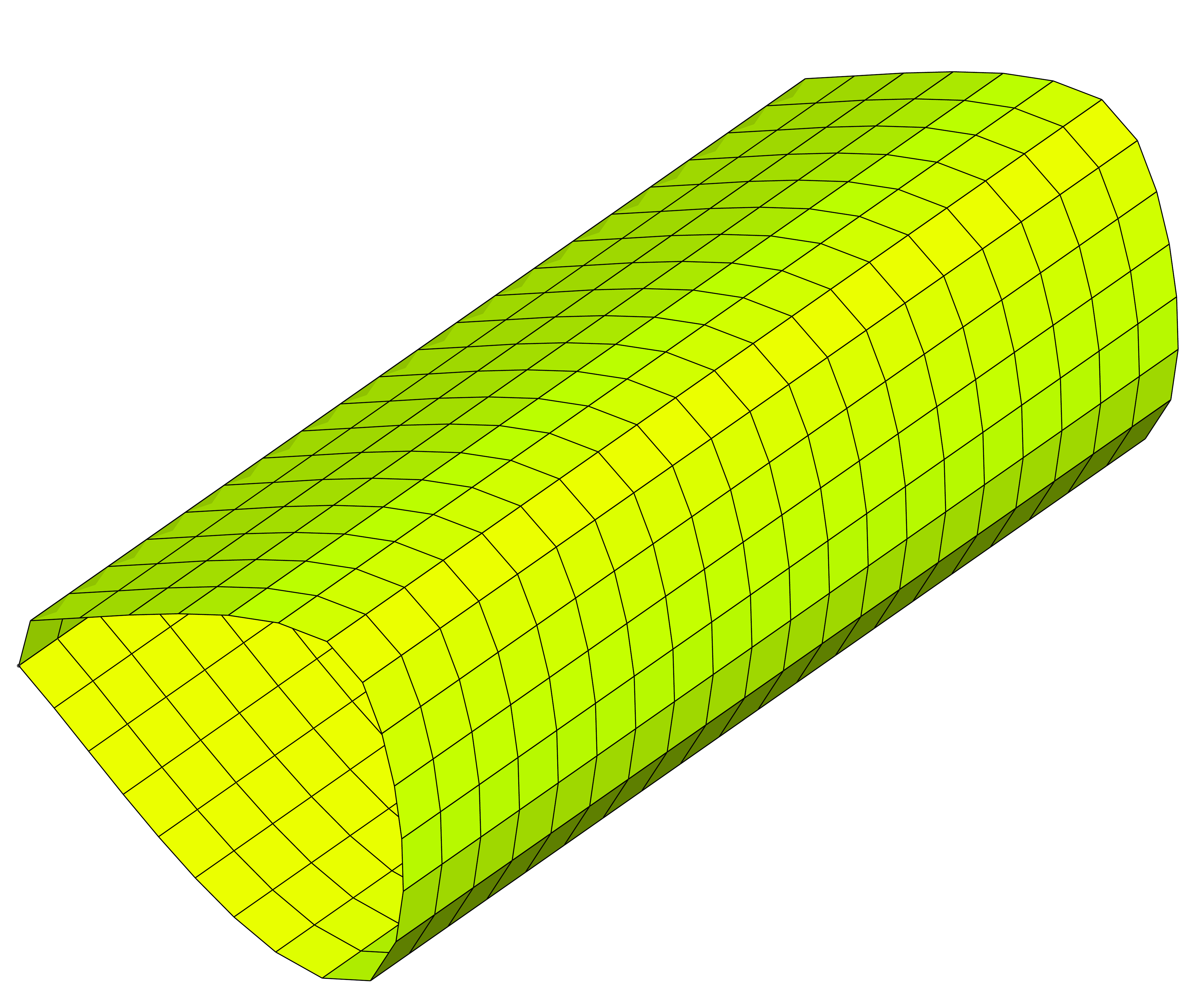}
    \includegraphics[width=0.45\textwidth]{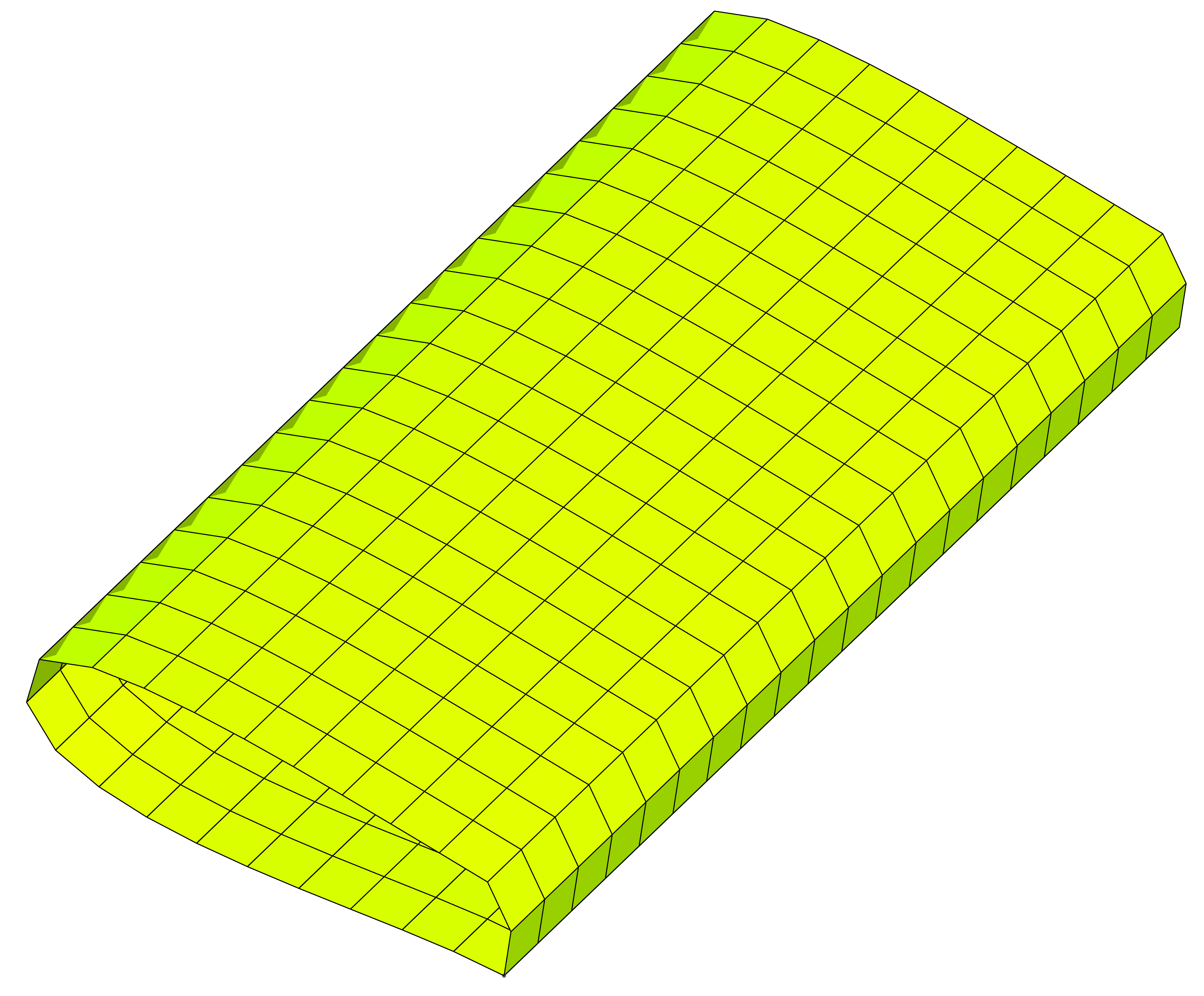}
    \includegraphics[width=0.45\textwidth]{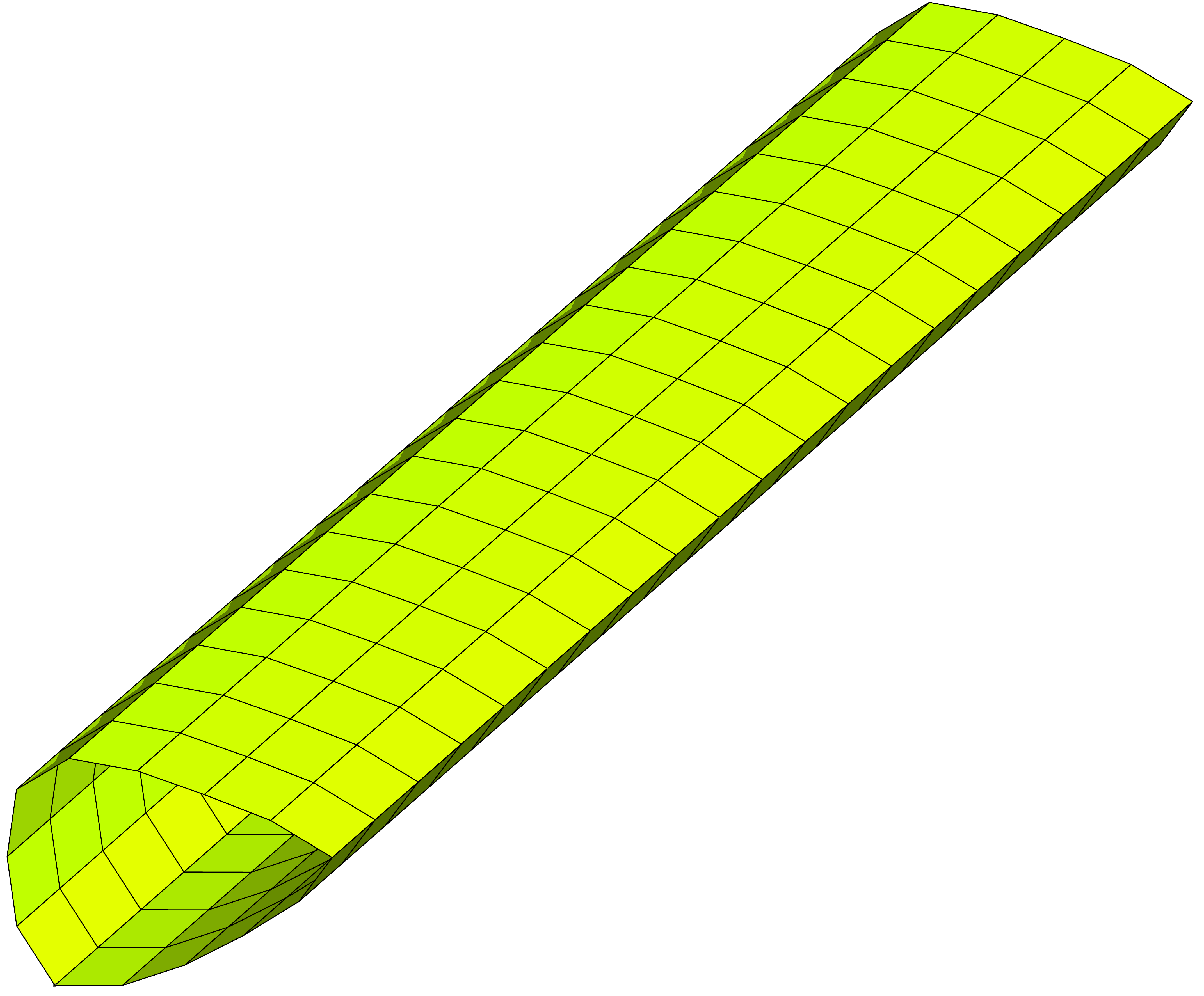}
    \includegraphics[width=0.45\textwidth]{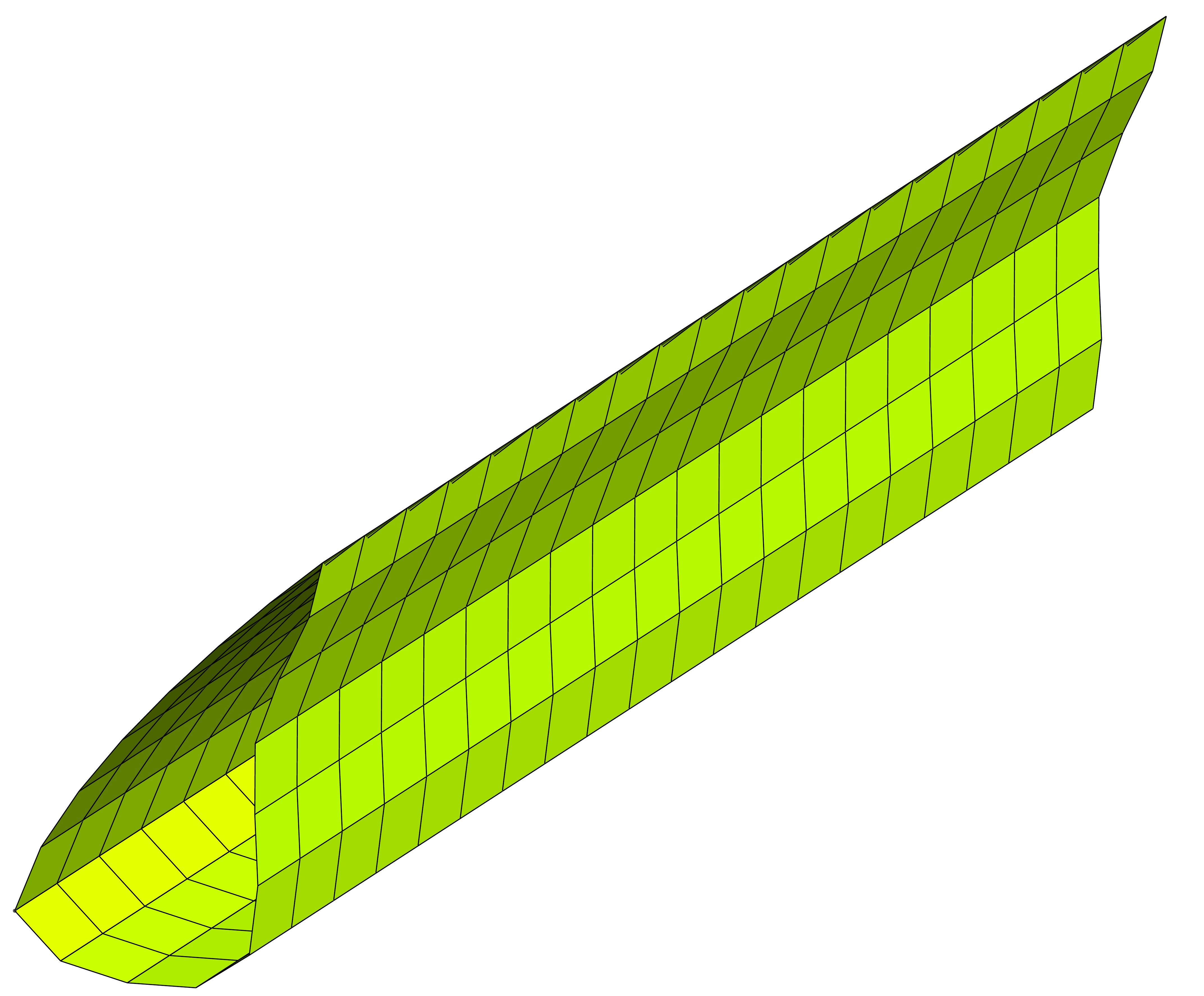}
    \includegraphics[width=0.45\textwidth]{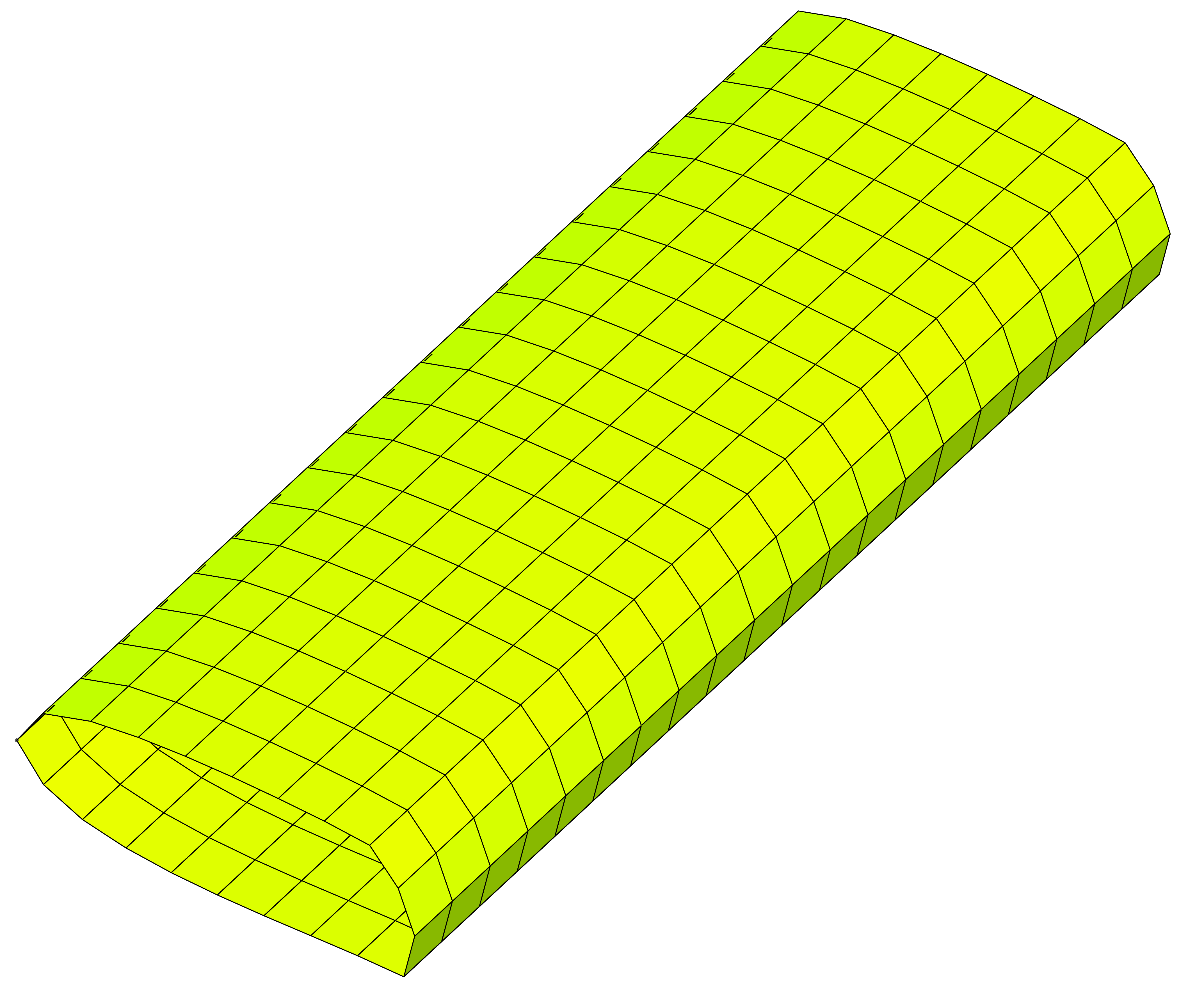}
    \includegraphics[width=0.45\textwidth]{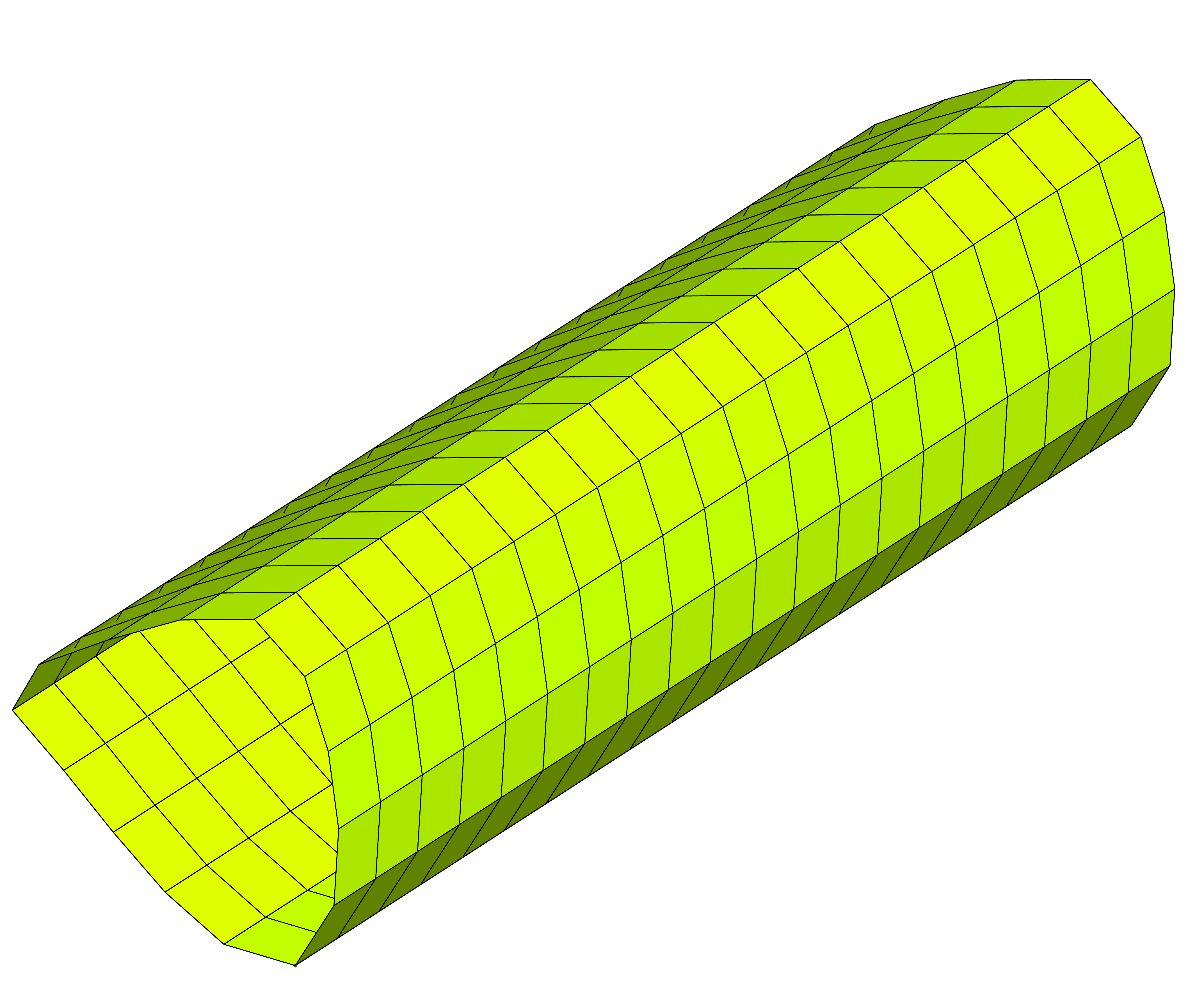}
    \caption{Surface meshes of a selection of the randomly generated geometries used in the study}
    \label{fig:shapes}
\end{figure}

Our dataset consists of 100 geometries $G_i, i \in [0, 99]$, randomly generated using a method based on Bezier curves by Viquerat et al. \cite{bezier_repo}. Each geometry uses 4 control points for the curves, chosen randomly in a square domain with characteristic length $L_m$, which enables the generation of convex as well as concave shapes. \figref{fig:shapes} showcases the diversity of the geometries created in this manner, including airfoil-like cross-sections, objects with concavities and objects with sharp corners.

Each geometry $G_i$ was placed in a $40L_m/7 \times 40L_m/7$ square domain, extruded for $20L_m/7$ in the spanwise direction. The domains were meshed using an automated procedure with c. 30,000 hexahedral and triangular prism elements each, with wake refinement applied. The flows were computed at a Reynolds number equal to $\Rey = u_\infty L_m / \nu$ = 500 using the \texttt{PyFR} solver \cite{pyfr} ($u_\infty$ is the free stream velocity and $\nu$ is the viscosity of the flow).  This solver is a flux reconstruction \cite{flux_reconstruction} based advection-diffusion equation solver using the artificial compressibility approach to solve the incompressible Navier-Stokes equations. It was chosen for its Python interface and GPU acceleration capabilities. 800 snapshots between $\tau^* = 5.71$ and $\tau^* = 17.14$ (when the flow is fully established) were recorded per geometry, where $\tau^* = u_\infty \tau / L_m$ is the time normalized by the large eddy turnover time $L_m/u_\infty$.

The snapshots collected were split into training and validation datasets, with all snapshots belonging to 20 randomly chosen geometries constituting the validation set, and the rest serving as the training set. This setup ensures that our model must have reasonable generalization performance to perform well on the validation set, as learning the reduced-order dynamics of specific flows (as in e.g. the Dynamic Mode Decomposition \cite{dmd}) is not sufficient to reconstruct flows past unseen geometries.

\subsection{Flow validation}

The solver settings used to generate the flow dataset were validated using the canonical case of the flow past a cylinder with diameter $D$ in a spanwise periodic domain. Two quantities were compared against reference data by Mittal and Balachandar \cite{balachandar-cylinder} at $\Rey = 300$, both averaged in the spanwise direction:

\begin{itemize}
    \item $<u> - u_\infty$: Time averaged $u$-velocity deficit
    \item $<u'u'>/u_\infty^2$: Streamwise-component of the Reynolds stress tensor
\end{itemize}

\begin{figure}
    \centering
    \includegraphics[width=0.40\textwidth]{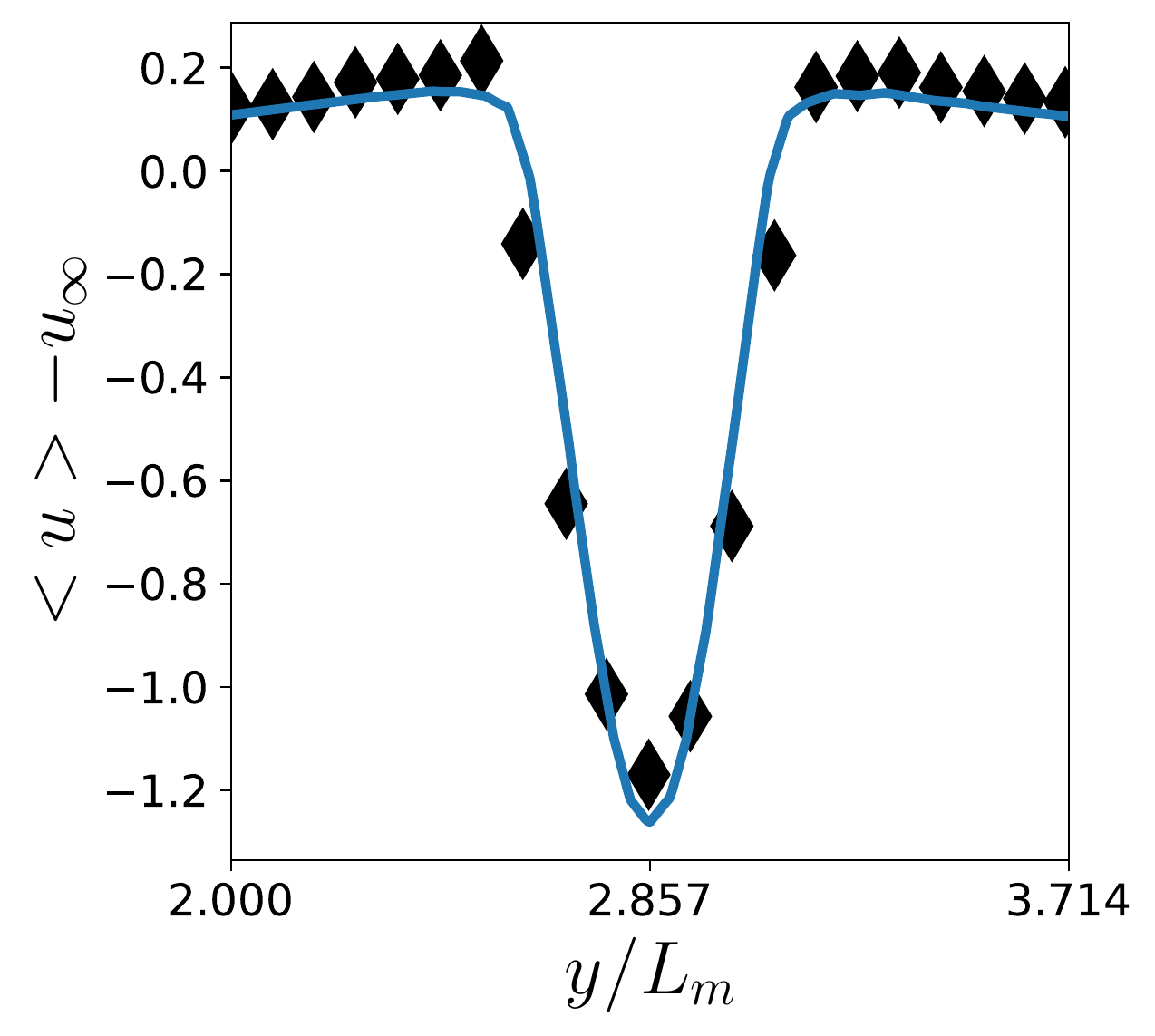}
    \includegraphics[width=0.40\textwidth]{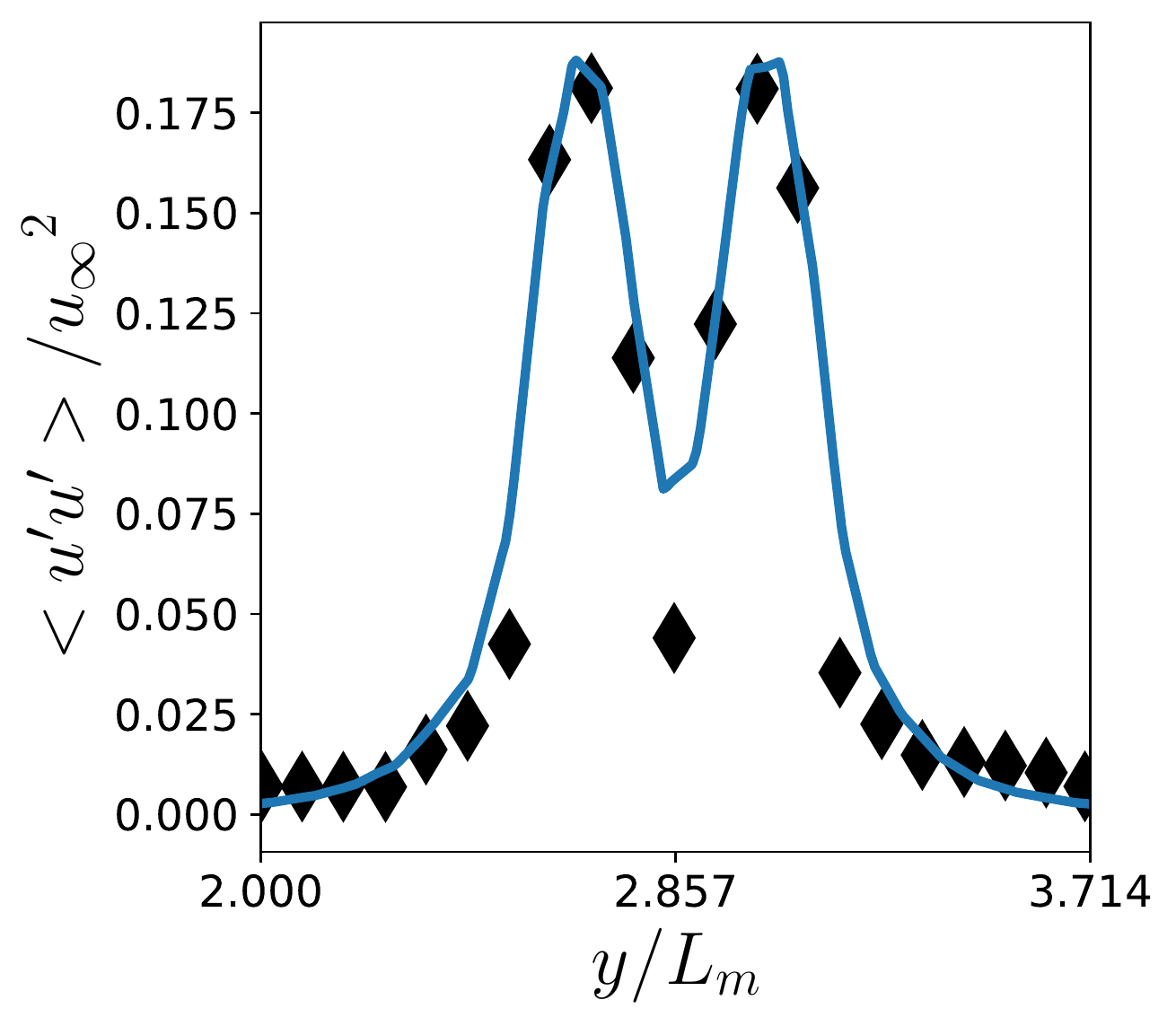}
    \includegraphics[width=0.40\textwidth]{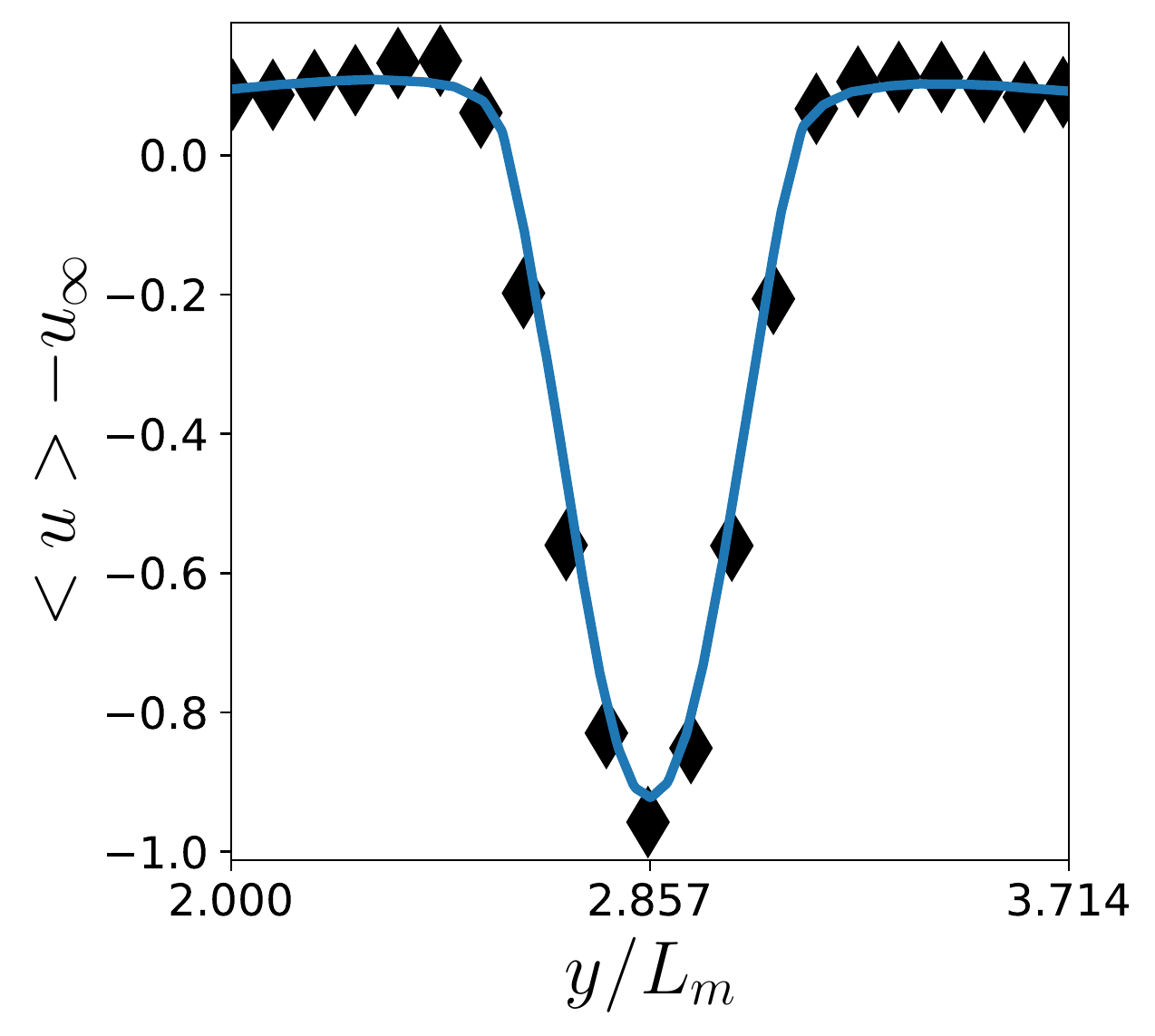}
    \includegraphics[width=0.40\textwidth]{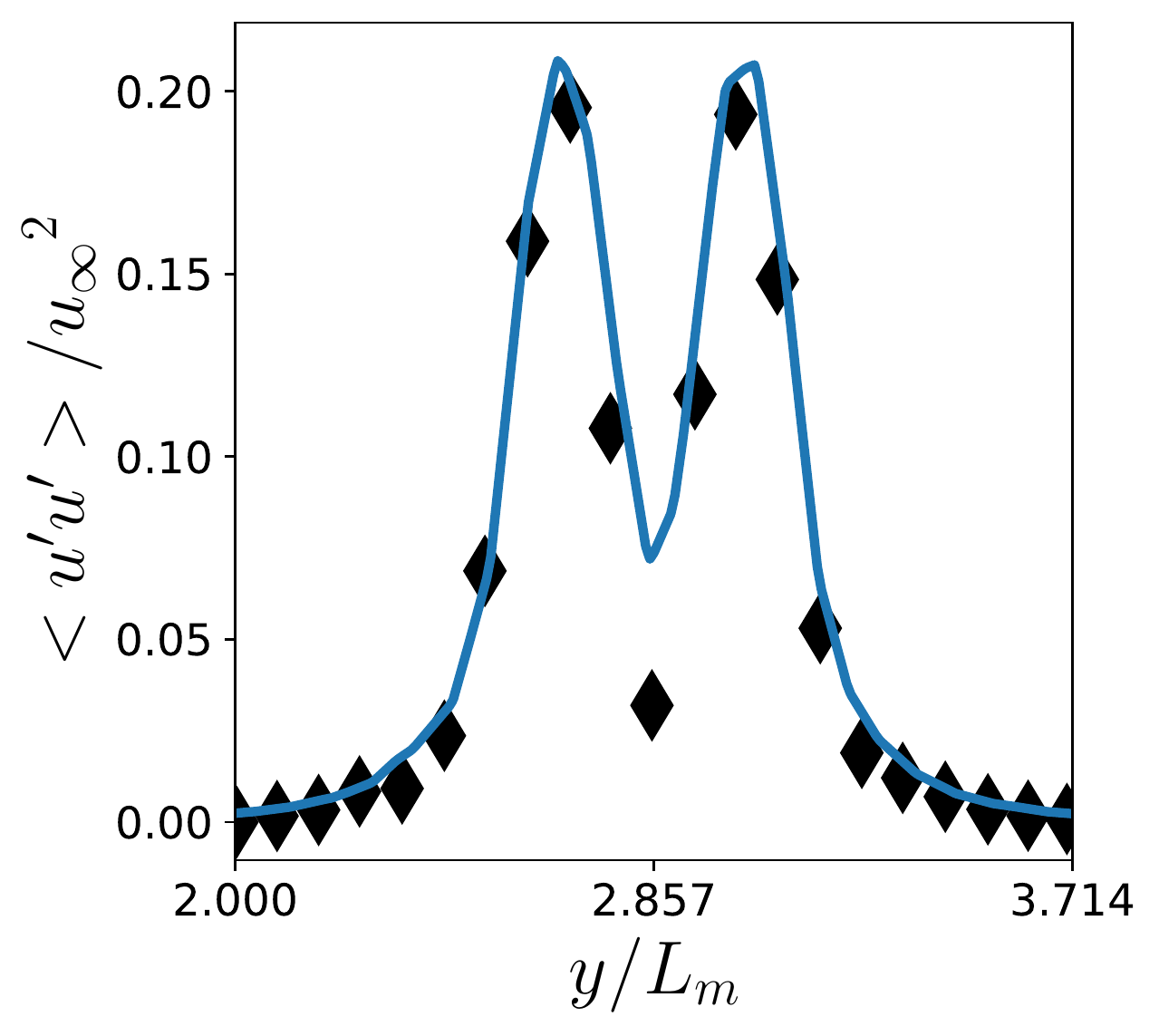}
    \includegraphics[width=0.40\textwidth]{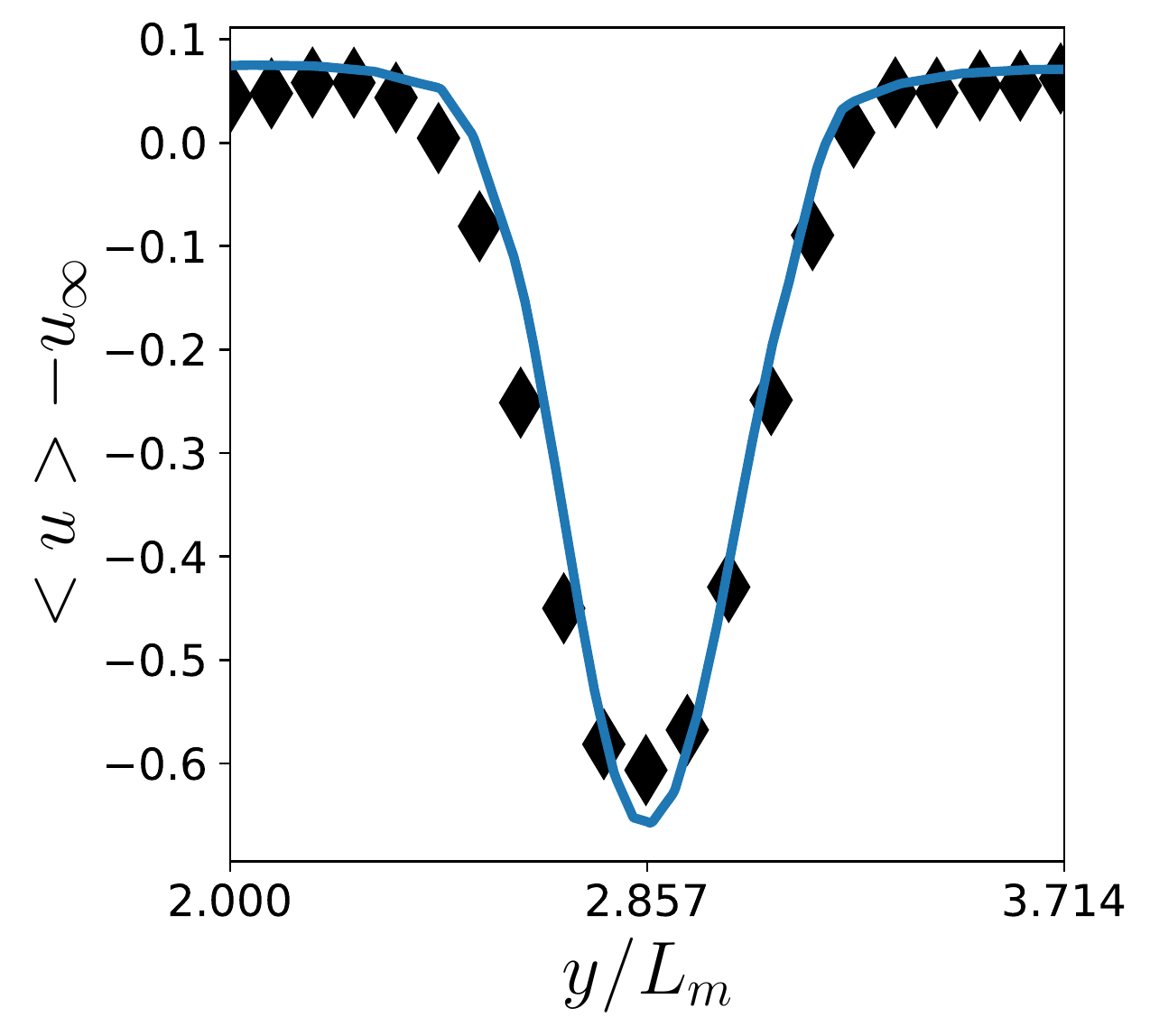}
    \includegraphics[width=0.40\textwidth]{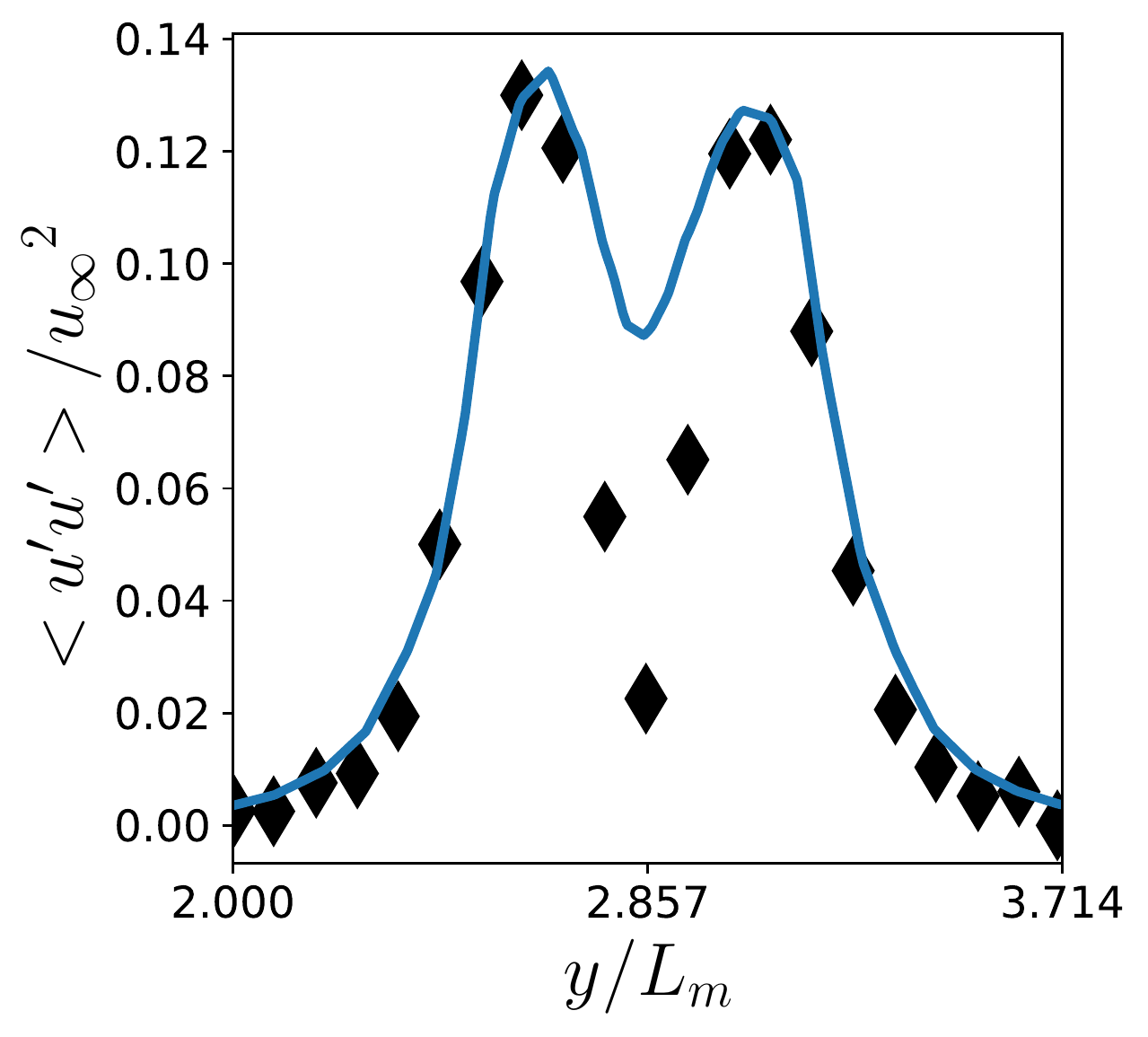}
    \includegraphics[width=0.33\textwidth]{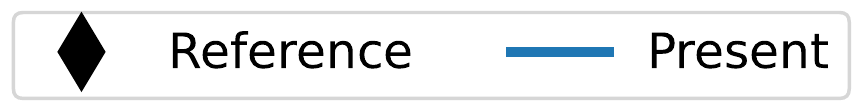}
    \caption{Comparison of results by Mittal and Balachandar \cite{balachandar-cylinder} with our results for $<u>-u_\infty$ (left) and $<u'u'>/u_\infty^2$ (right) for the $\Rey=300$ cylinder, $1.2D$ (top), $1.5D$ (middle) and $2.0D$ (bottom) downstream the centre of the cylinder.}
    \label{fig:validation}
\end{figure}

Our results were obtained using an automatically generated mesh, created with the same procedure as the one used to generate the meshes for the random geometries. The solver settings were kept identical to the settings used to obtain flow solutions for the random geometries, save for adjusting $u_\infty$ to obtain the correct Reynolds number. The solution took approximately 1 hour to complete on a single Nvidia A100 GPU. The two quantities are plotted at several downstream locations in \figref{fig:validation}.

The plots show good agreement between the reference values and our flow solution. $u$-velocity deficit profiles, including the peak deficit, are replicated with minimal errors at all three downstream locations. Streamwise Reynolds stress tensor results also show good agreement throughout most of the domain -- important features such as the dual peaks of the profile are closely followed -- although the dip between the two peaks is underestimated. 

As a further soundness check, we also compared the time averaged lift coefficient ($C_L$) and drag coefficient ($C_D$) values obtained from our simulation with published data. We recorded a $C_D$ value of 1.46, which is in line with the range of values between 1.22 and 1.50 in previous literature as compiled by Giannenas and Laizet \cite{cylinder_re300_reference_Cd}, and a $C_L$ of $2.8 \times 10^{-5}$ -- i.e.\ practically 0 for a single precision calculation -- as expected. Thus, as our meshing and solver settings produce results that are largely in line with previous literature for this validation case, we can be confident that our dataset consists of physically correct snapshots.

\subsection{Postprocessing}

As the final step in our data generation process, postprocessing was applied to the snapshots to obtain sets of inputs and outputs used to train the neural network models. Our postprocessing extends the methodology which was used in our previous works based on conformal mappings to incorporate geometry invariance in neural network based FR methods \cite{our_journal_article, our_tsfp_article}. 

\begin{figure}
    \centering
    \includegraphics{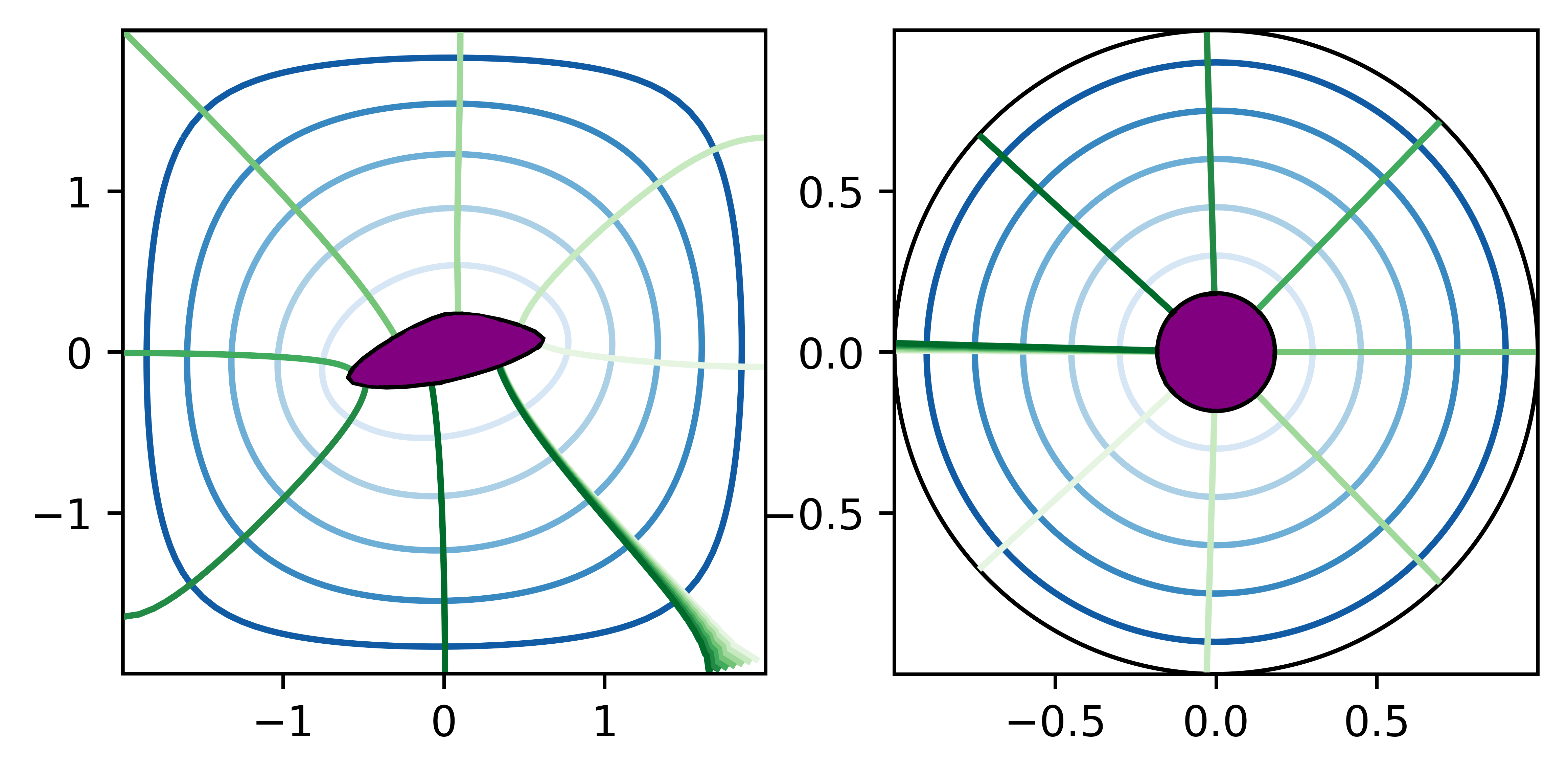}
    \caption{A random geometry (left) and its corresponding preimage (right). Blue and green contours depict the norm and argument in the annular domain, respectively.}
    \label{fig:mapping}
\end{figure}

In the aforementioned method, the fluid domain $F_i$ around each geometry $G_i$ is treated as a doubly connected 2D region, and a mapping $f$ between an annulus and $F_i$ is computed, as visualized in \figref{fig:mapping}. Subsequently, a grid equispaced in the radial and angular dimensions is generated in the annular domain, and the mapping is used to calculate the coordinates of the gridpoints in $F_i$, which constitute the sampling points of the ground truth dense fields. As a result, each slice of the grid along the radial direction is guaranteed to start on the surface of the geometry $G_i$ and end on the outer boundary of the fluid domain.

This work extends the method by incorporating the extrusion of the 2D cross section in the third dimension. In essence, this allows the dense flow fields around each geometry to be represented in cylindrical coordinates $(r, \theta, z)$. We choose a resolution of 64 grid points per dimension; hence, the dense fields are sampled on a $64 \times 64 \times 64$ grid.

We consider two experiments with different types of sensor inputs. In the first experiment, the sensor inputs consist of pressure and velocity probes. The former consist of 50 probe locations equispaced in the angular direction along the inner ring in the annular domain at 10 stations equispaced in the spanwise direction, for a total of 500 pressure sensors. The latter, also consisting of 500 sensors, are arranged in a $10 \times 10 \times 5$ grid, spanning a $L_m \times 5L_m/3 \times 20L_m/7$ box $L_m/7$ downstream of the trailing edge of each object. Using this setup, dubbed the "sparse" setup, we consider the reconstruction of the four primary variables in the Navier-Stokes equations -- the pressure $p$ and the three velocity components $u, v$ and $w$.

\begin{figure}
    \centering
    \includegraphics[width=\textwidth]{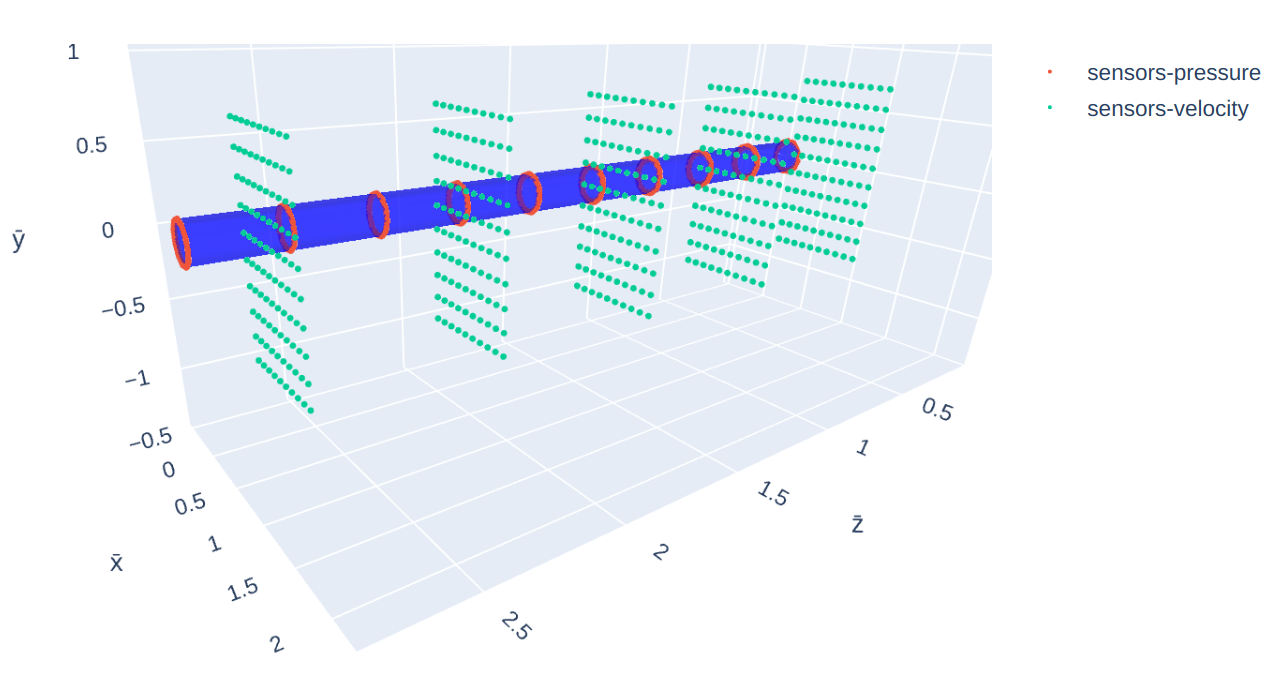}
    \includegraphics[width=\textwidth]{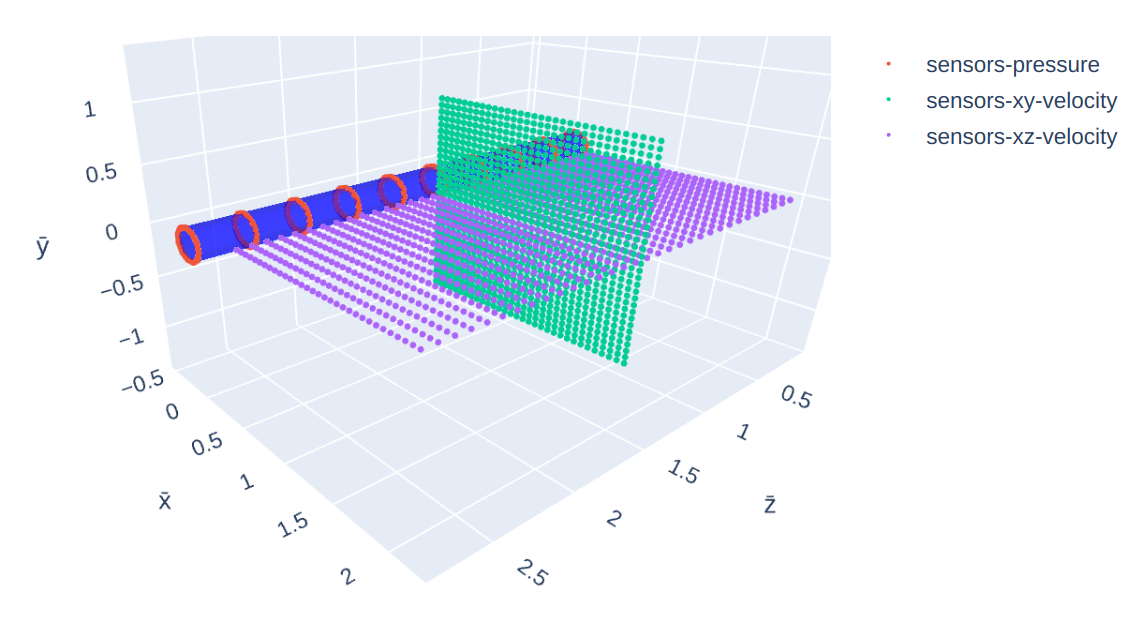}
    \caption{3D representation of the sparse (top) and plane (bottom) sensor setups on a circular cylinder. $\Bar{x}, \Bar{y}, \Bar{z}$ refer to the $x,y,z$ coordinates normalized by $L_m$.}
    \label{fig:sensor_setup}
\end{figure}

The second experiment retains the previous pressure sensor setup but uses plane measurements of velocity fields, called the "plane" setup. Two perpendicular planes with $32 \times 32$ velocity sensors each are considered; an $xy$-plane and an $xz$-plane, both downstream of the randomly generated objects. \textcolor{black}{This sensor setup has seen use in experiments \cite{two_plane_piv_experiment}, and hence is of interest within the context of certain setups involving PIV, an optical technique used in experiments capable of measuring the velocity field for turbulent fluid flows,\ and for which the pressure fields must be inferred from the velocity measurements \cite{piv_only_velocity}}. The two sensor setups are visualized in \figref{fig:sensor_setup}. 

\textcolor{black}{We would like to point out that our flow reconstruction framework has been designed to be trained with three-dimensional high-fidelity computational data. Experimental three-dimensional data (such as 3D PIV) could also potentially be used in the future for the training of our framework.}

\section{Model architecture and training}
\label{sec:model}

To carry out 3D flow reconstruction, we use a convolutional autoencoder-based architecture, dubbed FR3D. It consists of three parts: an encoder, a decoder and a latent embedder. The encoder 
\begin{equation*}
    \mathcal{E}: \mathbb{R}^{I} \rightarrow \mathbb{R}^{E}
\end{equation*}
compresses its input $\mathbf{x}$ into a latent space embedding $\mathbf{l}$, where $I << E$. Its architecture is a typical convolutional encoder architecture; each block consists of an initial convolution to double the number of channels, followed by sub-blocks applying batch normalization \cite{batchnorm}, a further convolution and a residual (skip) connection \cite{resnet} to the beginning of the sub-block, and finally downsampling via average pooling. Our setup involved four such blocks, with four sub-blocks per block. The activation function used for all intermediate convolutional layers is the leaky rectified linear unit (LeakyReLU) 
\begin{equation*}
    \mathrm{LeakyReLU}(x) = \max(\alpha x, x).
\end{equation*}

After the encoder, the decoder
\begin{equation*}
    \mathcal{D}: \mathbb{R}^E \rightarrow \mathbb{R}^I
\end{equation*}
de-compresses $\mathbf{l}$ into an approximation $\hat{\mathbf{x}}$ of $\mathbf{x}$. The decoder is similar to the encoder in structure, except the downsampling operations are replaced with upsampling operations via transpose convolutions.

\begin{figure}
    \centering
    \includegraphics[width=\textwidth]{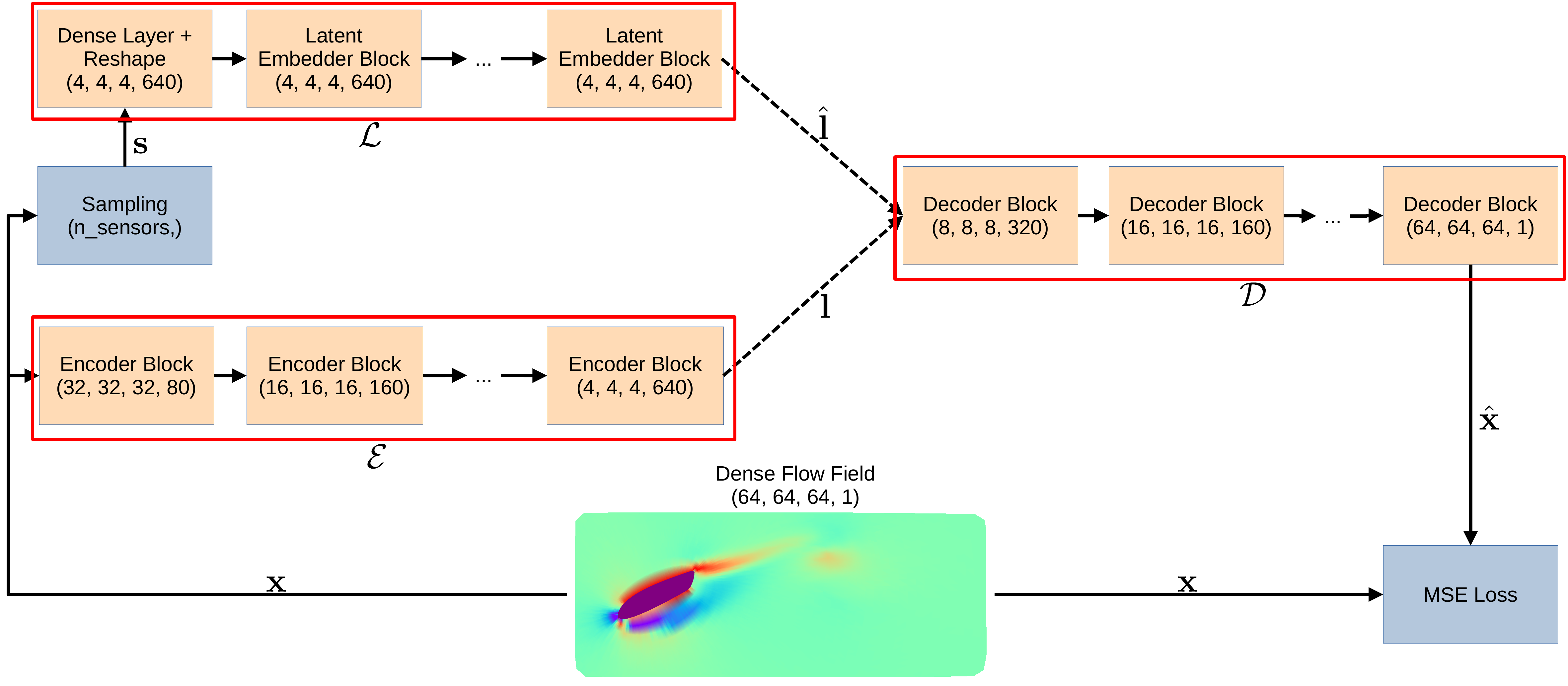}
    \caption{Our model architecture for 3D flow reconstruction. The integer tuples in each block indicate the shape of the tensor outputted by that block. The first three dimensions correspond to the spatial dimensions, and the final dimension is the channel dimension.}
    \label{fig:model}
\end{figure}

In order to make the autoencoder useful for the flow reconstruction task, a further submodel
\begin{equation*}
    \mathcal{L}: \mathbb{R}^S \rightarrow \mathbb{R}^E
\end{equation*}
is necessary, which we call the latent space embedder. The latent space estimates latent space embeddings $\hat{\mathbf{l}}$ from the the sensor inputs $\mathbf{s}$, which may be then used to compute $\hat{\mathbf{x}}$. It consists of a dense layer and several convolutional layers. The overall FR3D architecture is summarized in \figref{fig:model}, and the parameter counts are displayed in \tabref{tab:parameters}.  

{The choice of architecture was motivated by two factors. First, a convolutional approach was adopted due to such architectures' strong relative performance in previous multi-geometry FR scenarios \cite{our_journal_article}. The autoencoder structure, meanwhile, was chosen to control the number of parameters in the fully connected layer ingesting the sensor inputs; many NN architectures used for FR incorporate a fully connected layer which accepts the low-dimensional sensor inputs and outputs a vector with dimensionality equal to the high-fidelity field. As the number of parameters of a fully connected layer scales linearly with the dimensionality of the output, and considering that the high-fidelity fields in 3D FR  have a higher dimension compared to more commonly studied 2D FR scenarios, the autoencoder approach can substantially cut the number of parameters in the fully connected layers by constraining their outputs to the autoencoder's latent space.}

\begin{table}[]
\caption{Parameter counts of the overall FR3D model and its submodels in the sparse sensor configuration.}
\label{tab:parameters}
\centering
\begin{tabular}{@{}lcc@{}}
\toprule
                                & Submodel parameters & Total parameters             \\ \midrule
Encoder         & 124,910,960         & \multirow{3}{*}{402,163,801} \\
Decoder         & 128,917,481         &                              \\
Latent embedder & 148,335,360         &                              \\ \bottomrule
\end{tabular}
\end{table}

The training of the autoencoder was conducted using the $\textproc{Adam}$ optimization algorithm with an initial learning rate of $10^{-4}$, {using the mean squared error (MSE) as the loss function $L$. An alternative approach using a generative adversarial network (GAN) \cite{gan} instead of the more traditional MSE loss was also explored, however it was discarded in favor of the MSE loss function due to exhibiting lower accuracy, particularly for the $v$-velocity. \ref{sec:fr3d:gan} provides more information about the GAN approach.} 

Optimization of the FR3D model using the MSE loss is done as follows: first, for each batch in the dataset, an optimization step is taken for the weights of $\mathcal{E}$ and $\mathcal{D}$. Then, the weights of the decoder are 'frozen' and a further optimization step is taken for $\mathcal{L}$. This procedure, involving the dense field $\mathbf{x}$, the sensor inputs $\mathbf{s}$ and the weights $\mathbf{w}_E, \mathbf{w}_D, \mathbf{w}_L$ of $\mathcal{E}, \mathcal{D}, \mathcal{L}$ respectively, is summarized in \algoref{algo:fr3d:training}.

\begin{algorithm}
\begin{algorithmic}[1]
\Function{Optimize}{$\mathbf{w}_E,\mathbf{w}_D,\mathbf{w}_L,\mathbf{s},\mathbf{x}$}
\State $\mathbf{l} := \mathcal{E}(\mathbf{x}, \mathbf{w}_E)$
\State $\hat{\mathbf{x}} := \mathcal{D}(\mathbf{l}, \mathbf{w}_D)$
\State $\mathbf{w}_E \leftarrow \textproc{Adam}(\nabla_{\mathbf{w}_E} L(\mathbf{x}, \hat{\mathbf{x}}), \mathbf{w}_E)$
\State $\mathbf{w}_D \leftarrow \textproc{Adam}(\nabla_{\mathbf{w}_D} L(\mathbf{x}, \hat{\mathbf{x}}), \mathbf{w}_D)$
\State $\hat{\mathbf{l}} := \mathcal{L}(\mathbf{s}, \mathbf{w}_L)$
\State $\hat{\mathbf{x}} \leftarrow \mathcal{D}(\hat{\mathbf{l}}, \mathbf{w}_D)$
\State $\mathbf{w}_L \leftarrow \textproc{Adam}(\nabla_{\mathbf{w}_L} L(\mathbf{x}, \hat{\mathbf{x}}), \mathbf{w}_L)$
\EndFunction
\end{algorithmic}
\caption{Optimization of the FR3D model using a single batch of data}
\label{algo:fr3d:training}
\end{algorithm}

The models and the training procedure were implemented using Tensorflow \cite{tensorflow} version 2.9 running on a server with two Nvidia A100 40GB GPUs and an AMD EPYC 7443 24-core CPU. The number of training epochs was determined using an early stopping mechanism based on the validation loss, which automatically stops training when the validation loss level does not decline after a set number of epochs. To expedite the training process, once the weights of $\mathcal{E}$ and $\mathcal{D}$ were obtained during training for the sparse sensor setup, they were reused for the plane sensor setup. The procedure was carried out separately for all flow variables (pressure $p$ and velocity components $u,v,w$), each taking approximately 48 hours to converge.

\section{Results}
\label{sec:results}

Below, we present the results obtained by applying the FR3D model trained using the procedure in \secref{sec:model} to the validation dataset, consisting of 20 geometries not encountered during training. First, we display the results obtained for reconstructing the pressure and velocity from sparse sensors in \secref{sec:results:sparse}, with Q-criterion contours also accurately reconstructed from the predicted velocity fields. Next, we demonstrate that the FR3D model can also be extended to estimate pressure fields from velocity data sampled on perpendicular planes in \secref{sec:results:planes}. Finally, we demonstrate that the pressure and velocity predictions from the FR3D model can be used to accurately estimate the time evolution of the drag and lift coefficients in \secref{sec:results:clcd}. 

\subsection{Reconstruction from sparse sensors}
\label{sec:results:sparse}

We begin our analysis of the results using the sparse sensor case. \tabref{tab:sparse-error} provides an overview of the model's overall performance via the mean absolute percentage error (MAPE) and mean squared error (MSE) metrics averaged over the entire validation dataset. To focus on the regions of highest interest in the domain, the error metrics are computed for sampling points inside a box with extents $[-L_m, 8L_m/3] \times [-L_m, L_m]$ in the $x$- and $y$-directions relative to the centroid of each object and covering the entire spanwise direction. 
\begin{table}[h!]
\caption{Mean absolute percentage (MAPE) and mean squared (MSE) error levels achieved by the FR3D model on the validation dataset for reconstruction from sparse sensors. The "Input to $\mathcal{D}$" column refers to whether the model was run with the sensors (using the latent embedder submodel $\mathcal{L}$, inference configuration) or the ground truth field (using the encoder submodel $\mathcal{E}$) as the inputs.}
\label{tab:sparse-error}
\begin{tabular}{@{}llcccc@{}}
\toprule
Var.                  & Input to $\mathcal{D}$ & MAPE \tablefootnote{Percentage error figures are filtered to remove gridpoints with ground truth values less than 2\% of the maximum absolute ground truth value in a snapshot}    & Min-max MAPE \tablefootnote{"Min-max" refers to error figures with min-max normalization applied based on the ground truth field.} & MSE                   & Min-max MSE           \\ \midrule
\multirow{2}{*}{$p$} & $\mathcal{L}$   & 9.50\%  & 6.33\%       & $6.21 \times 10^{-3}$ &  $8.76 \times 10^{-4}$ \\
                          & $\mathcal{E}$   & 3.68\%  & 2.34\%       & $1.86 \times 10^{-3}$ & $3.07 \times 10^{-4}$ \\ \midrule
\multirow{2}{*}{$u$}        & $\mathcal{L}$   & 4.41\%  & 2.68\%       & $9.06 \times 10^{-3}$ & $2.20 \times 10^{-4}$ \\
                          & $\mathcal{E}$   & 2.59\%  & 1.36\%       & $2.69 \times 10^{-3}$ & $6.37 \times 10^{-5}$ \\ \midrule
\multirow{2}{*}{$v$}        & $\mathcal{L}$   & 16.33\% & 3.07\%       & $6.31 \times 10^{-3}$ & $1.99 \times 10^{-4}$ \\
                          & $\mathcal{E}$   & 11.96\% & 1.88\%       & $2.08 \times 10^{-3}$ & $6.52 \times 10^{-5}$ \\ \midrule
\multirow{2}{*}{$w$}        & $\mathcal{L}$   & 35.06\% & 3.52\%       & $3.82 \times 10^{-3}$ & $5.15 \times 10^{-4}$ \\
                          & $\mathcal{E}$   & 33.75\% & 2.69\%       & $2.21 \times 10^{-3}$ & $2.97 \times 10^{-4}$ \\  \bottomrule
\end{tabular}
\end{table}

The error figures show that, overall, our model generalizes well to the validation set. The raw MAPE values for $p$ and $u$ are both below 10\% for previously unseen geometries, which is in line with our previous work with 2D geometries \cite{our_tsfp_article, our_journal_article}, despite the substantially greater challenge of 3D flows which contain more complicated structures orientated in various directions. In terms of absolute errors and normalized percentage errors, the predictions for $v$ and $w$ are also at a similar level of accuracy. However, we draw attention to the large discrepancy between MAPE and Min-max MAPE figures for $v$ and $w$. This is caused by the fact that, due to our choice of boundary conditions, the mean ground truth values for $p$ and $u$ are distributed around a mean of 1, while those of $v$ and $w$ are distributed around 0. Due to this, though the absolute error levels (i.e. the numerator of the percentage error expression) are broadly similar for all four variables, the percentage error metrics for $v$ and $w$ are much higher since the denominator of the percentage error expression is much smaller for those variables.

\begin{figure}
    \centering
    \includegraphics[width=0.6\textwidth]{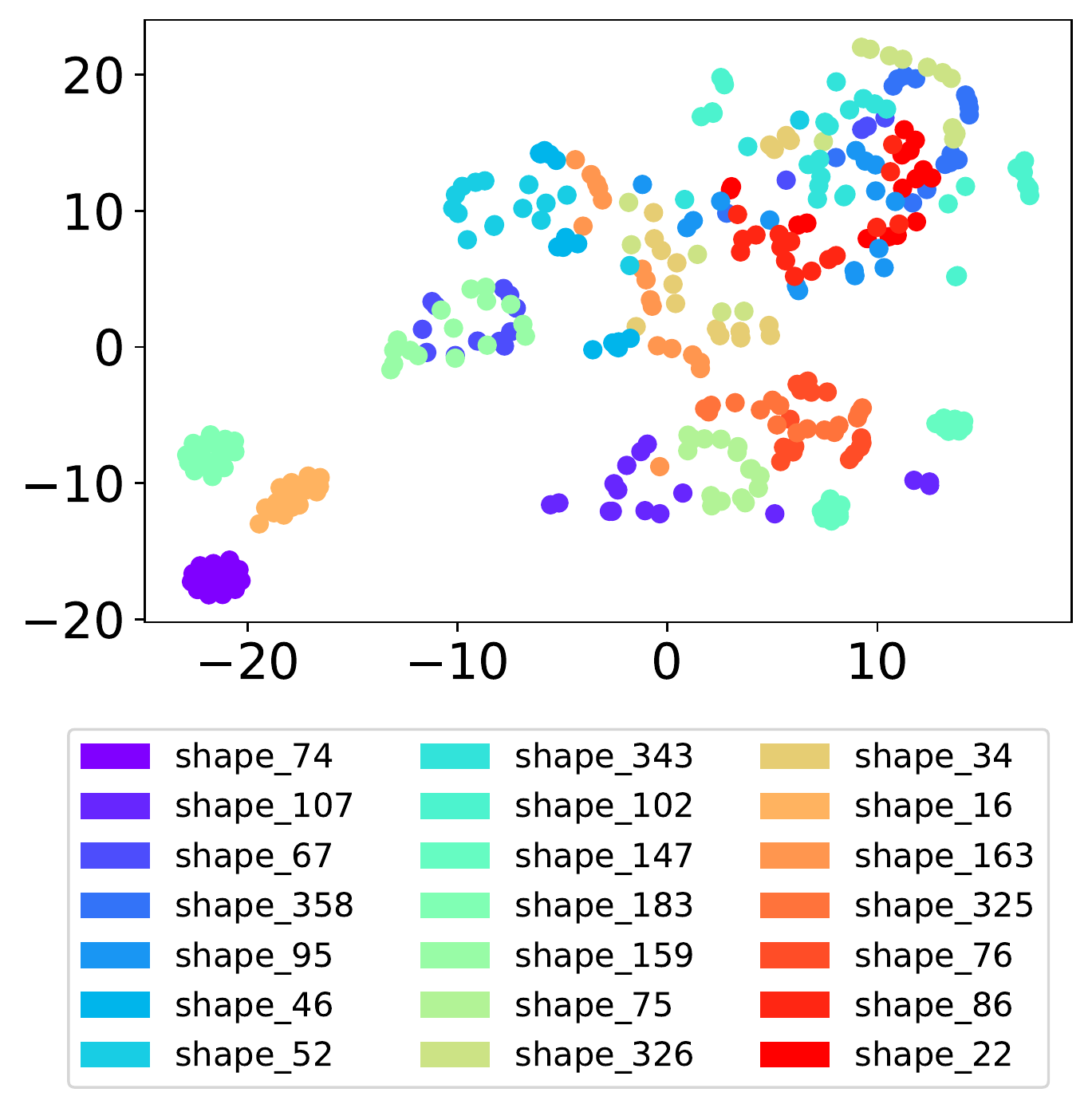}
    \caption{{t-SNE visualization of the FR3D model's latent space colored by geometry, generated using the validation data.}}
    \label{fig:fr3d:tsne}
\end{figure}

{Further evidence displaying the FR3D model's generalization performance can be seen in \figref{fig:fr3d:tsne}, which visualizes the (encoder-computed) latent space vectors associated with validation snapshots via the t-SNE \cite{tsne} method. The latent space vectors show that FR3D's latent space is able to cluster the snapshots associated with different geometries together even for geometries unseen during training, which suggests that the latent space is well-conditioned for multi-geometry flow reconstruction. }

\begin{figure}[h!]
    \centering
    \includegraphics[width=0.46\textwidth]{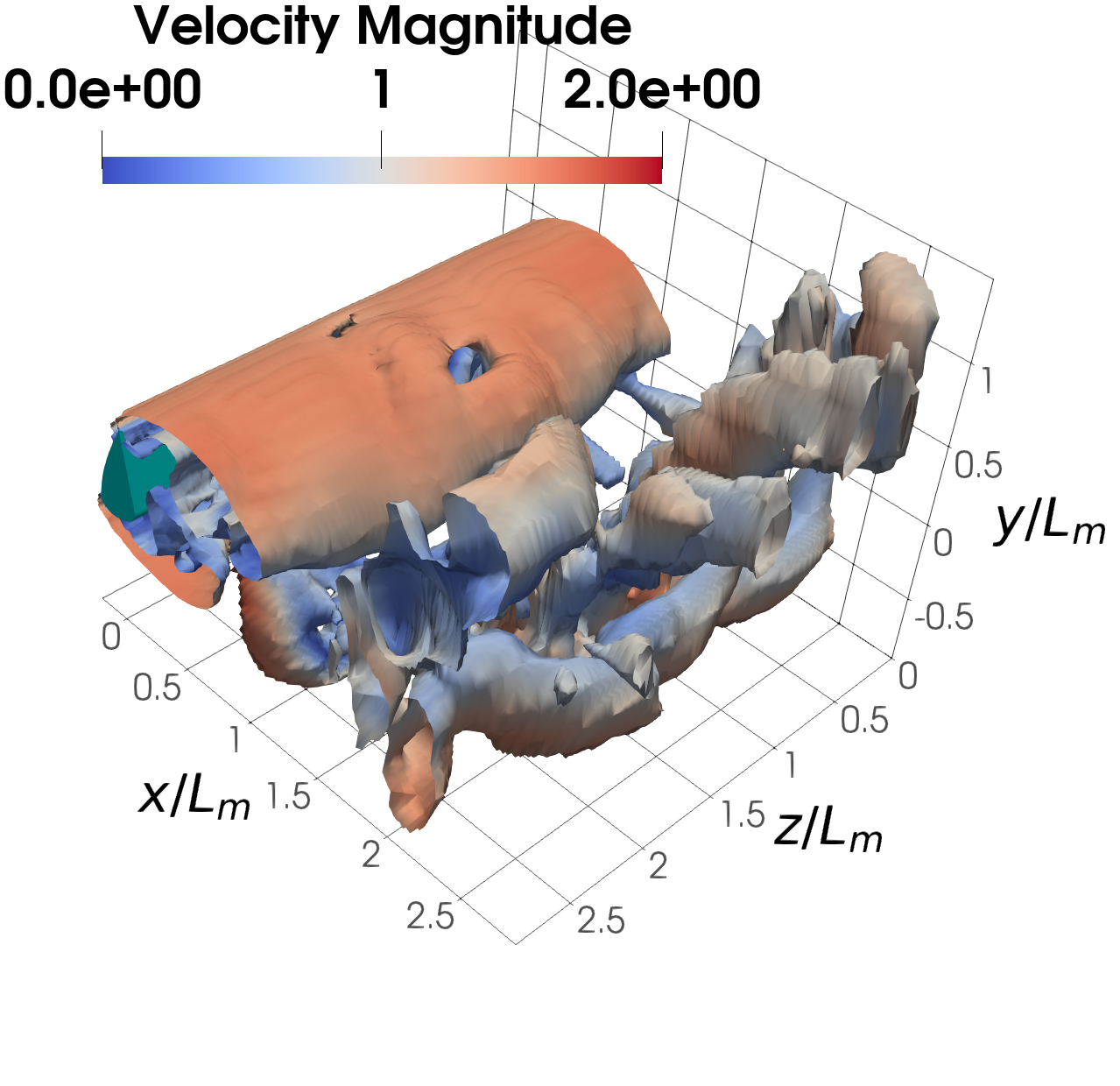}
    \includegraphics[width=0.46\textwidth]{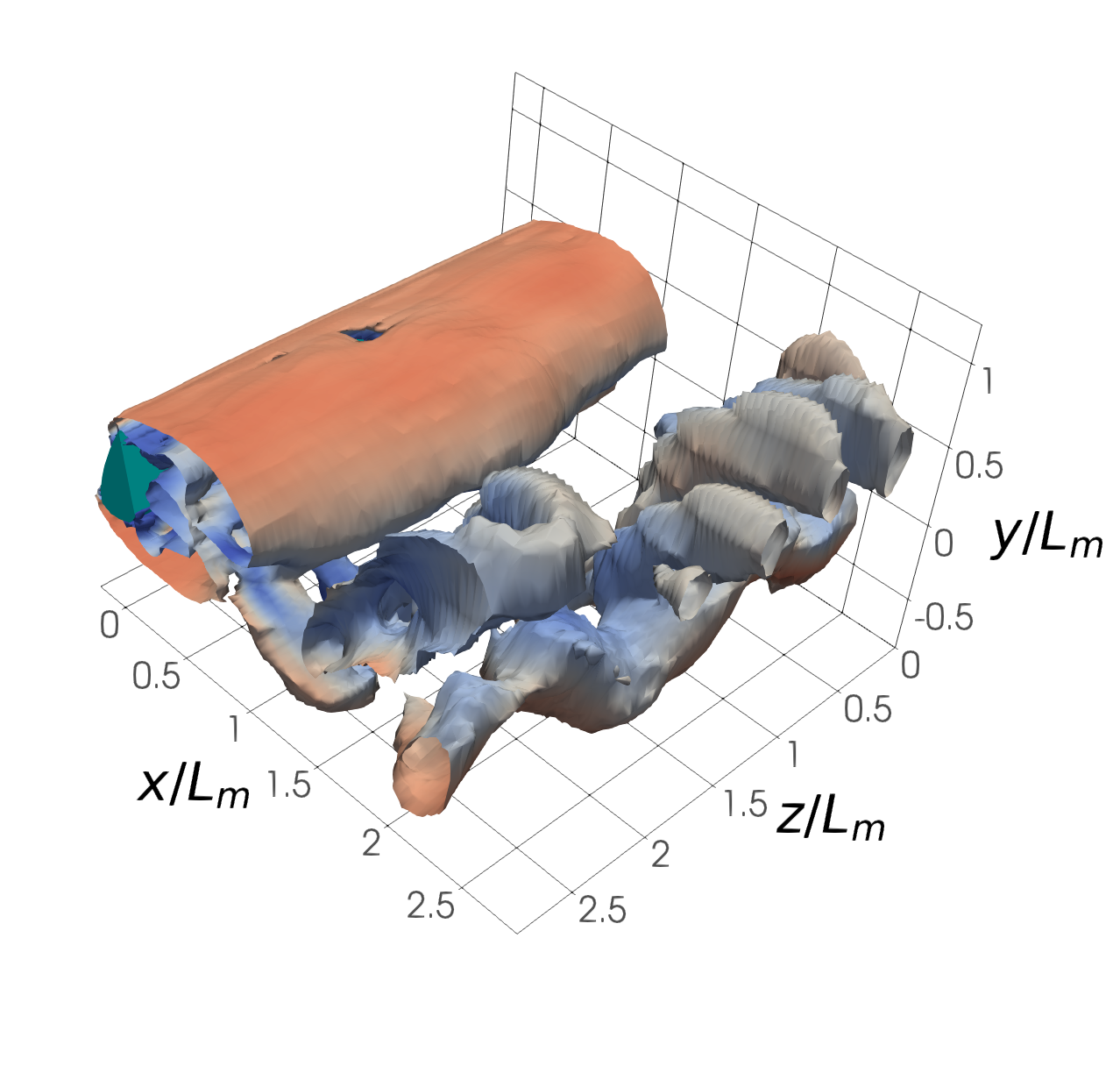}
    \includegraphics[width=0.46\textwidth]{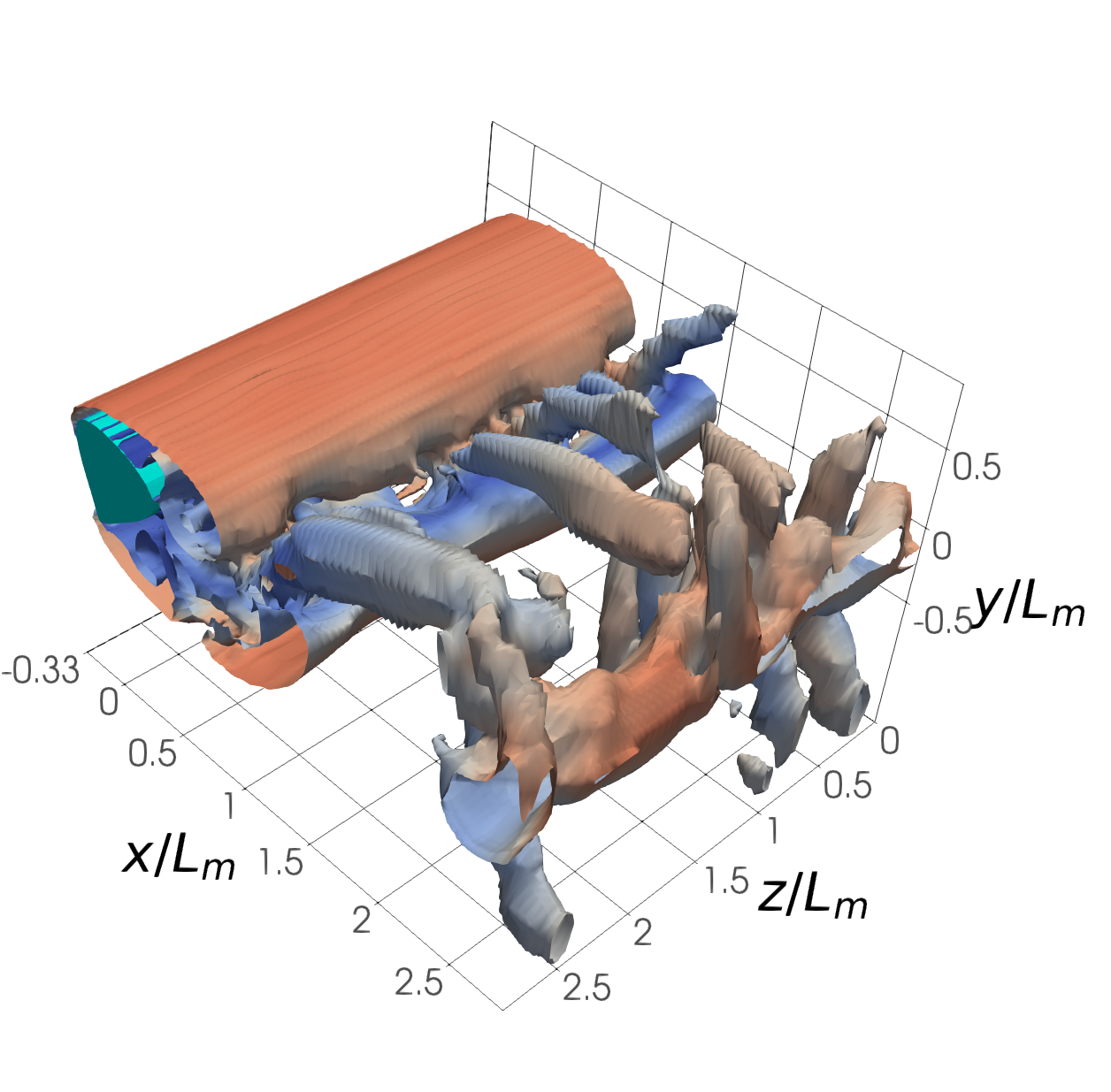}
    \includegraphics[width=0.46\textwidth]{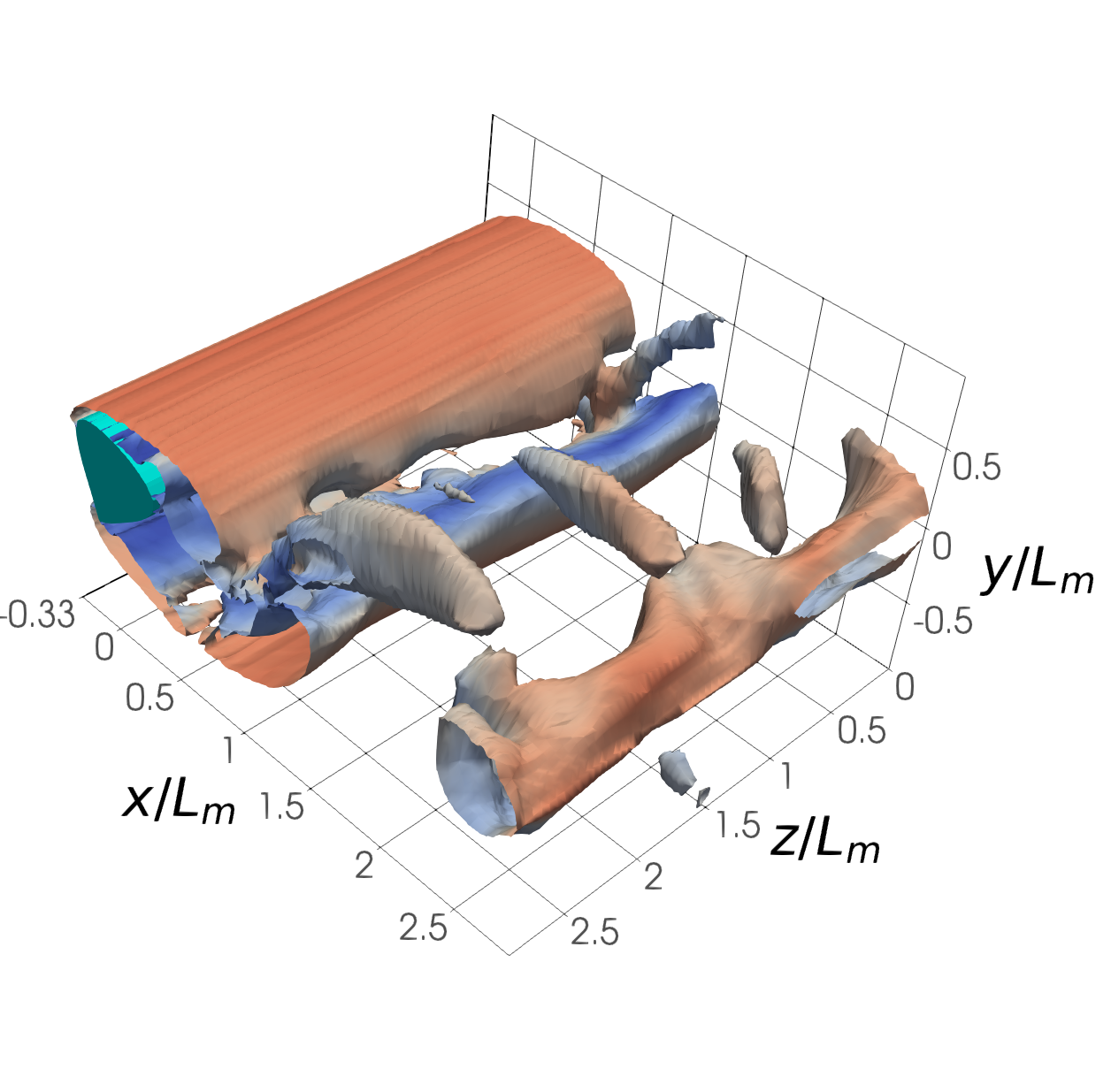}
    \includegraphics[width=0.46\textwidth]{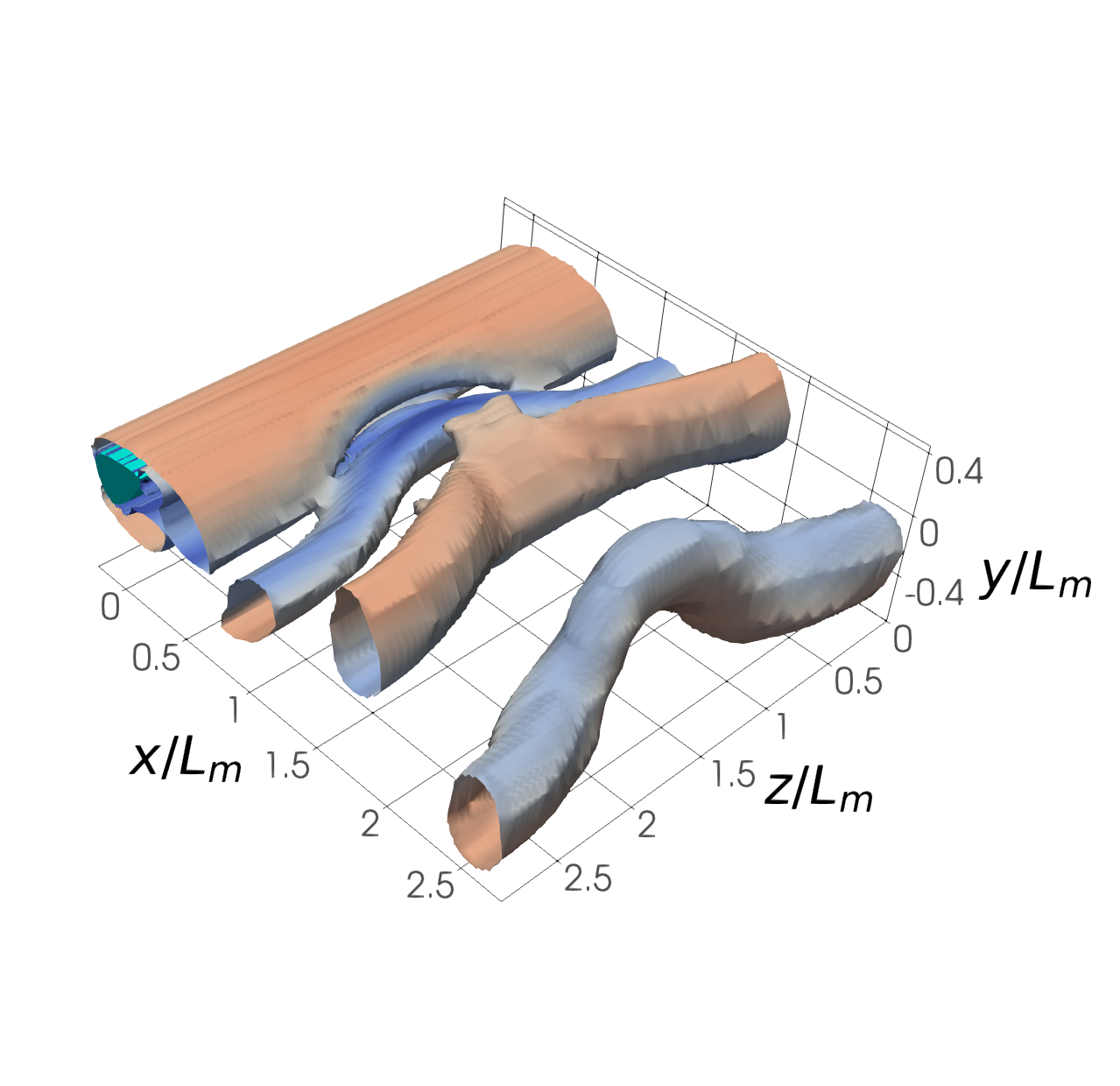}
    \includegraphics[width=0.46\textwidth]{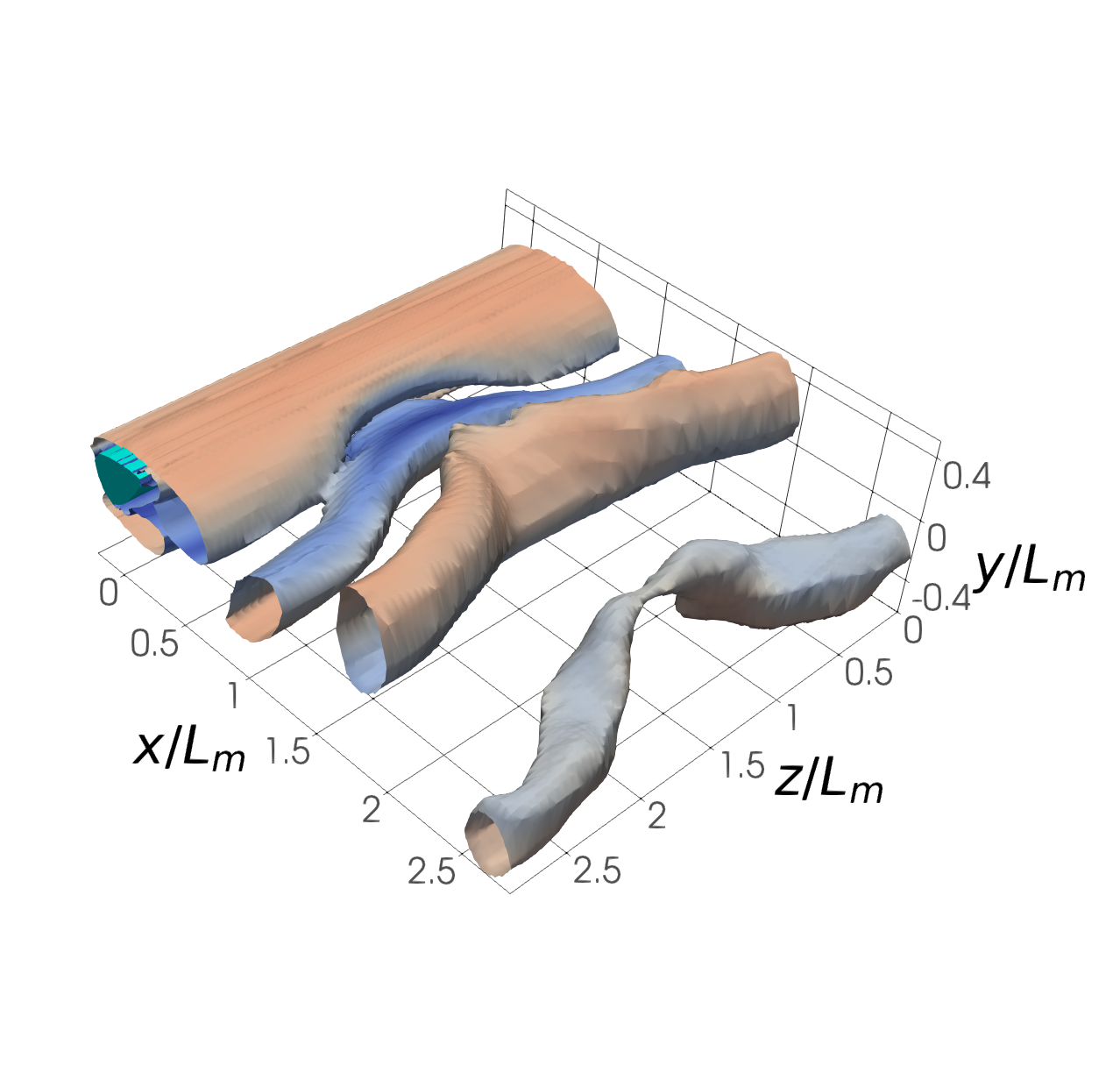}
    \caption{Ground truth (left) and predicted (right) $Q$-criterion $(Q=0.05)$ iso-contours from three randomly chosen snapshots belonging to different geometries.}
    \label{fig:sparse-qcrit}
\end{figure}

\begin{figure}[h!]
    \centering
    \includegraphics[width=0.46\textwidth]{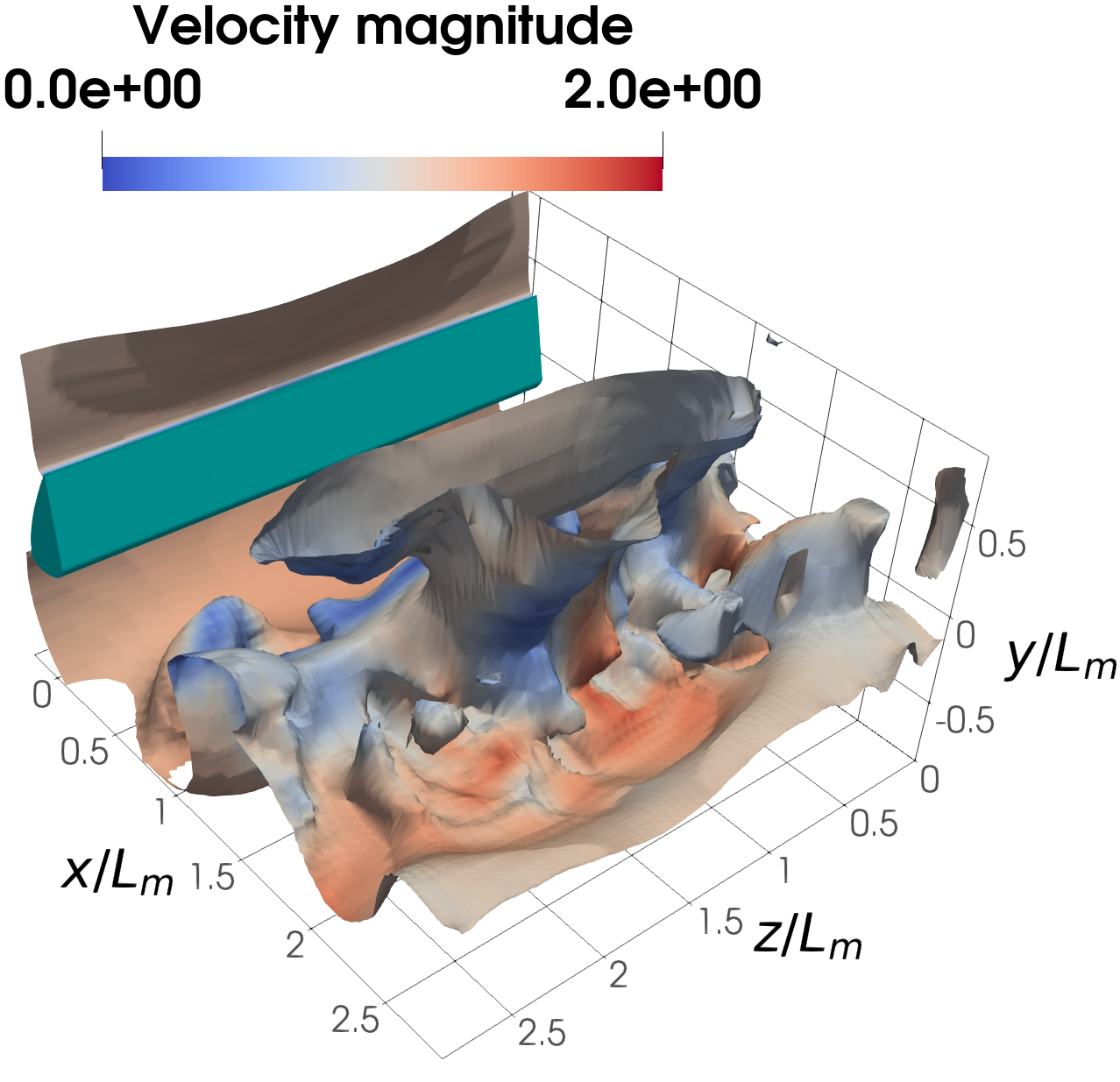}
    \includegraphics[width=0.46\textwidth]{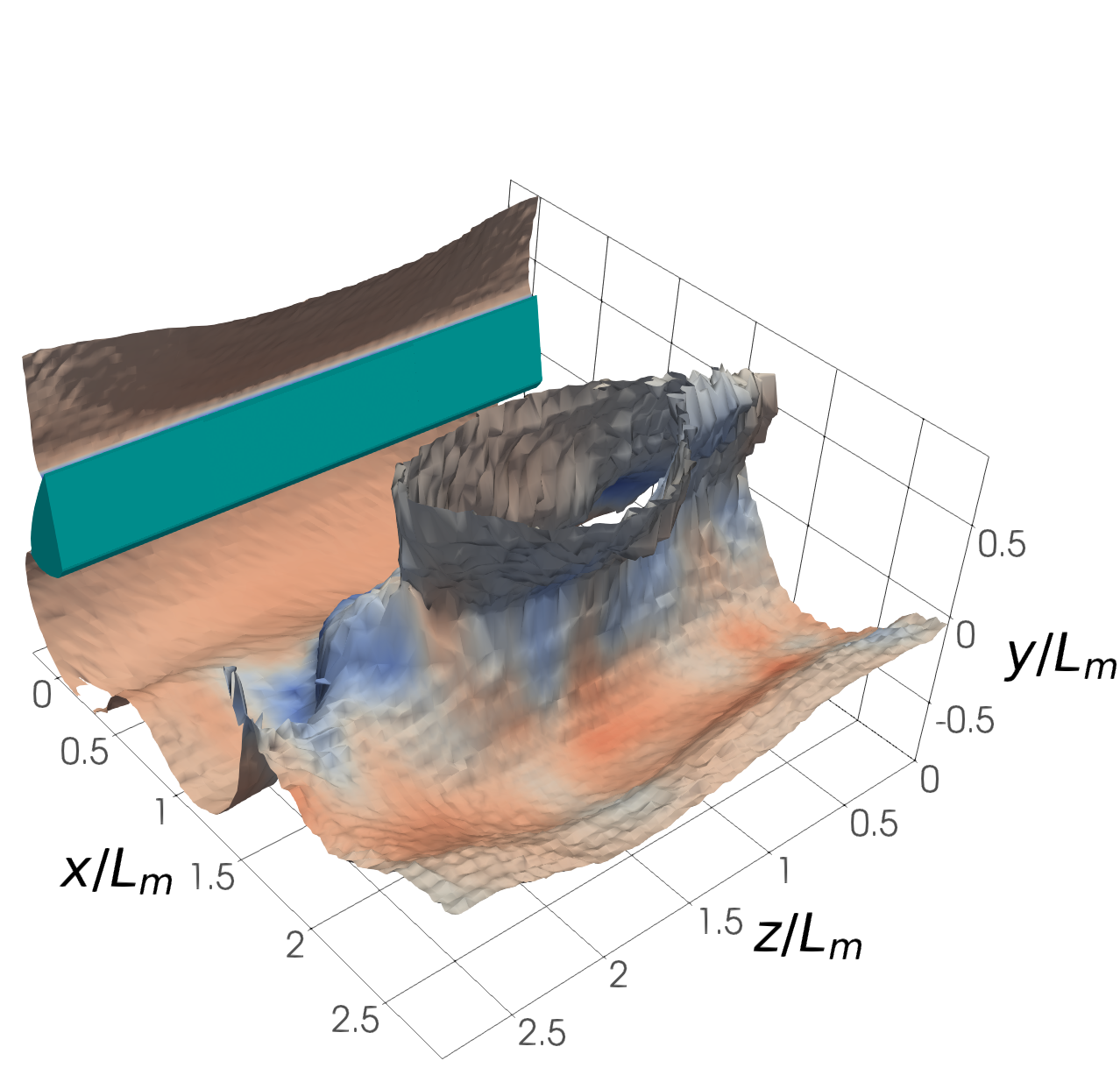}
    \includegraphics[width=0.46\textwidth]{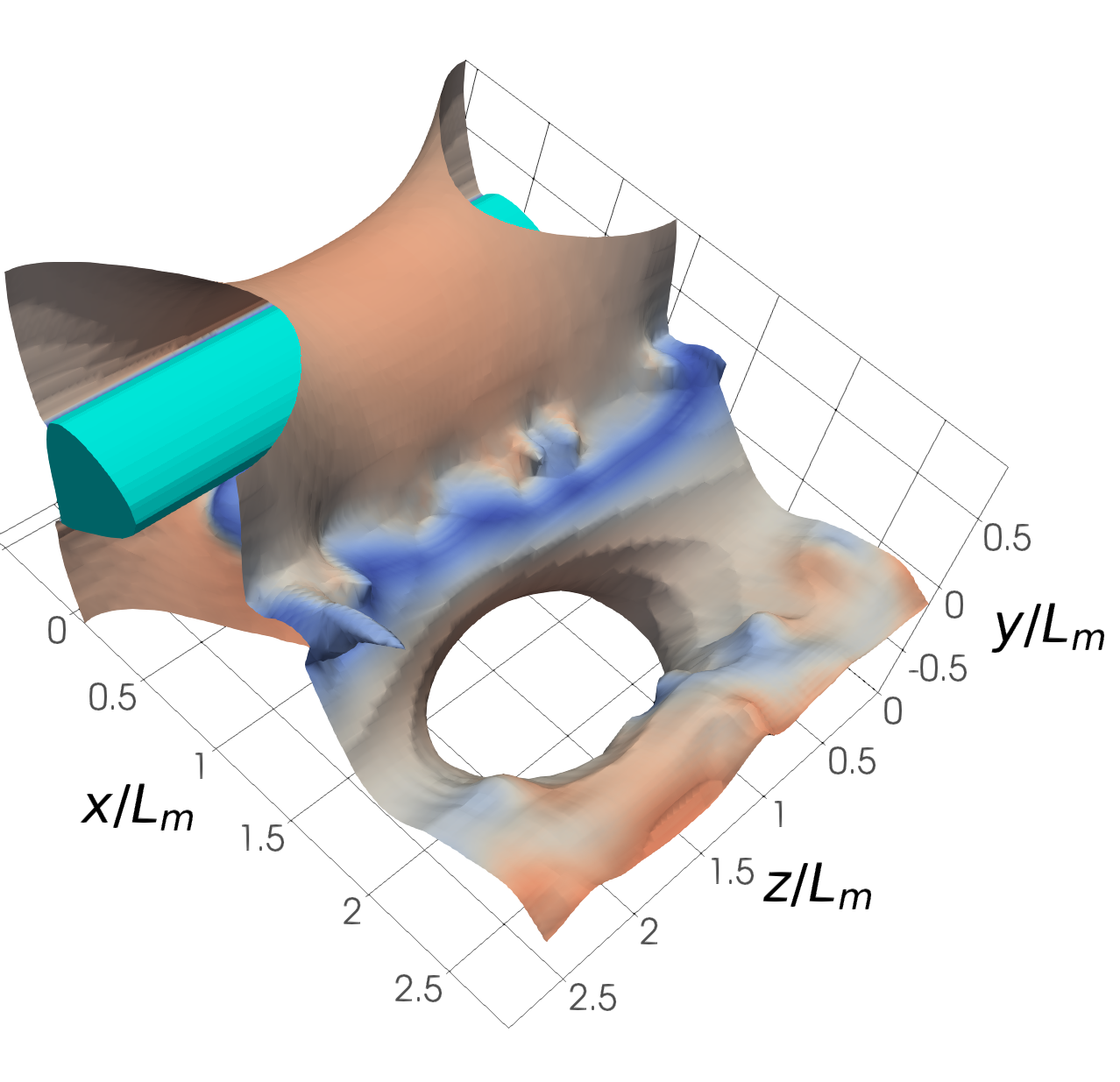}
    \includegraphics[width=0.46\textwidth]{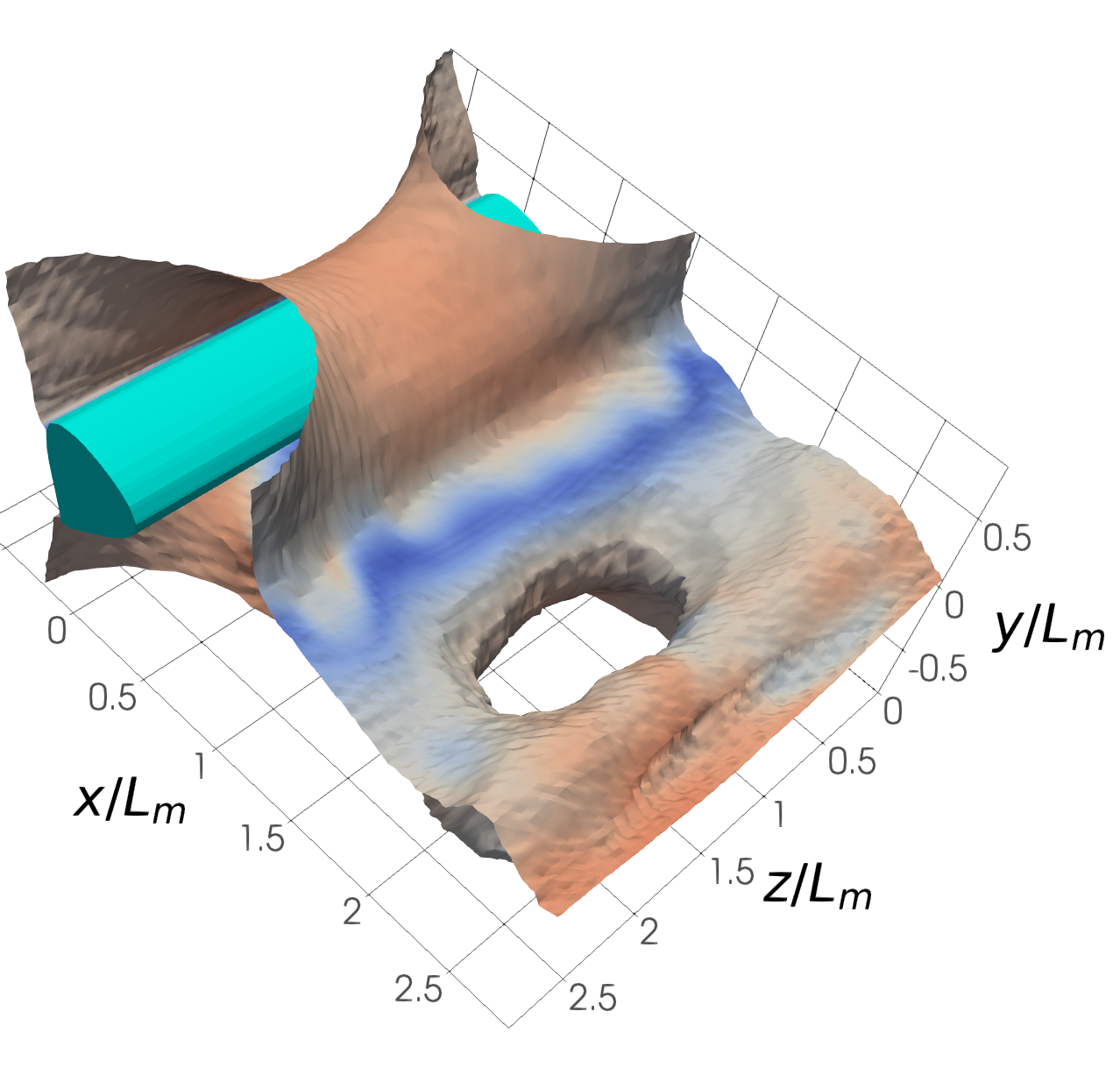}
    \includegraphics[width=0.46\textwidth]{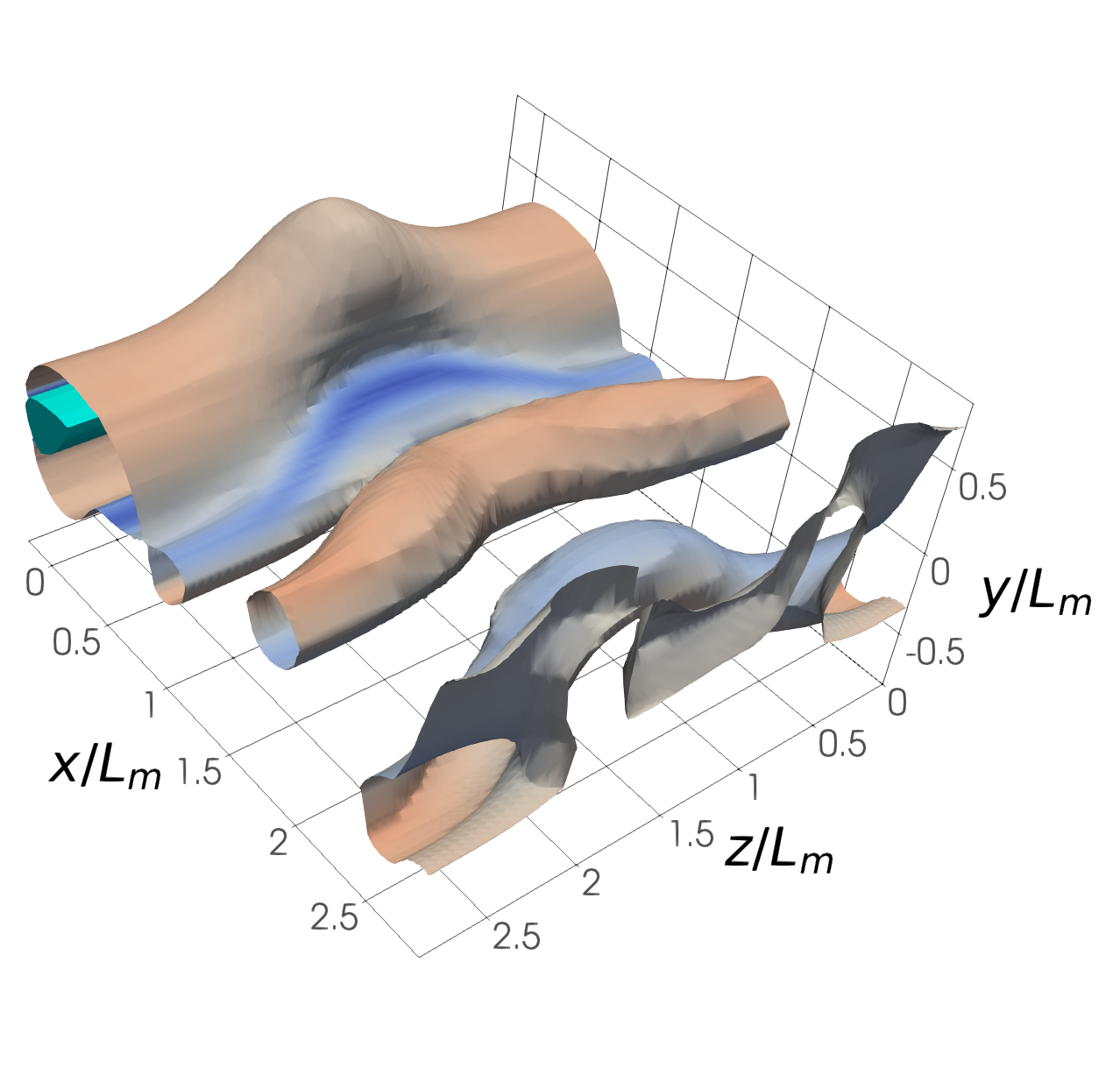}
    \includegraphics[width=0.46\textwidth]{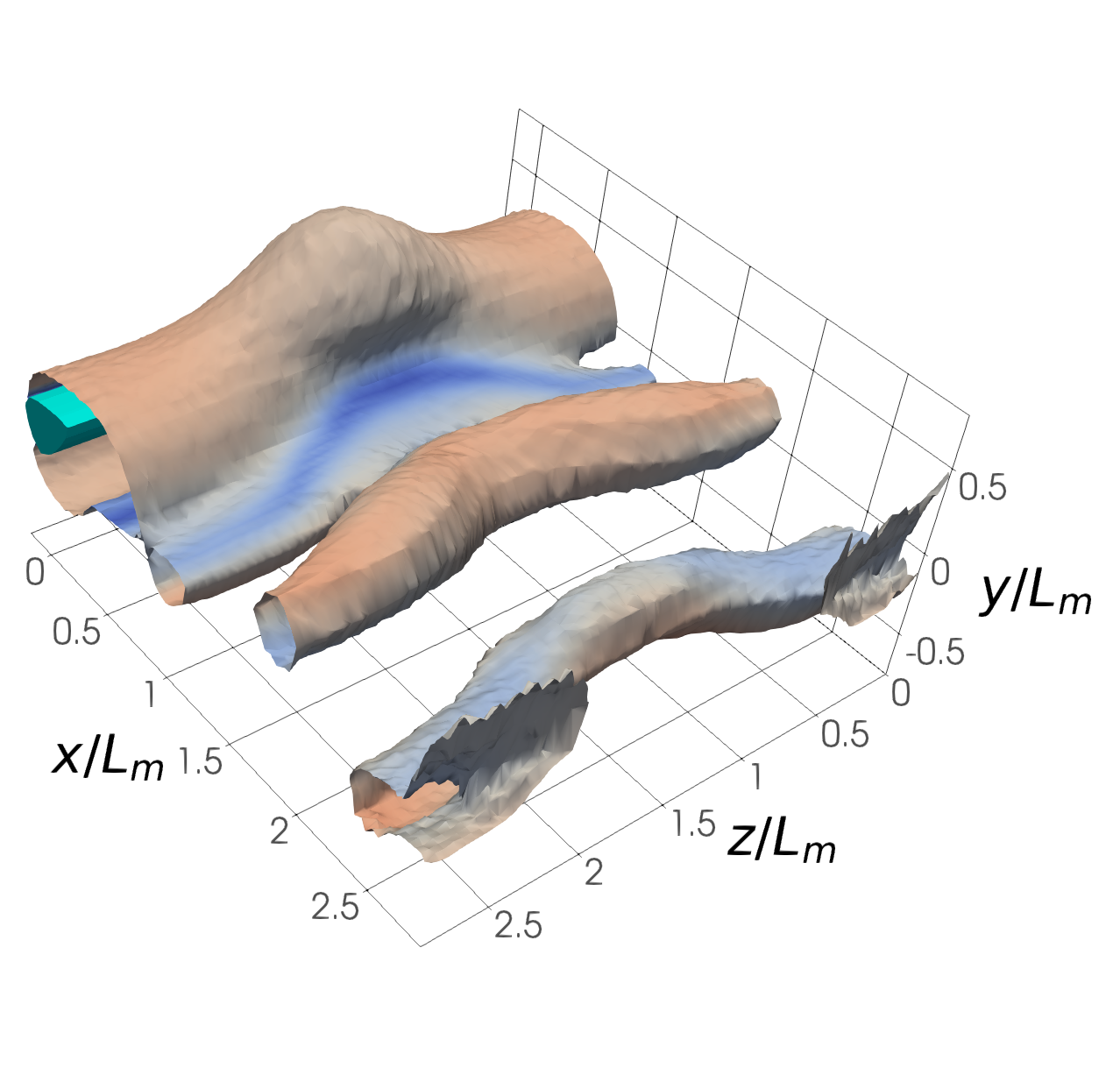}
    \caption{Ground truth (left) and predicted (right) pressure (top, bottom: $p=0.85$, middle: $p=0.75$) iso-contours coloured by velocity magnitude from three randomly chosen snapshots belonging to different geometries, using the model trained on the sparse sensor setup.}
    \label{fig:sparse-p}
\end{figure}

\begin{figure}[h!]
    \centering
    \includegraphics[height=0.23\textheight]{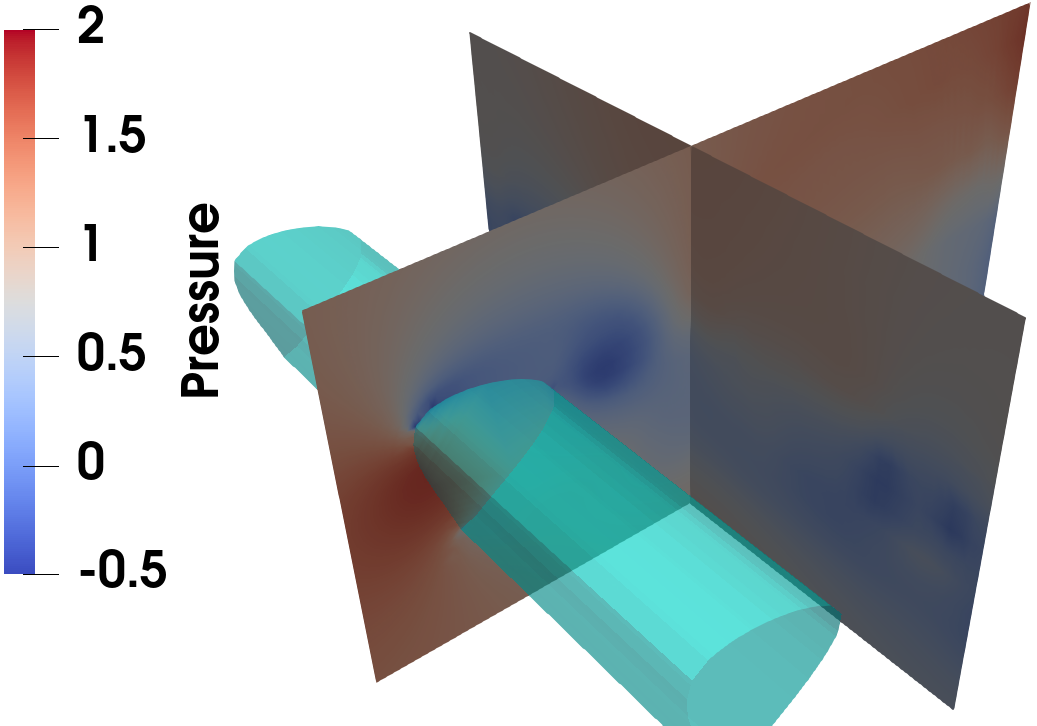}
    \includegraphics[height=0.23\textheight]{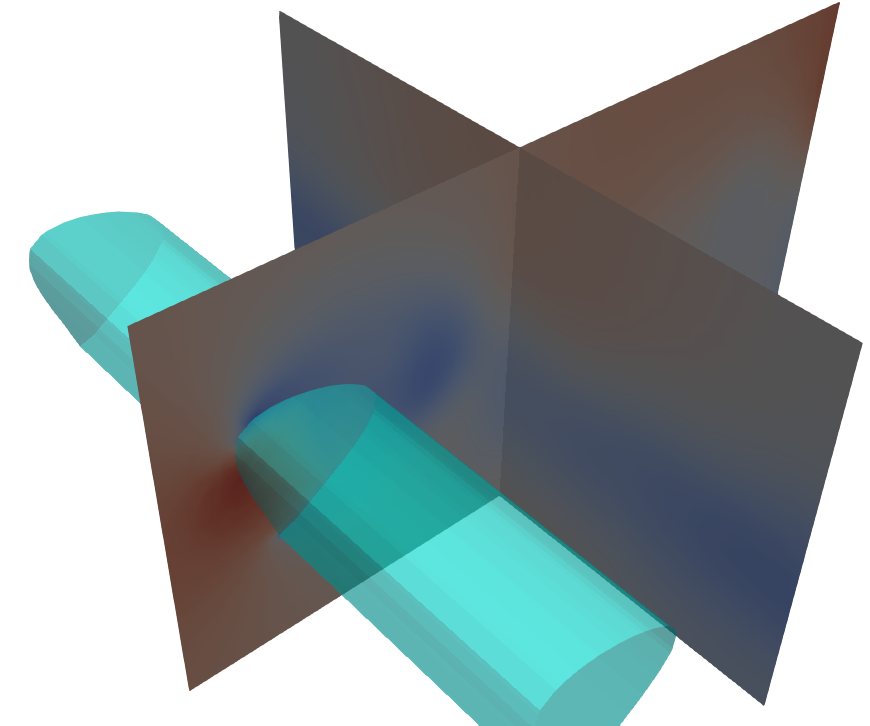}
    \includegraphics[height=0.23\textheight]{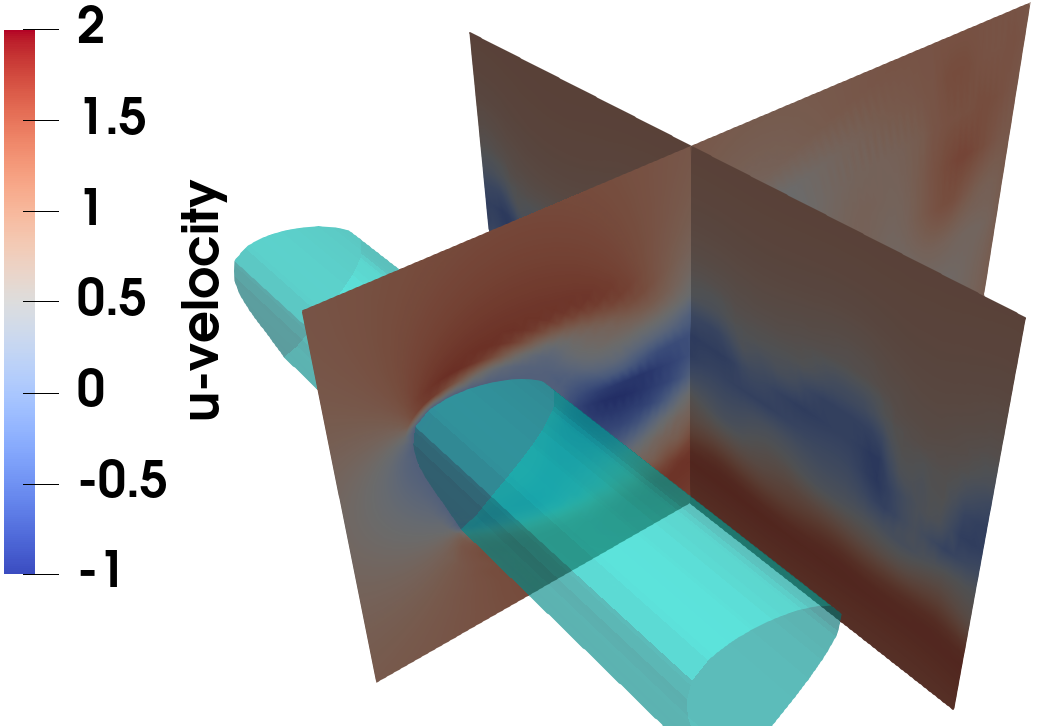}
    \includegraphics[height=0.23\textheight]{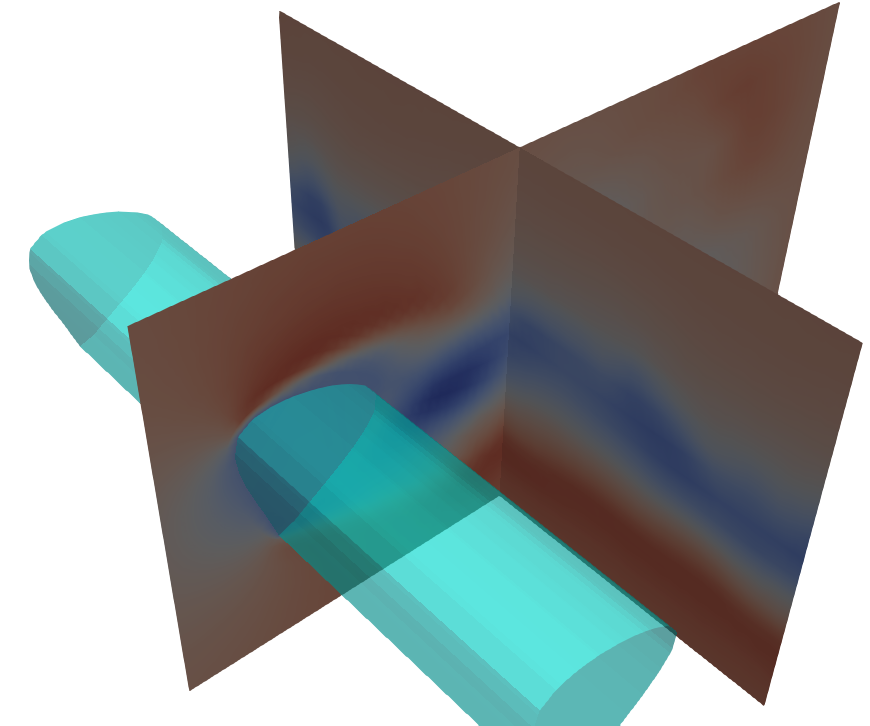}
    \includegraphics[height=0.23\textheight]{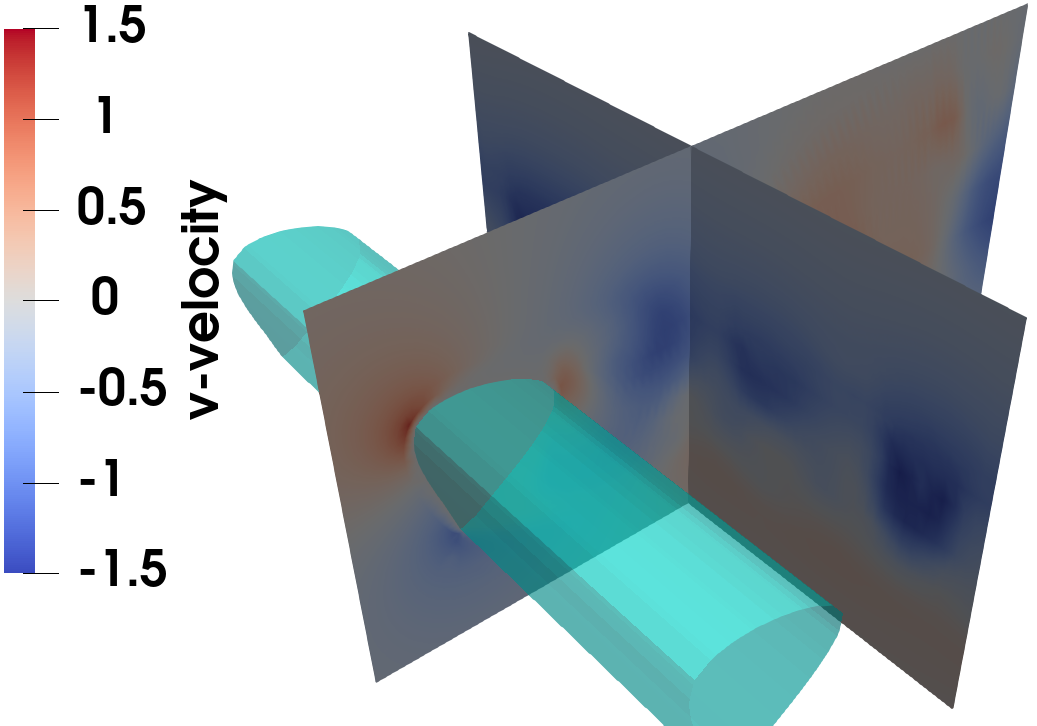}
    \includegraphics[height=0.23\textheight]{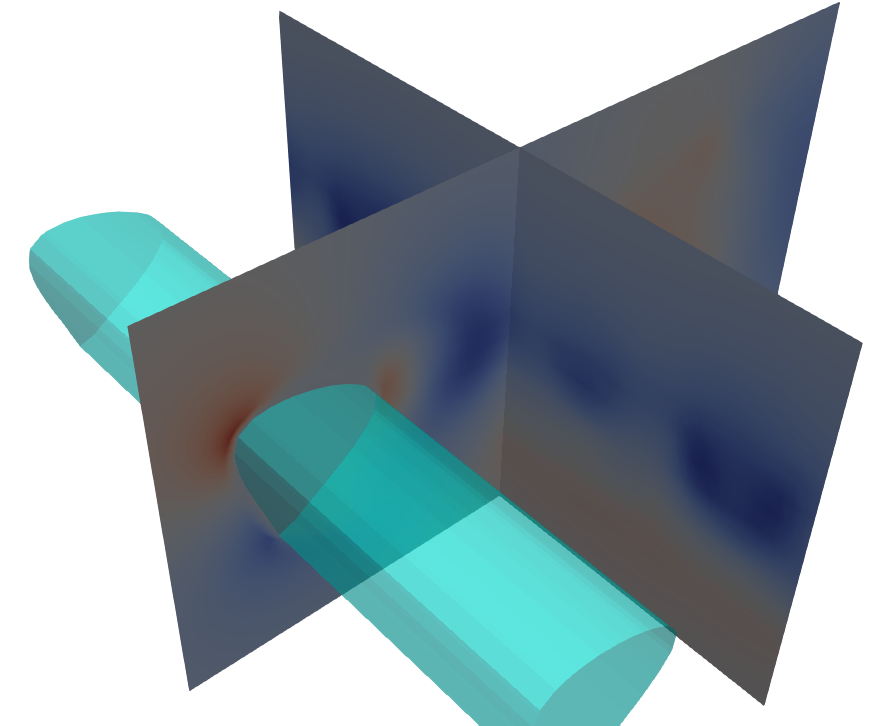}
    \includegraphics[height=0.23\textheight]{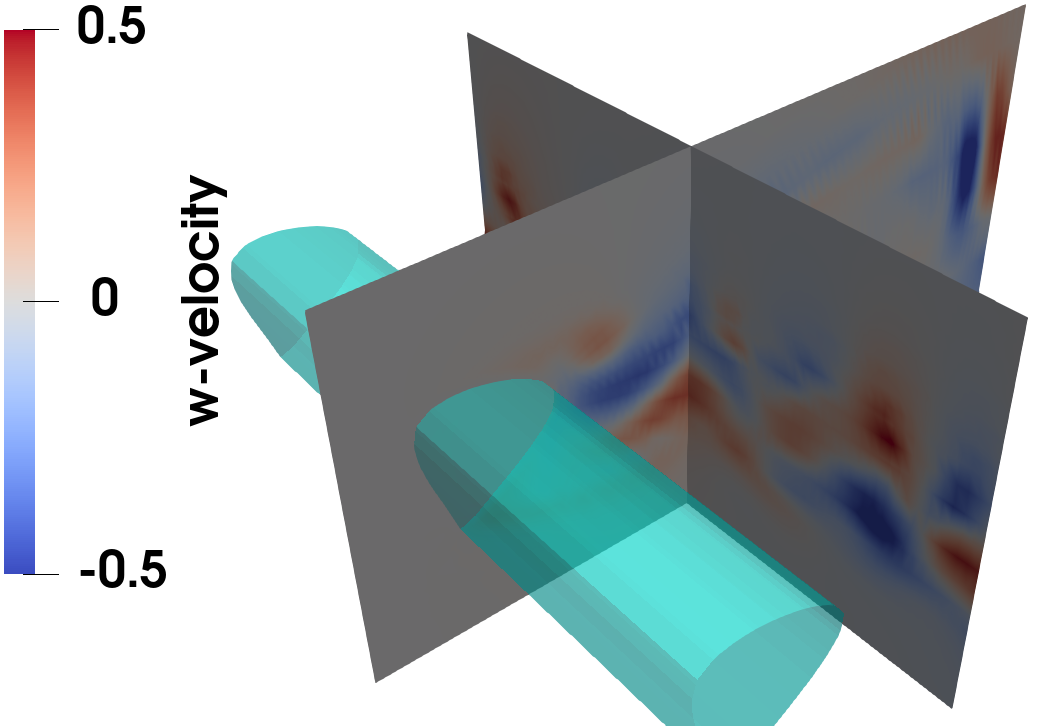}
    \includegraphics[height=0.23\textheight]{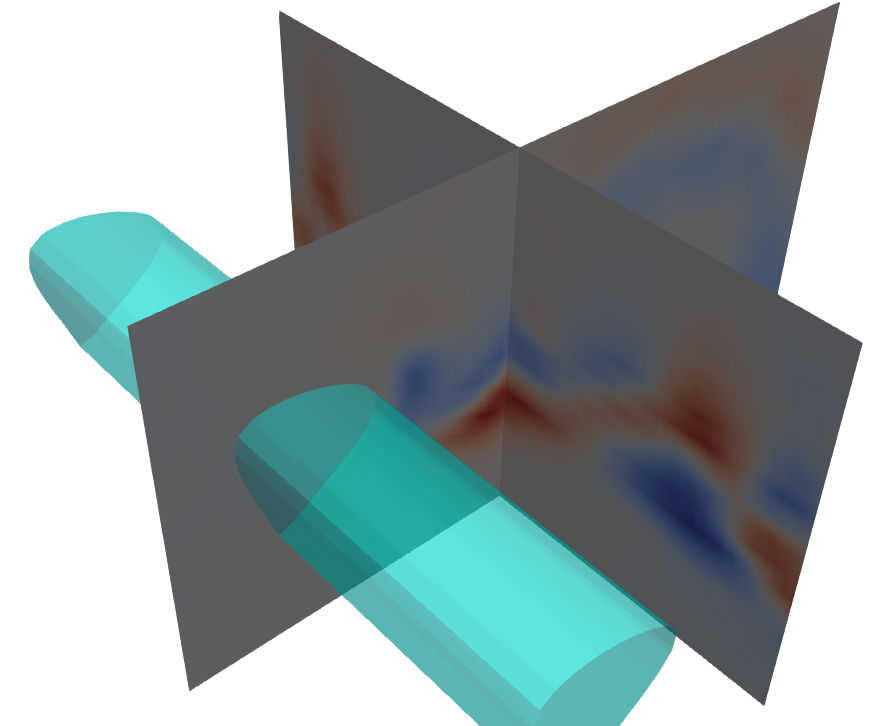}
    \caption{Ground truth (left) and predicted (right) slices of pressure and velocity fields for the middle snapshot in Figures \ref{fig:sparse-qcrit} and \ref{fig:sparse-p}.}
    \label{fig:sparse-slices}
\end{figure}

Figures \ref{fig:sparse-qcrit}, \ref{fig:sparse-p} and \ref{fig:sparse-slices} provide further qualitative insights into the strong performance of the model via iso-contours of the Q-criterion, iso-contours of the pressure, and slices of both pressure and velocity components (respectively), for three randomly chosen snapshots exhibiting varying degrees of spanwise effects. 3D visualisations of these quantities from experimental data are often difficult as previously discussed -- e.g.\ the amplification of sensor noise when computing velocity gradients presents a serious challenge for plotting the Q-criterion from experimental measurements. Thus, our results lay the groundwork for a step-function improvement in visualisation of results from fluid dynamics experiments via flow reconstruction, all without the need for complicated  post-processing techniques.

The contour plots indicate that the model is able to reconstruct the key details of the flow field. {First, we draw attention to} the shed vortices downstream of the object. {In the top snapshot, the spanwise structure is placed at approximately $(2.1L_m, -0.2L_m)$ on the $xy$-plane; its shape is elliptic with a width of $~0.2L_m$ and a height of $~0.4L_m$. The middle snapshot shows two spanwise structures, one far downstream at $(2.2L_m, -0.1L_m)$ with a diameter of $~0.5L_m$, and a second newly forming one at $(L_m,-0.1L_m)$. Finally, the bottom snapshot shows three structures, placed at $(0.8L_m,-0.1L_m)$, $(1.3L_m,0.2L_m)$ and $(2.6L_m, -0.2L_m)$. The first and the third are roughly circular with $~0.25L_m$ diameters, while the middle one is elliptical with a horizontal axis of $~0.2L_m$ and a vertical axis of $~0.4L_m$. Furthermore, these structures are slightly "bent" about the middle of the domain $(z = 1.5L_m)$ in the $x-$ and $z-$ directions respectively by about $0.4L_m$.} The locations and sizes of these structures are correctly reconstructed in all three cases {and major features such as the "bends" in the bottom snapshot are present, although minor inaccuracies are present such as the thin section of the bend in the furthest downstream structure in the bottom snapshot.}

Larger scale streamwise structures are also accurately reconstructed, particularly the hairpin structure {spanning the box $[L_m,2.2L_m]\times[-0.3L_m,0.5L_m]\times[2.1L_m,2.8L_m]$} in the first snapshot and the finger-like structures {with size $~1.2L_m$ aligned across the $x$-axis placed at $(0.5L_m, 1.5L_m)$ and $(0.5L_m, 2.5L_m)$ on the $yz$-plane} connecting the two shed vortices in the second snapshot. Hence, our model is capable of replicating flows with different intensities of spanwise effects, which manifest as streamwise aligned vortices, spanwise aligned vortices, or a combination of both.

Pressure contours in \figref{fig:sparse-p} corroborate the observations from Q-criterion contours; the first two pressure snapshots display that the model is capable of reconstructing three dimensional features (the chimney-like structure in the first snapshot and the hole in the second snapshot) of the pressure field, while the third snapshot shows that the location and intensity of the shed vortices are correctly replicated. However, two areas of improvement stand out in both contour plots: first, compared to the ground truth, the predicted surfaces are noisier and less smooth. Second, some flow features at smaller length scales are missing; this is pronounced especially for the first snapshot in the figures.

\clearpage

\clearpage

\clearpage
\clearpage

\clearpage
\clearpage

\begin{figure}
    \centering
    \includegraphics[width=0.49\textwidth]{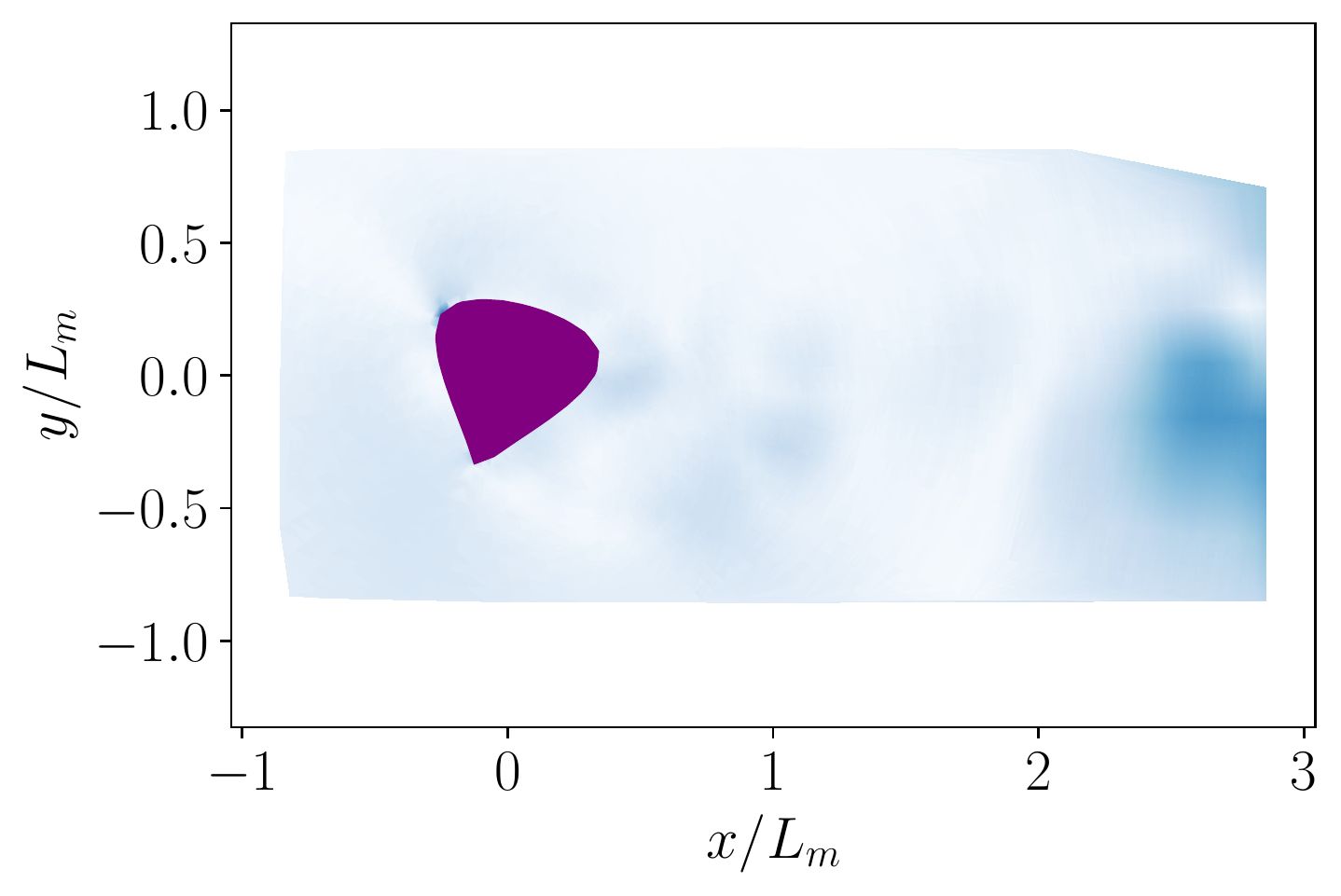}
    \includegraphics[width=0.49\textwidth]{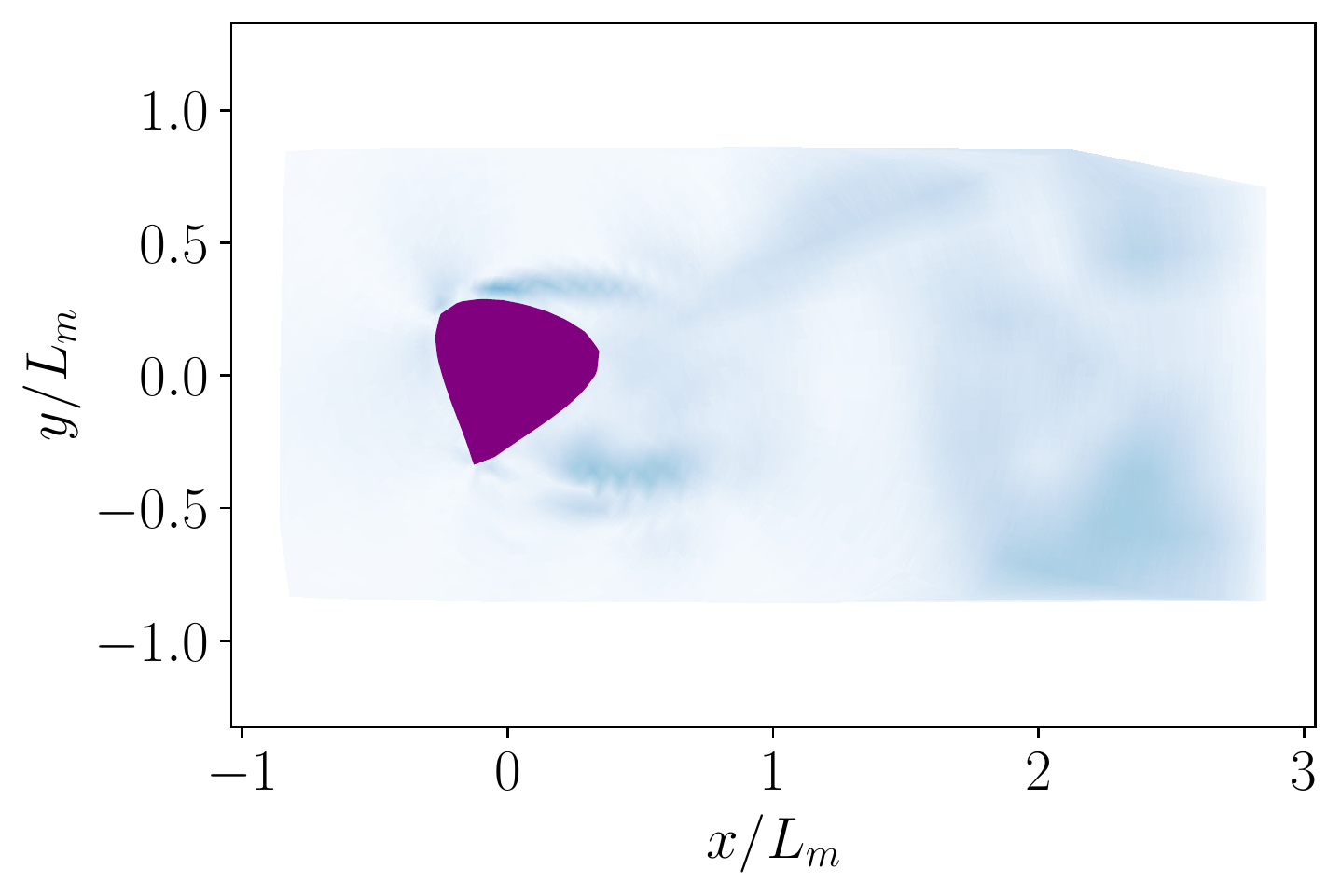}
    \includegraphics[width=0.49\textwidth]{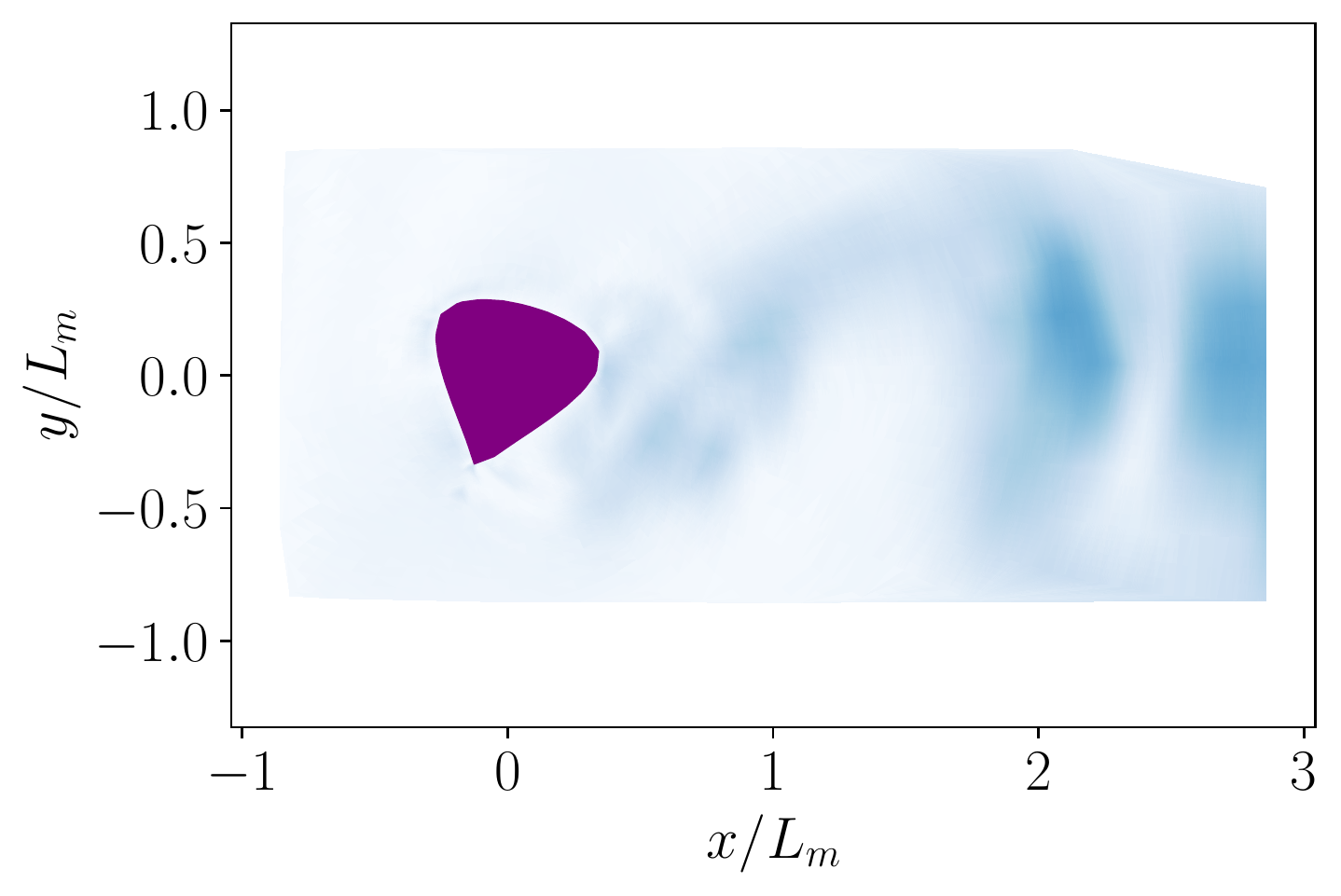}
    \includegraphics[width=0.49\textwidth]{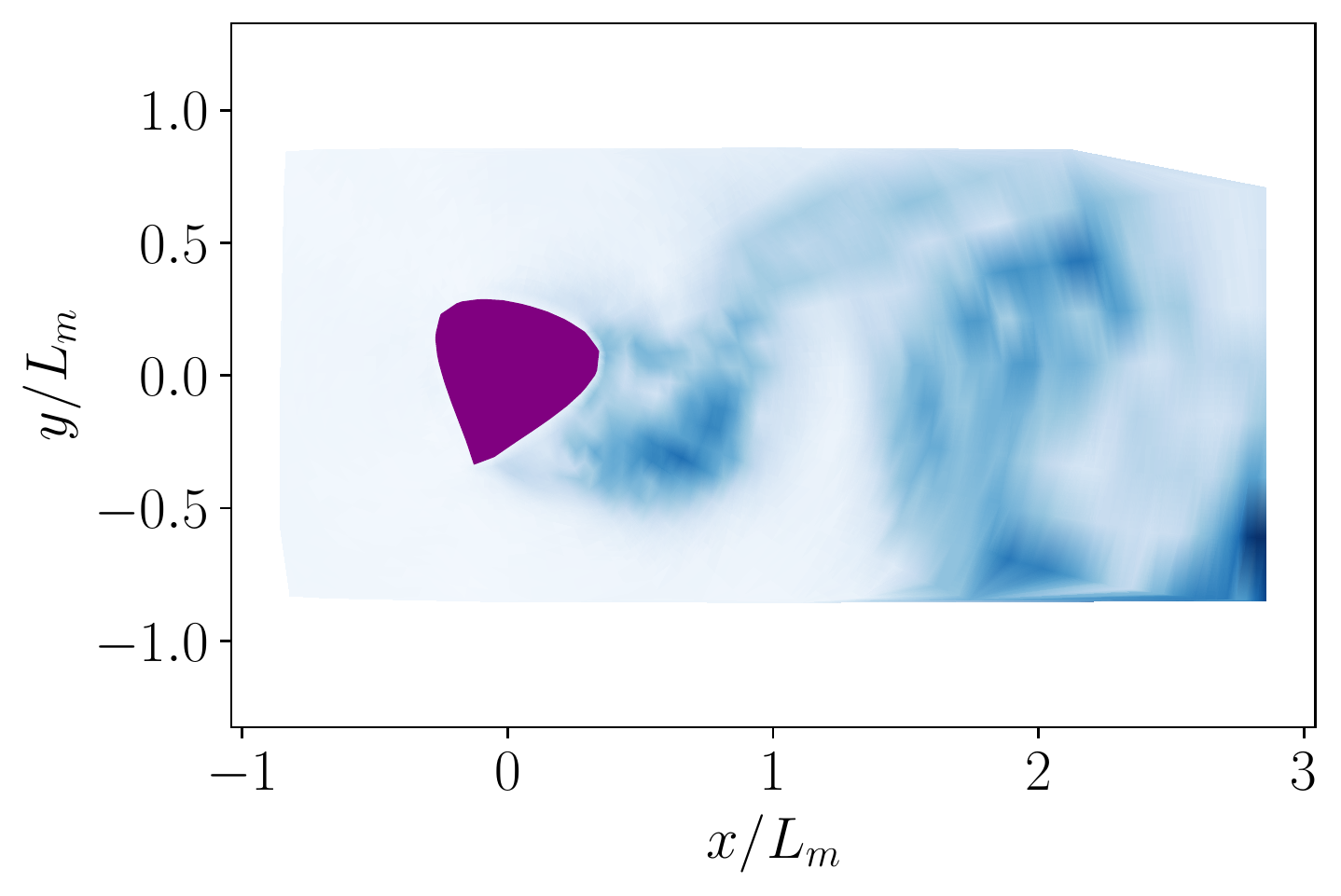}
    \includegraphics[width=0.5\textwidth]{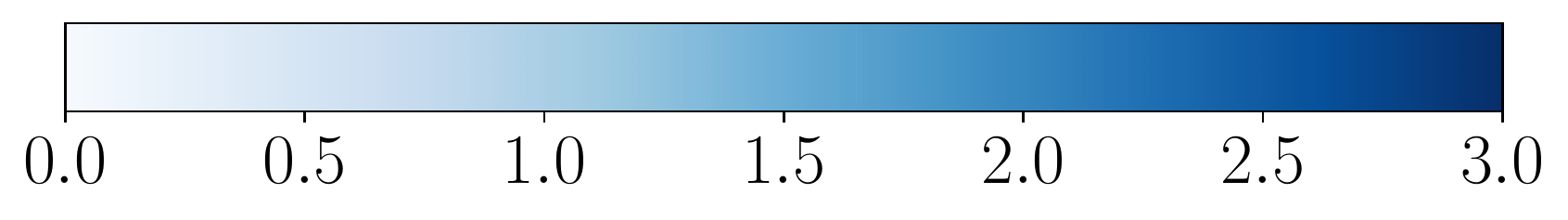}
    \caption{Spanwise-averaged absolute error fields for the pressure, $u$-velocity, $w$-velocity and $v$-velocity (clockwise from top left), \textcolor{black}{normalized by the standard deviations of the respective field variables,} for the middle snapshot in Figures \ref{fig:sparse-qcrit} and \ref{fig:sparse-p}.}
    \label{fig:fr3d:spanwise-averaged-error}
\end{figure}

Complementing the contour plots, \figref{fig:sparse-slices} provides greater detail regarding the middle snapshot in Figures \ref{fig:sparse-qcrit} and \ref{fig:sparse-p}, which is depicting a spanwise vortex that has just been shed from the trailing edge of the object, connected via streamwise structures to a second spanwise vortex further downstream. The core of the newly shed vortex in the snapshot is reconstructed clearly as shown by the $xy$-planes of the $p$ and $u$ plots, identifiable by the low pressure region in the $p$ plot and the recirculation region in the $u$ plot. Meanwhile, the $yz$-planes of the $v$ and $w$ provide further evidence of the high quality of the reconstructed streamwise structures; the negative $v$ regions and the rapidly alternating $w$ regions are accurately reconstructed and correspond to the finger-like structures observed in \figref{fig:sparse-qcrit}. {\figref{fig:fr3d:spanwise-averaged-error} depicts 2D maps of the spanwise-averaged absolute error for the predicted fields. Errors are chiefly coincident with the coherent structures in the flow, and do not increase with distance from sensor locations, highlighting the capability of the FR3D architecture to predict the flow accurately in the entire problem domain.}

\subsection{Reconstruction from plane measurements}
\label{sec:results:planes}

Next, we showcase the results using the plane sensor setup. \tabref{tab:plane-error} displays error metrics with this sensor setup, similar to \tabref{tab:sparse-error} for the sparse sensor setup. The error levels of are slightly higher overall with this sensor setup, with the largest relative rise in error encountered in predictions of $u$. Overall, the similarity between the error levels of the plane setup and the error levels with the sparse setup demonstrates the flexibility of our model in regards to the sensor setup, which is highly important for its future potential applications since its main envisioned future use --physical experiments-- can involve widely varying sensor setups.

\begin{table}[h!]
\caption{Mean absolute percentage (MAPE) and mean squared (MSE) error levels achieved by the FR3D model on the validation dataset for reconstruction from plane velocity measurements. Error metrics using the encoder input are not provided, as they are identical to the values in \tabref{tab:sparse-error}.} 
\label{tab:plane-error}
\begin{tabular}{@{}llcccc@{}}
\toprule
Var.                   & Input to $\mathcal{D}$ & MAPE    & Min-max MAPE & MSE                   & Min-max MSE           \\ \midrule
$p$  & $\mathcal{L}$   & 10.19\%  & 6.55\%      & $6.55 \times 10^{-3}$ & $9.03 \times 10^{-4}$ \\ \midrule
$u$  & $\mathcal{L}$   & 7.76\%  & 5.17\%       & $1.34 \times 10^{-2}$ & $9.66 \times 10^{-4}$ \\ \midrule
$v$  & $\mathcal{L}$   & 18.15\% & 3.08\%       & $6.96 \times 10^{-3}$ & $2.00 \times 10^{-4}$ \\ \midrule
$w$  & $\mathcal{L}$   & 38.24\% & 5.77\%       & $4.16 \times 10^{-3}$ & $7.43 \times 10^{-4}$ \\ \bottomrule
\end{tabular}
\end{table}

\begin{figure}
    \centering
    \includegraphics[width=0.46\textwidth]{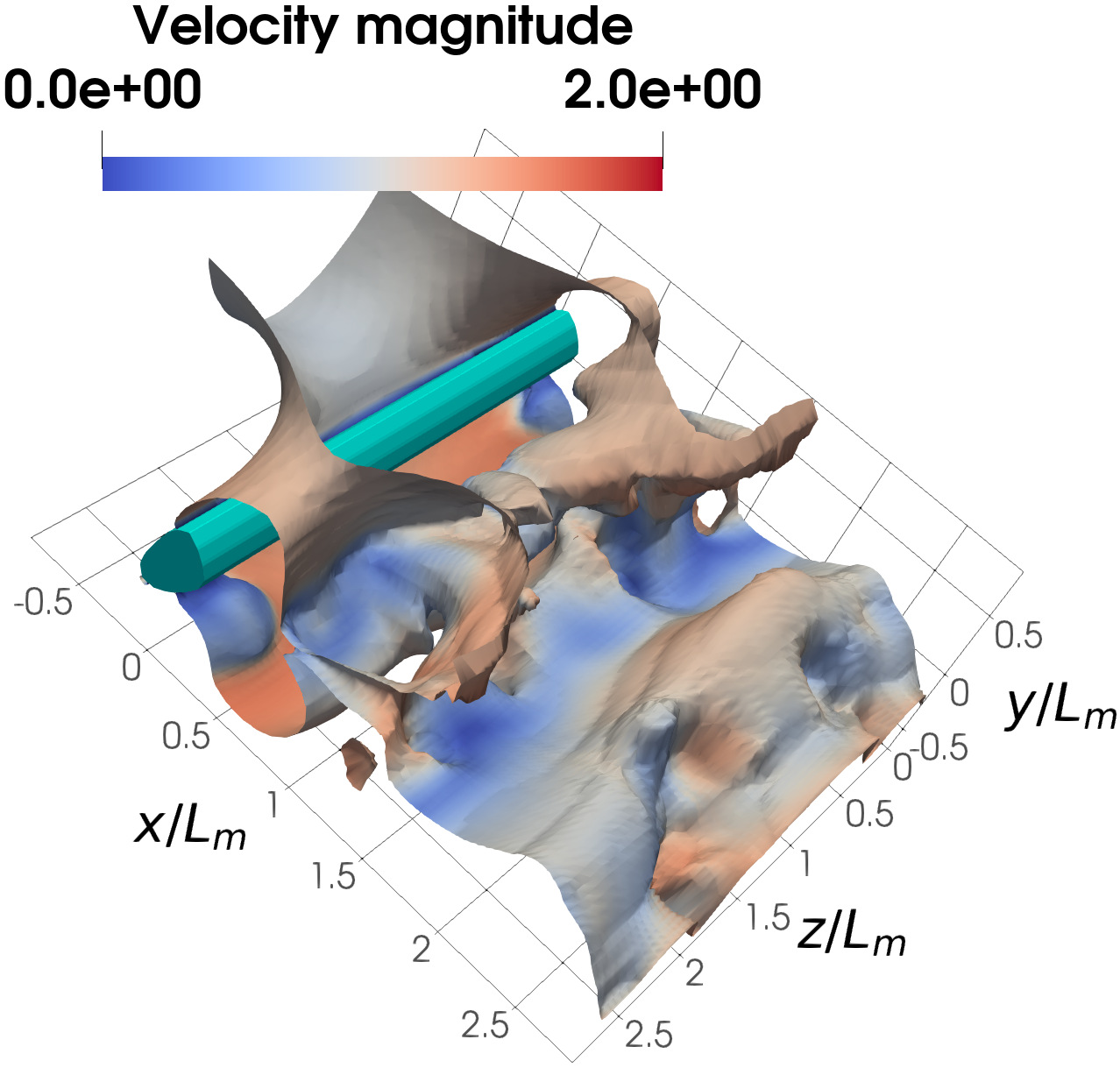}
    \includegraphics[width=0.46\textwidth]{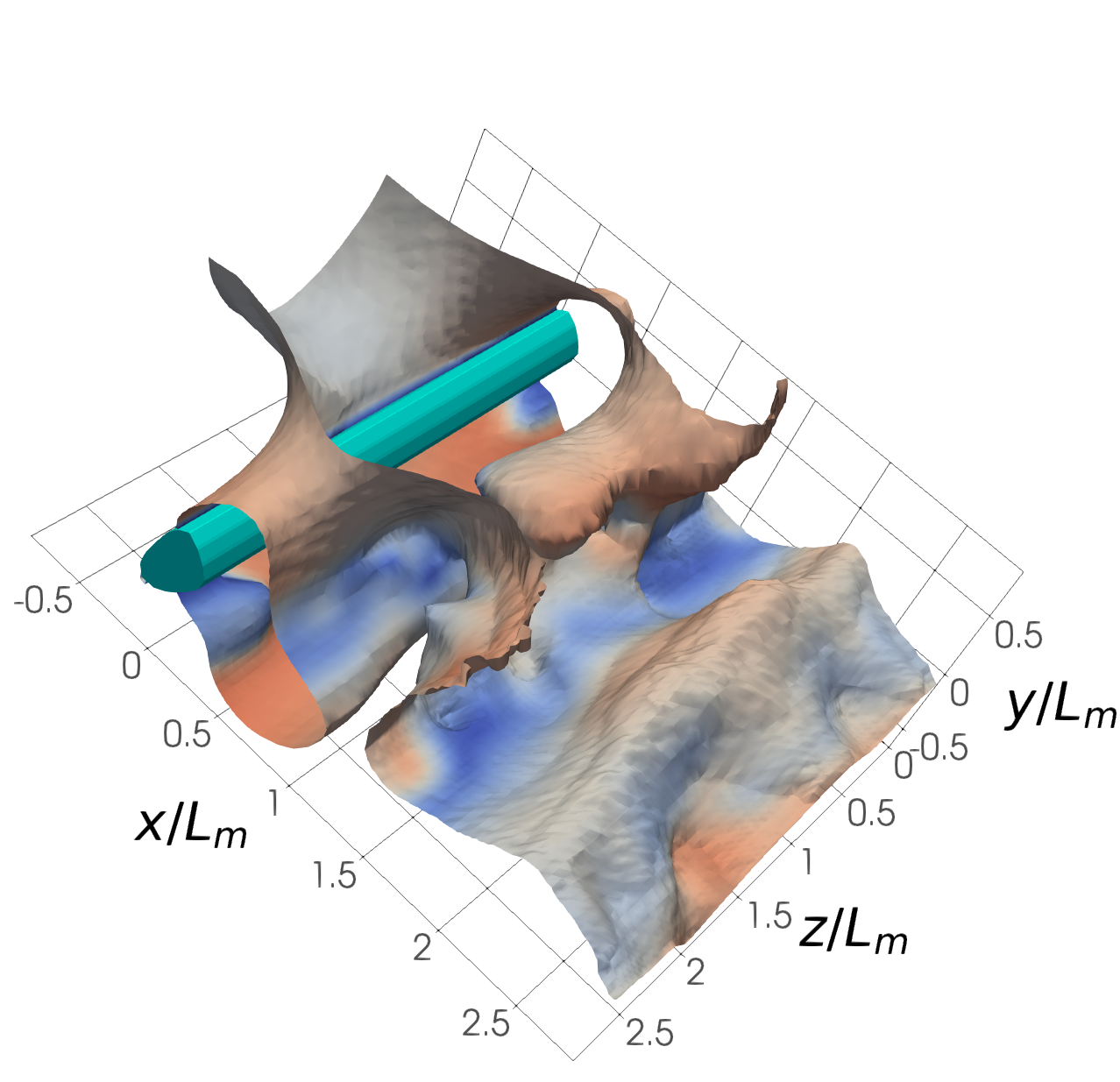}
    \includegraphics[width=0.46\textwidth]{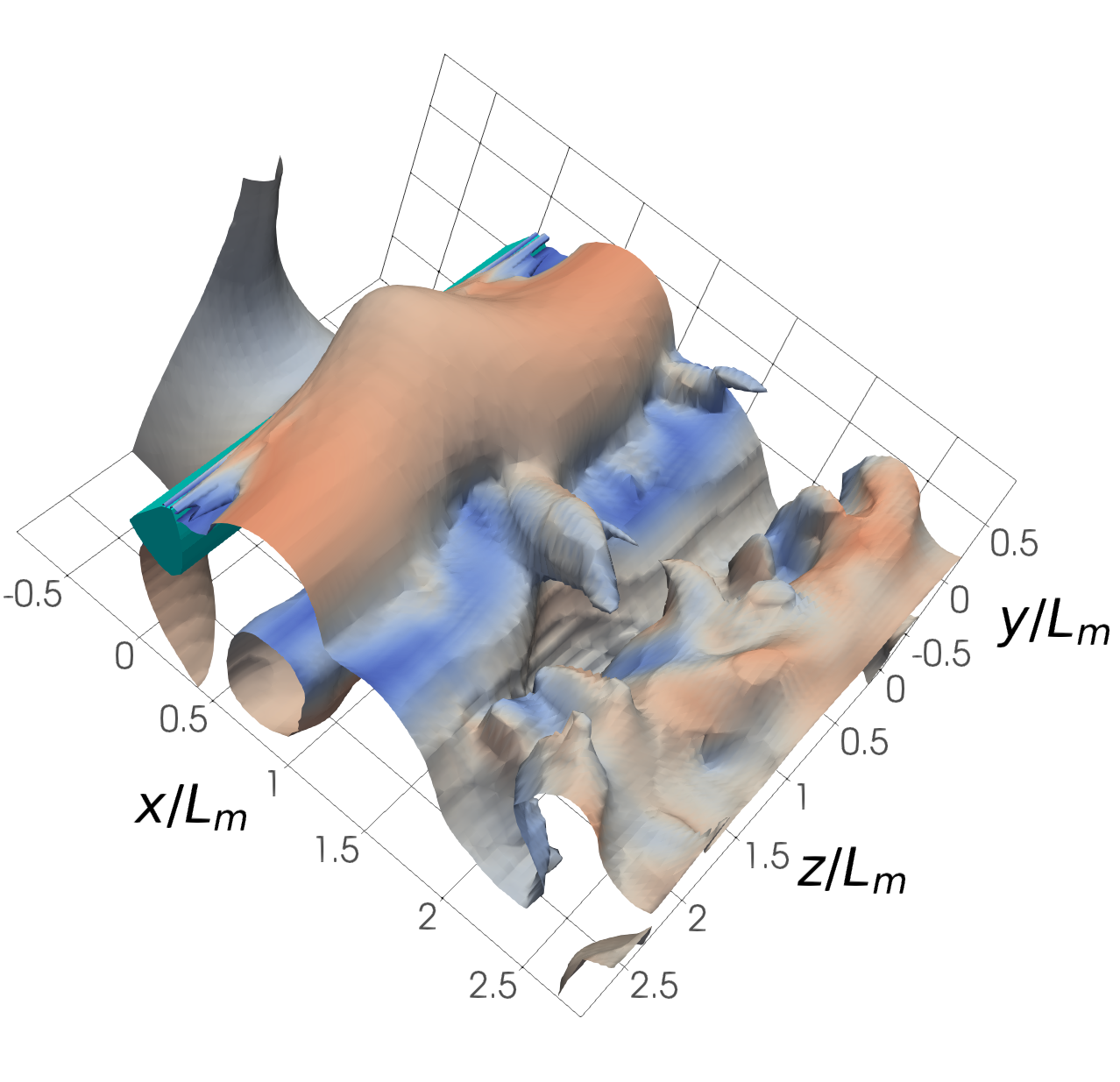}
    \includegraphics[width=0.46\textwidth]{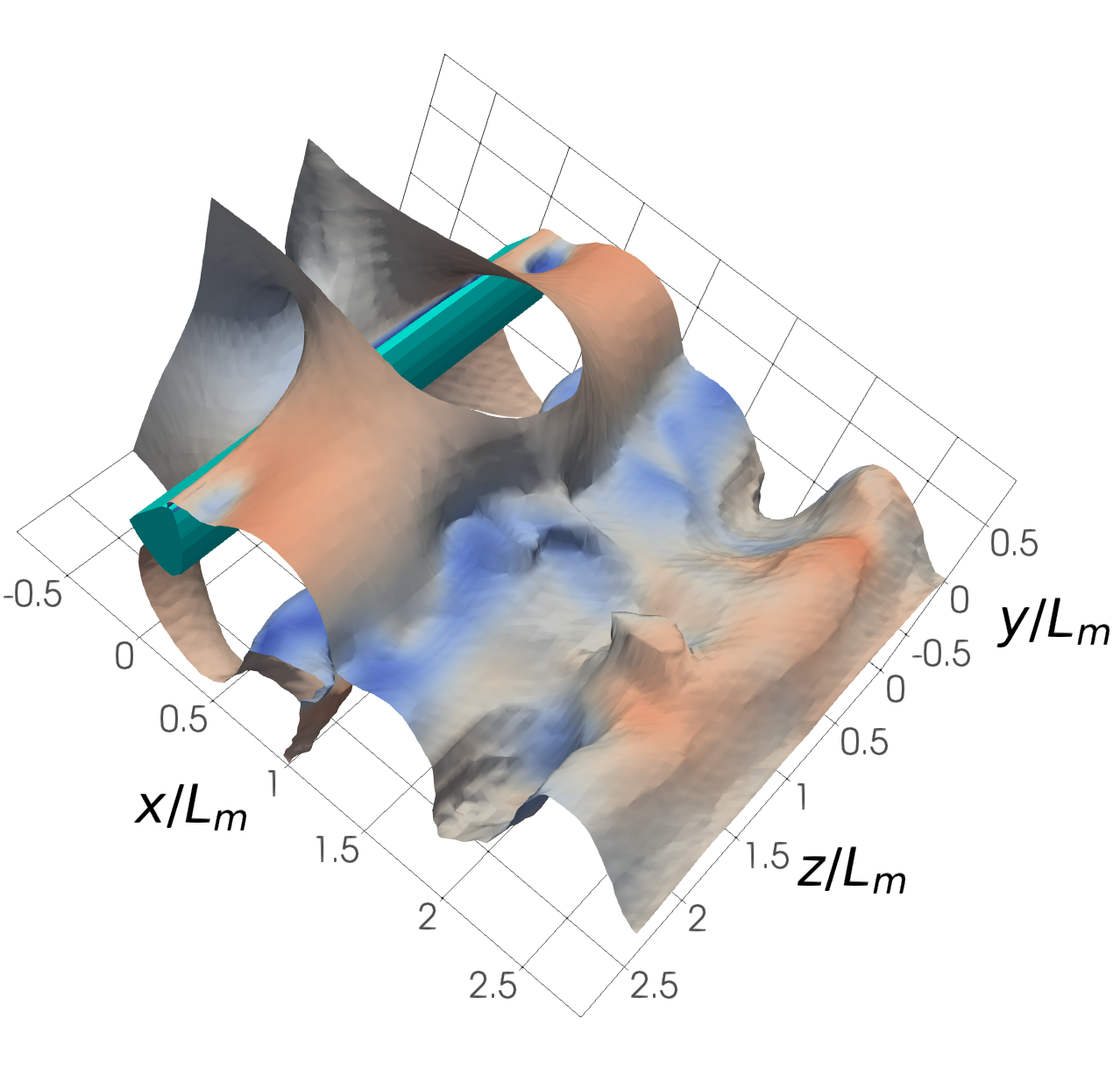}
    \includegraphics[width=0.46\textwidth]{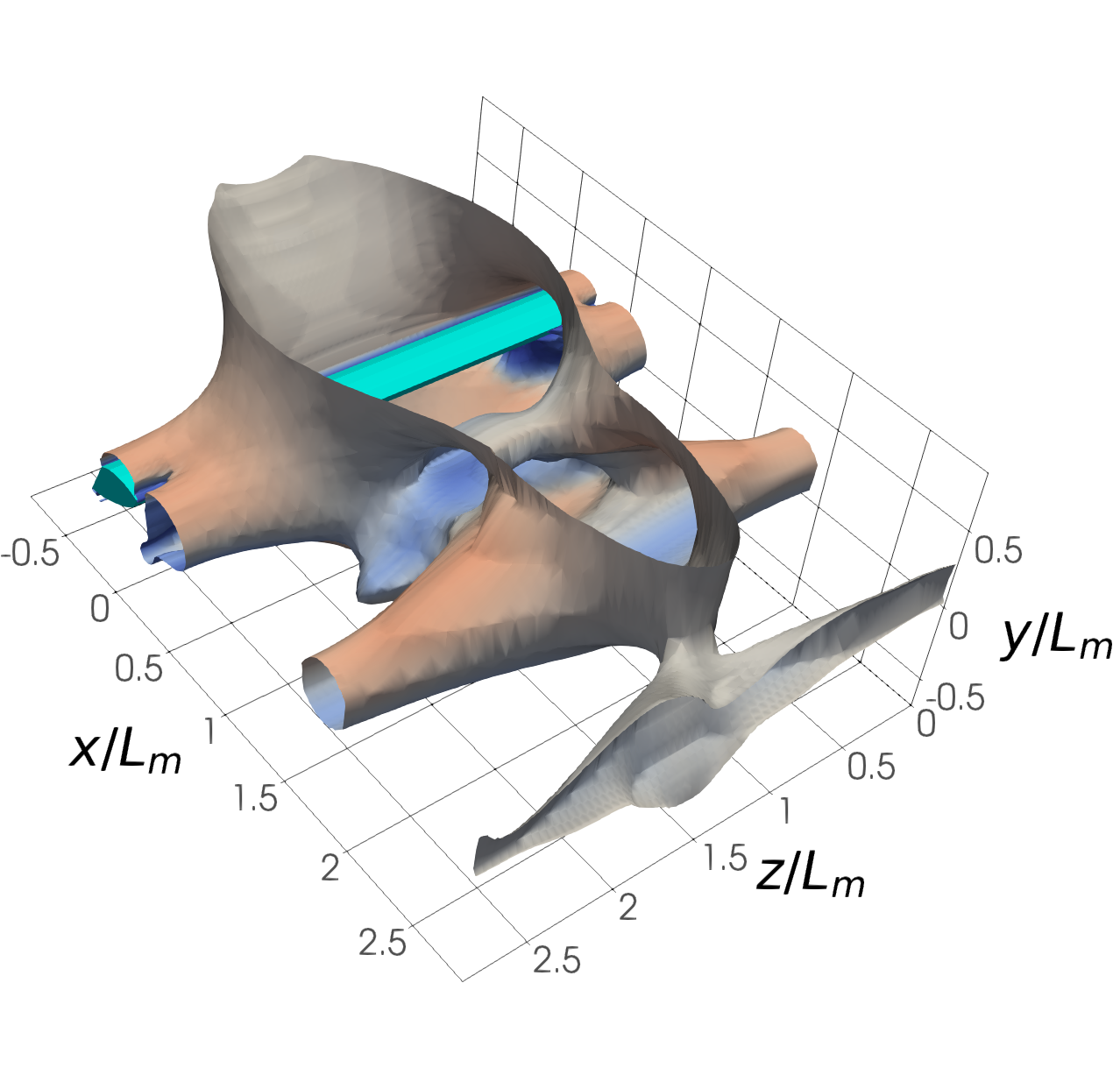}
    \includegraphics[width=0.46\textwidth]{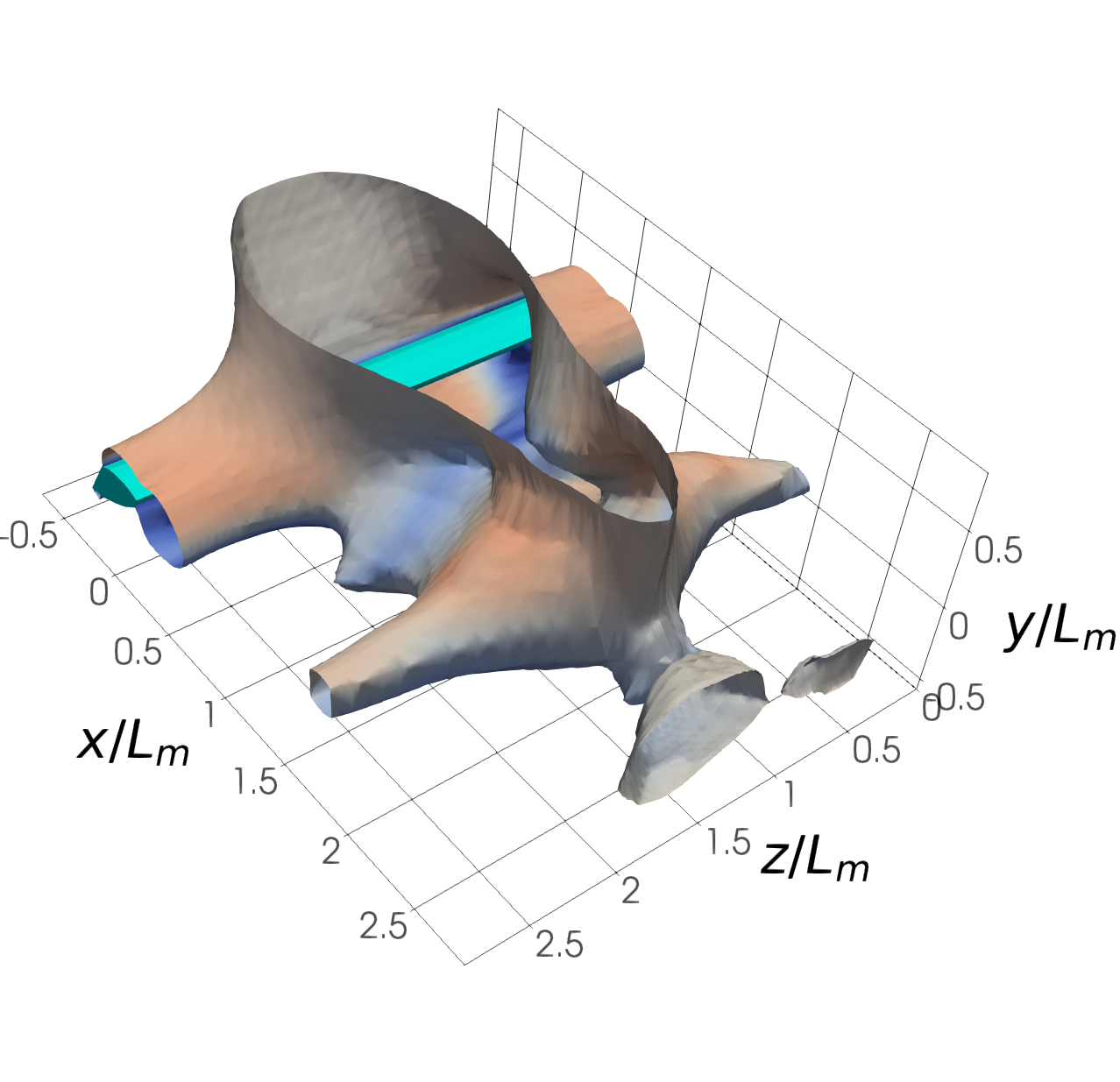}
    \caption{Ground truth (left) and predicted (right) pressure (top: $p=0.75$; middle, bottom: $p=0.85$) iso-contours coloured by velocity magnitude from three randomly chosen snapshots belonging to different objects, using the model trained on the plane sensor setup.}
    \label{fig:plane-p}
\end{figure}

Pressure contours for three further geometries, using predictions made for this sensor setup, can be found in \figref{fig:plane-p}. Similar to the pressure contour results for the sparse sensor setup in \figref{fig:sparse-p}, major features of the pressure field are accurately reconstructed with this sensor setup as well. Recovering pressure from plane measurements of velocity is often a challenge in experimental settings involving PIV \cite{piv_only_velocity}. Thus, the present results suggest the FR3D model has the potential to help overcome challenges associated with recovering pressure fields from PIV experiments.
\clearpage
\clearpage

\subsubsection{Estimation of lift and drag}
\label{sec:results:clcd}

Since the FR3D model is capable of accurately reconstructing the pressure and velocity fields, these results can be also applied to estimate the instantaneous lift and drag coefficients $C_L$ and $C_D$ experienced by the investigated geometries. Our choice of sampling points, utilizing the aforementioned conformal mapping approach instead of the more traditional Cartesian approach, makes this task substantially easier, as the need for interpolating the pressure and velocity fields onto the object surfaces is removed. This is due to the fact that the sampling points which lie on the inner ring of the annulus (cf. \figref{fig:mapping}) always lie on the original geometry's surface when the conformal mapping is applied.

To compute lift and drag forces, we adopt the approach of computing the body forces (pressure force and skin friction) through the integration of the pressure and the shear stresses across the object surface. The integration is carried out through the finite element method, approximating the surface of the geometry as a collection of quadrilateral surfaces. The dense field sampling points which lie on the object surface serve as the vertices of each quadrilateral. Since we do a simple extrusion in the z-direction, each quadrilateral is a rectangle. Defining a Cartesian coordinate system$(\xi, \eta)$ on the quadrilateral, with the origin on one of the vertices of the quadrilateral, the standard bilinear basis functions
\begin{equation*}
\begin{split}
    N_1 = (1-\xi)(1-\eta), \; N_2 = \xi (1-\eta), \\
    N_3 = (1-\xi) \eta, \; N_4 = \xi \eta,
\end{split}
\end{equation*}
can be used to approximate the pressure distribution on quadrilateral $j$ as 
\begin{equation*}
    \tilde{p}_j = \sum_i \hat{p}_{ij} N_i,
\end{equation*}
where $\hat{p}_{ij}$ is the pressure prediction on vertex $i$ of quadrilateral $j$. Consequently, the pressure force on the quadrilateral (with surface normal $\hat{\mathbf{n}}_j$) can be approximated by integrating $\Tilde{p}_j$ along the surface of the quadrilateral:
\begin{equation}
    \label{eq:fe_pressure_force_integral}
    \Vec{f}_{p, j} = \int \Tilde{p}_j \hat{\mathbf{n}}_j \: \mathrm{d}A.
\end{equation}

For a rectangle, the result of \equref{eq:fe_pressure_force_integral} is simply the average of the pressure values at its vertices, times its area
\begin{equation}
    \label{eq:fe_pressure_force}
    \Vec{f}_{p, j} = \hat{\mathbf{n}}_j A_j \sum_{i=1}^4 \frac{1}{4} \hat{p}_{ij} = \Vec{A}_j \sum_{i=1}^4 \frac{1}{4} \hat{p}_{ij}.
\end{equation}

In practice, $\Vec{A}_j$ can be directly computed by taking the cross product of the vectors which run along the edges of the rectangle, and these vectors can be easily computed given the rectangle's vertex coordinates. The overall pressure force can be computed by adding the pressure force on each quadrilateral $\Vec{f}_p = \sum_j \Vec{f}_{p,j}$.

The skin friction force $\Vec{f}_{f}$ can be computed by repeating the above procedure, substituting the wall shear stresses for the pressures. The shear forces can be approximated using the standard formula
\begin{equation*}
    \mu \frac{\partial u_{||}}{\partial n},
\end{equation*}
where $\mu$ is the dynamic viscosity, $u_{||}$ is the velocity parallel to the surface and $n$ is the coordinate normal to the surface. The gradient term can be approximated as 
\begin{equation}
\label{eq:shear_stress_approximation}
    \frac{\partial u_{||}}{\partial n} \approx \frac{u_{||}(n=\delta n)-u_{||}(n=0)}{\delta n} = \frac{u_{||}(n=\delta n)}{\delta n}
\end{equation}
The annular sampling method also greatly simplifies the computation of the approximation in \equref{eq:shear_stress_approximation}, as it ensures that each sampling point on the object surface (slice index 0 across the first dimension of the $64 \times 64 \times 64$ prediction array) has a corresponding sampling point above it in the wall-normal direction (slice index 1 across the first dimension of the same array).

\begin{figure}[]
    \centering
    \includegraphics[width=0.49\textwidth]{images/shape_34.pdf}
    \includegraphics[width=0.49\textwidth]{images/shape_183.pdf}
    \includegraphics[width=0.49\textwidth]{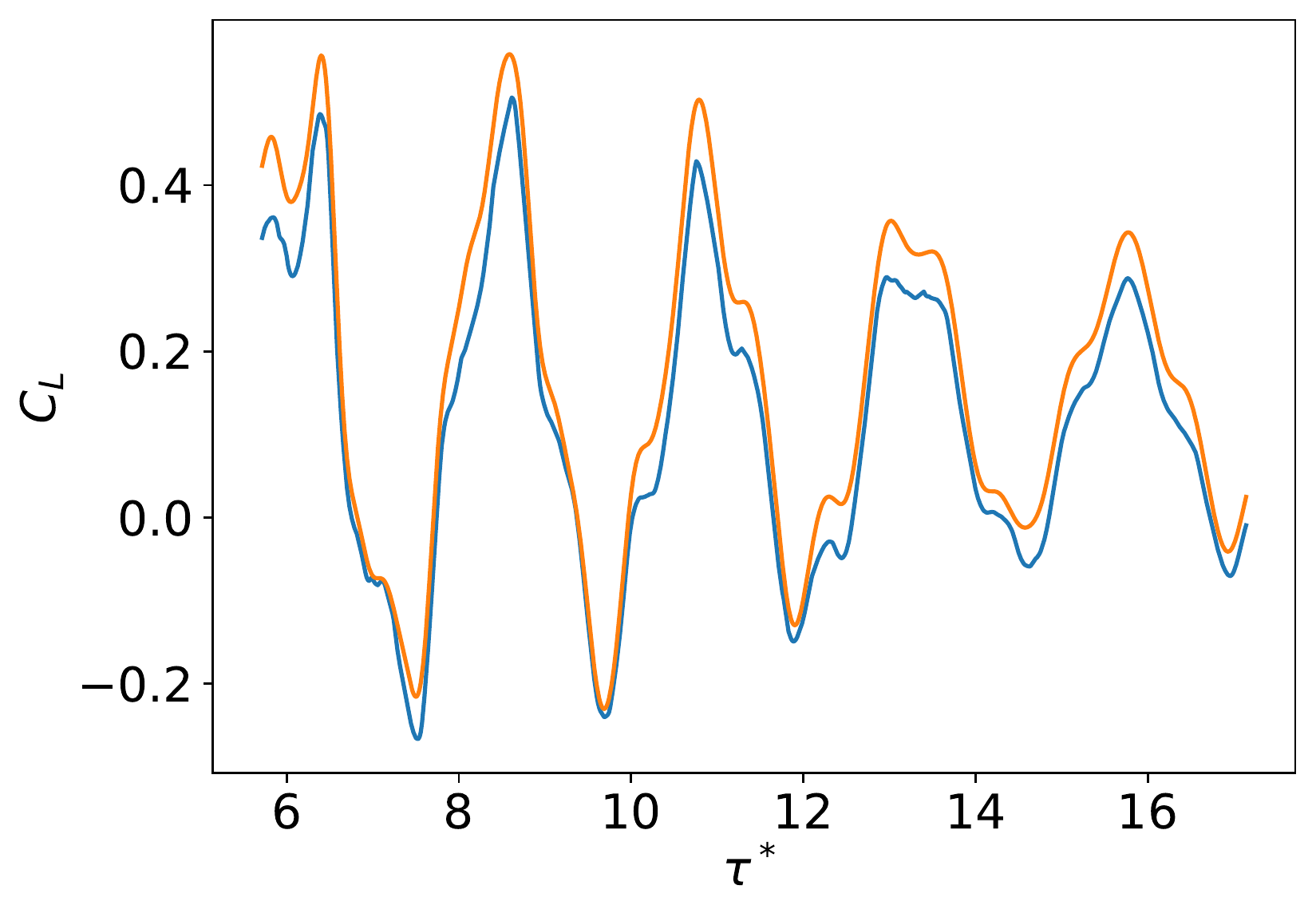}
    \includegraphics[width=0.49\textwidth]{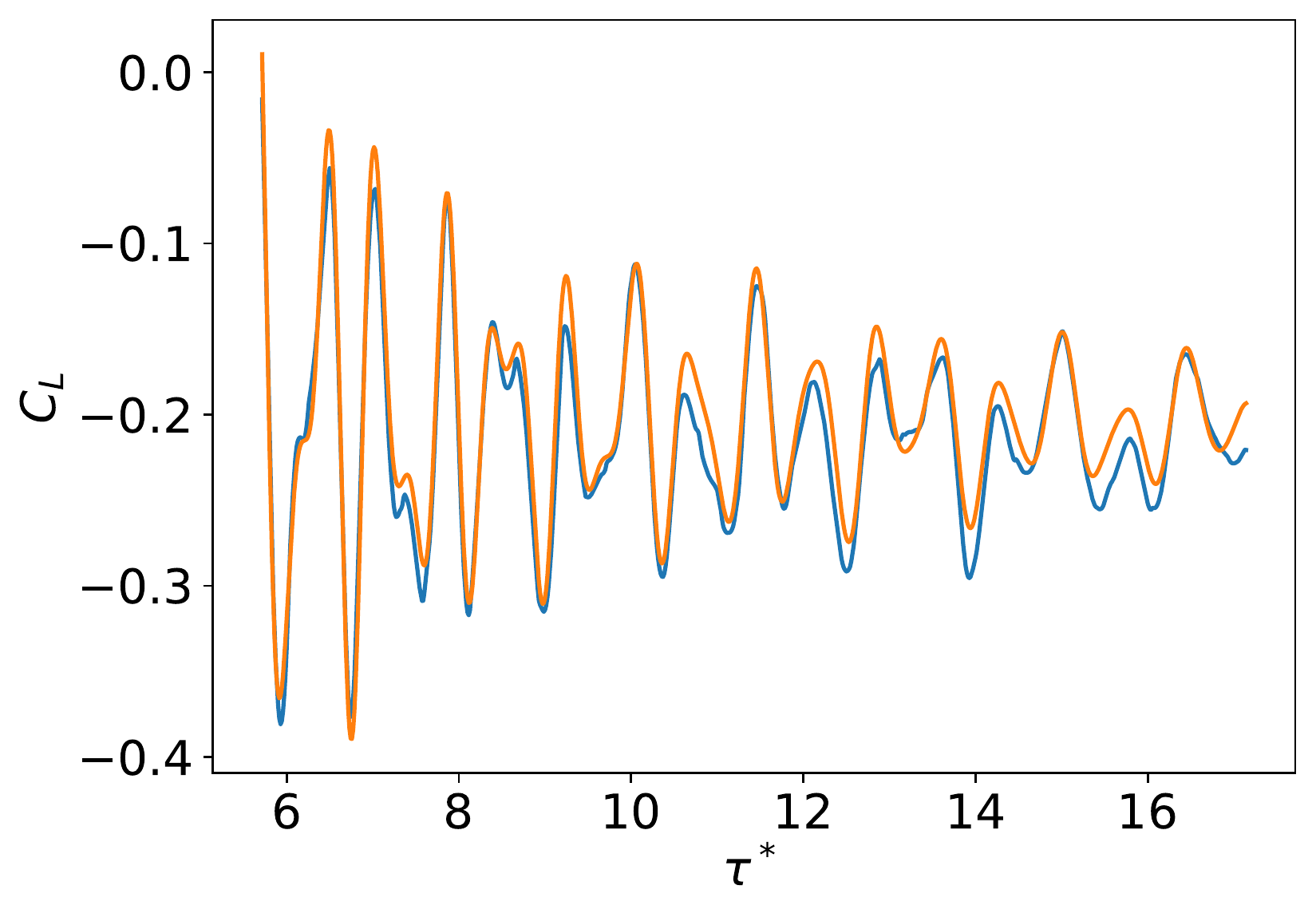}
    \includegraphics[width=0.49\textwidth]{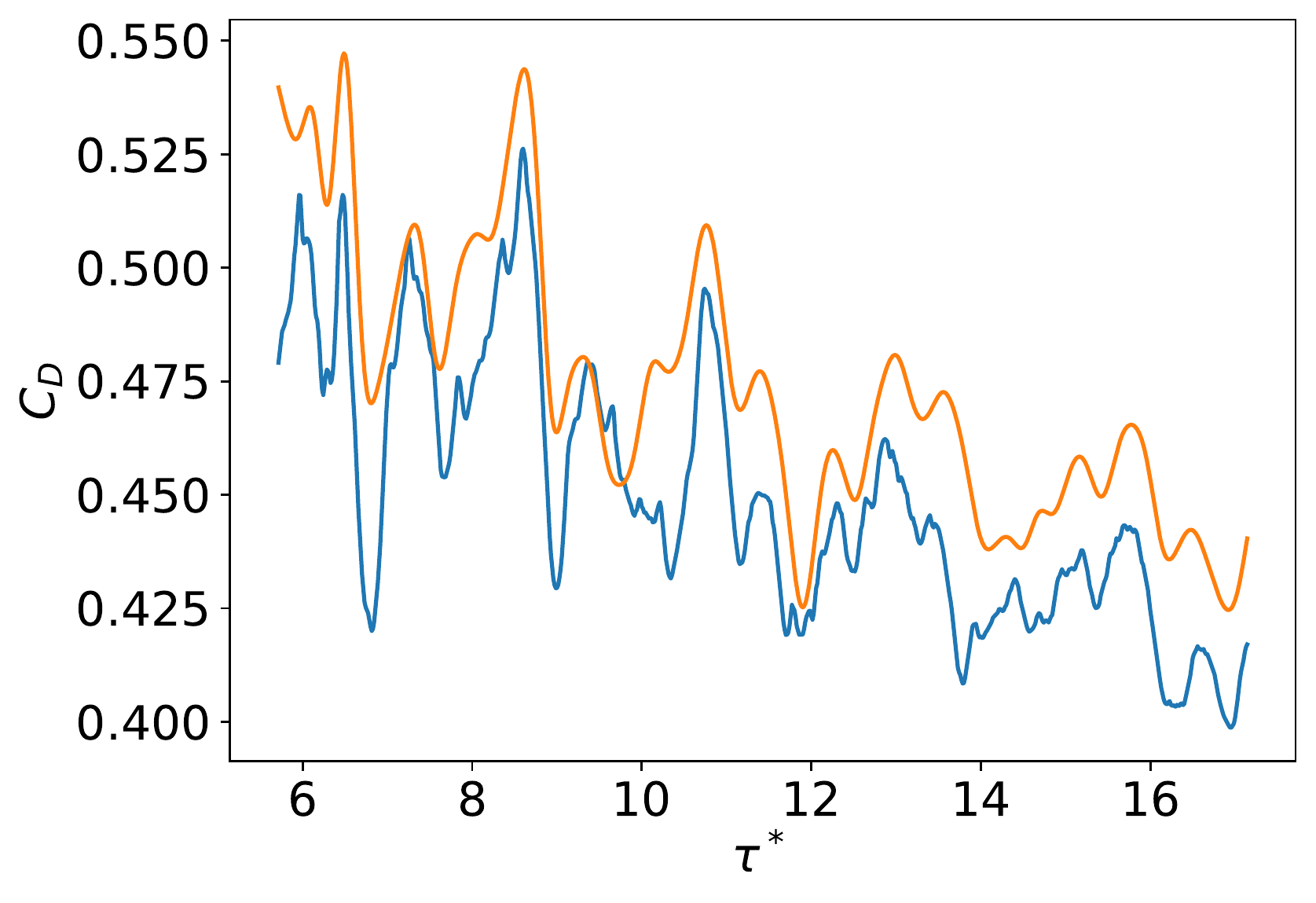}
    \includegraphics[width=0.49\textwidth]{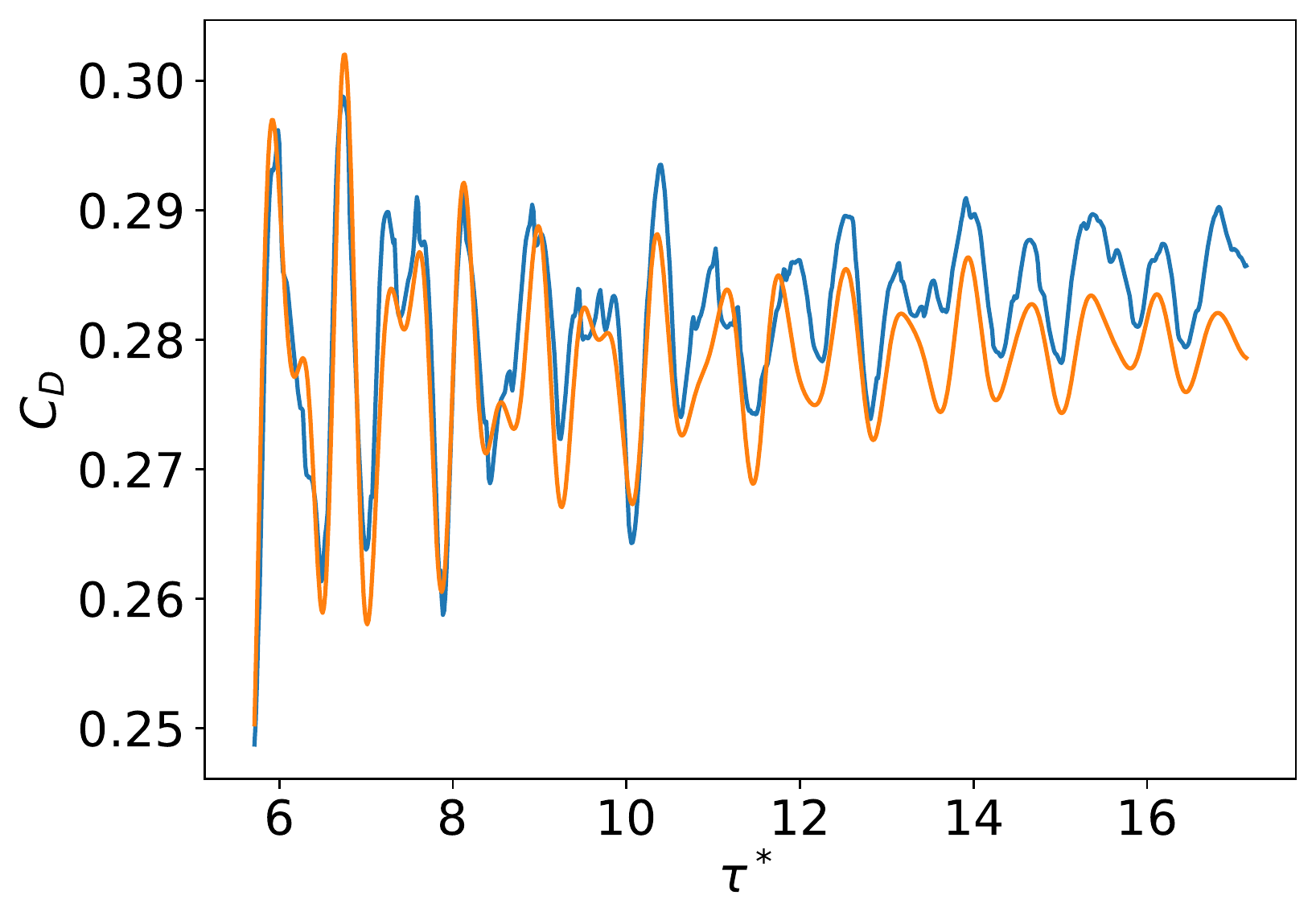}
    \includegraphics[width=0.3\textwidth]{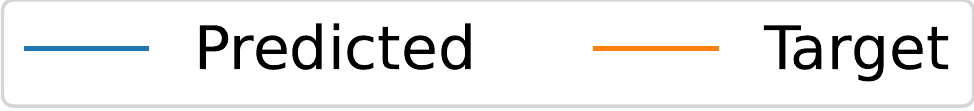}
    \caption{Time evolutions of the lift (middle) and drag (bottom) coefficients for two of the random geometries (left and right) in the test dataset. Predictions were made using the model trained using the sparse sensor setup, and the target values were computed using the body force results outputted by PyFR.}
    \label{fig:clcd_time_evolution}
\end{figure}

Once both the skin friction forces $\Vec{f}_f$ and pressure forces $\vec{f}_p$ are known, the overall body forces can be simply computed as $\Vec{f} = \Vec{f}_f + \Vec{f}_p$. This method, combined with the FR3D model, leads to very accurate lift and drag predictions as seen in the error metrics averaged over the entire test dataset in \tabref{tab:clcd_error} and the time evolutions of $C_L$ and $C_D$ plotted for two of the random test geometries in \figref{fig:clcd_time_evolution}.

\begin{table}[h!]
\caption{Mean absolute percentage error (MAPE) and mean squared error (MSE) metrics averaged over the test dataset.}
\label{tab:clcd_error}
\centering
\begin{tabular}{@{}lccc@{}}
\toprule
Sensor setup & Coefficient & MAPE    & MSE                  \\ \midrule
\multirow{2}{*}{Sparse} & $C_L$ & 9.16\%  & $9.91 \times 10^{-4}$ \\
                        & $C_D$ & 4.31\%  & $4.86 \times 10^{-4}$ \\ \midrule
\multirow{2}{*}{Planes} & $C_L$ & 7.18\%  & $6.91 \times 10^{-4}$ \\
                        & $C_D$ & 3.43\%  & $2.77 \times 10^{-4}$ \\ \bottomrule
\end{tabular}
\end{table}

The lift and drag estimation capability presented by the FR3D model presents a substantial leap over similar approaches in previous literature; for example, Chen et al. \cite{mgfr_1} demonstrated the capability to predict $C_D$ with percentage errors of 3.43\%, but only for steady 2D flows at Reynolds number equal to 10. In comparison, our model is capable of achieving similar levels of error, but for unsteady 3D flows for Reynolds numbers 50 times larger. The results are especially impressive when using the plane sensor setup, where we match the drag error levels reported by Chen et al. \cite{mgfr_1}, despite the substantially more difficult task at hand.

We observed that our lift and drag predictions are especially accurate on high aspect ratio shapes resembling airfoils, such as the geometry on the right column in \figref{fig:clcd_time_evolution}. The level of accuracy on bluff bodies is still reasonably high, as seen in the bluff body on the left column in the same figure. However, we did see a slight degradation in accuracy on shapes with concave sections, such as the middle right geometry in \figref{fig:shapes} for which the time-averaged $C_L$ and $C_D$ prediction errors were 14\% and 8\% respectively. This is due to the substantially more complicated flow patterns occurring in such geometries, and the relative rarity of such geometries in the dataset.

\section{Conclusion and future work}
\label{sec:conclusion}

In this study, we described the performance of FR3D, a convolutional autoencoder neural network model, on the reconstruction of three-dimensional flows past objects with varying cross-sections, given measurements of the flow field from sparse sensors and plane measurements. Using a conformal mapping technique to achieve geometry invariance, we demonstrated that the FR3D architecture is capable of reconstructing instantaneous pressure and velocity fields of flows past such geometries with min-max normalized percentage error levels under 10\% for geometries not encountered during training at a fixed Reynolds number. The reconstructions are of a quality sufficient to accurately replicate the major features of Q-criterion and pressure iso-contours.

Subsequently, we applied the FR3D model to a \textcolor{black}{scenario in which velocity measurements are available in two orthogonal planes}, whereby it is attempted to recover the flow variables -- including pressure -- from the measurements of the velocity components downstream of the object. The FR3D model also performed well in this additional scenario, reconstructing the dense pressure fields with percentage errors just above 10\%.

Finally, using the reconstructed fields, we demonstrated that the lift and drag coefficients can be estimated within 10\% of the ground truth values using both sensor setups, going as low as 3.43\% when estimating drag coefficients using the plane measurement setup.

In the future, we aim to extend this work by investigating:

\begin{itemize}
    \item \textbf{Noisy measurements:} The FR3D model was trained and evaluated using high-fidelity values obtained via computation. In contrast,  a large degree of uncertainty exists in real-world measurements and ensuring that FR models are robust to noise will be necessary before they can be utilized in real laboratory environments, as opposed to computational studies.
    \item \textbf{More complex geometries:} The present work focused on investigating only obejcts extruded in the spanwise direction, in order to leverage our previous work on applying Schwarz-Christoffel mappings for training a model that can handle different geometries well. Different mapping techniques, such as boundary conforming curvilinear coordinate systems \cite{curvilinear_coords_3d}, or different neural network model architectures such as graph neural networks that do not depend on regular grids, will be necessary to achieve geometry invariance for a broader class of objects.
    \item \textbf{Varying Reynolds numbers:} Since the dataset in this work consisted solely of flows at a fixed Reynolds number, the model is not expected to perform well at other Reynolds numbers. Extending the model to perform well for  a wide range of Reynolds numbers, possibly through adding a `physics-informed' loss function component, will greatly boost its usefulness. {Reconstructing three-dimensional flows at high Reynolds numbers, relevant to real-life applications might be possible with hybrid approaches,  combining different neural network architectures. For example, combining CNNs with recurrent neural networks (RNNs) could better capture both spatial and temporal dependencies in the flow field.}
    \item \textbf{State-of-the-art generative models:} Recent advances in generative models for images, such as Diffusion architectures \cite{stablediffusion}, constitute promising directions for substantial advances in flow reconstruction.
\end{itemize}

\section{Acknowledgements}

This work was supported by a PhD scholarship by the Department of Aeronautics, Imperial College London, and an Academic Hardware Grant by NVIDIA.

{The code used to generate the data and train the models used in this study is available at \href{https://github.com/aligirayhanozbay/fr3D}{https://github.com/aligirayhanozbay/fr3D}}

\appendix

\section{Results using the GAN approach}
\label{sec:fr3d:gan}

Flow reconstruction models, such as the FR3D model introduced in the present work, can be broadly described as `generative' deep learning models. Generative adversarial networks (GANs) \cite{gan} constitute a method of training generative deep learning models by introducing a second neural network called a `discriminator'. The discriminator is tasked with classifying whether a particular output belongs to the ground truth dataset or was generated by the generator. Throughout the training process, the generator tries to `fool' the discriminator; the discriminator outputs probabilities regarding whether the outputs of the generator were generated or not, and the generator tries to minimize that probability by using the discriminator as a component of its loss function.

Training the FR3D model with a GAN approach was attempted using a procedure based on \algoref{algo:fr3d:training}. The new approach first trains the FR3D encoder $\mathcal{E}$, decoder $\mathcal{D}$ and latent space embedder $\mathcal{L}$ in a similar fashion, but adds a binary cross-entropy loss component
\begin{equation}
    L_\mathcal{C}(\mathbf{t}, \mathbf{y}) = -(\mathbf{y} \ln(\mathbf{t}) + (1 - \mathbf{y}) \ln(1-\mathbf{t})),
\end{equation}
for the outputs of the discriminator $\mathcal{C}$. The outputs from the decoder are then used to train the discriminator with a batch of data consisting of both the decoder outputs and the ground truth snapshots. This procedure is outlined in \algoref{algo:fr3d:gan_training}.

\begin{algorithm}
\begin{algorithmic}[1]
\Function{OptimizeGAN}{$\mathbf{w}_E,\mathbf{w}_D,\mathbf{w}_L, \mathbf{w}_C, \mathbf{s},\mathbf{x}$}
\State $\mathbf{t}_\mathcal{D} := \textproc{Zeroes}(\mathbf{x})$
\State $\mathbf{l} := \mathcal{E}(\mathbf{x}, \mathbf{w}_E)$
\State $\hat{\mathbf{x}} := \mathcal{D}(\mathbf{l}, \mathbf{w}_D)$
\State $\mathbf{w}_E \leftarrow \textproc{Adam}(\nabla_{\mathbf{w}_E} L(\mathbf{x}, \hat{\mathbf{x}}) + L_\mathcal{C}(\mathbf{t}_\mathcal{D},  \mathcal{C}(\hat{\mathbf{x}}, \mathbf{w}_C)), \mathbf{w}_E)$
\State $\mathbf{w}_D \leftarrow \textproc{Adam}(\nabla_{\mathbf{w}_D} L(\mathbf{x}, \hat{\mathbf{x}}) + L_\mathcal{C}(\mathbf{t}_\mathcal{D},  \mathcal{C}(\hat{\mathbf{x}}, \mathbf{w}_C)), \mathbf{w}_D)$
\State $\hat{\mathbf{l}} := \mathcal{L}(\mathbf{s}, \mathbf{w}_L)$
\State $\hat{\mathbf{x}} \leftarrow \mathcal{D}(\hat{\mathbf{l}}, \mathbf{w}_D)$
\State $\mathbf{w}_L \leftarrow \textproc{Adam}(\nabla_{\mathbf{w}_L} L(\mathbf{x}, \hat{\mathbf{x}}) + L_\mathcal{C}(\mathbf{t}_\mathcal{D},  \mathcal{C}(\hat{\mathbf{x}}, \mathbf{w}_C)), \mathbf{w}_L)$
\State $\mathbf{t}_\mathcal{C} := \textproc{Concat}(\textproc{Ones}(\mathbf{x}), \textproc{Zeroes}(\mathbf{x}))$
\State $\mathbf{w}_C \leftarrow \textproc{Adam}(\nabla_{\mathbf{w}_C} L_\mathcal{C}(\mathbf{t}_\mathcal{C}, \mathcal{C}(\textproc{Concat}(\hat{\mathbf{x}}, \mathbf{x}), \mathbf{w}_C)) , \mathbf{w}_C)$
\EndFunction
\end{algorithmic}
\caption{{Optimization of the GAN variant of the FR3D model using a single batch of data. $\textproc{Zeroes}(\cdot)$ and $\textproc{Ones}(\cdot)$ are functions which output a vector of 0 or 1 values, respectively, with the number of entries equal to the number of samples in the input. $\textproc{Concat}(\cdot, \cdot)$ is a function which concatenates two tensors along their first dimension.}}
\label{algo:fr3d:gan_training}
\end{algorithm}

The NN architecture chosen for the discriminator $\mathcal{C}$ was identical to the encoder, however the pool size was increased to 4 and a flatten operation followed by a dense layer using the sigmoid activation function was appended to classify the inputs. The results from training the FR3D model with this GAN setup are summarized in \tabref{tab:fr3d:gan_results}. 

The GAN approach achieves largely similar error levels to MSE for the pressure field and $w$, however also exhibits severe degradation in performance for $u$ and $v$. As a result, it was observed that the FR3D model trained with the GAN approach had substantially worse performance than the baseline MSE version in terms of aerodynamic force coefficient predictions, with the mean percentage errors for $C_L$ and $C_D$ increasing to 23.39\% and 14.29\% respectively. Predictions of the location and intensity of vortices/coherent structures in the wakes are also similarly worse, with the Q-criterion contours not displaying many of the features otherwise reconstructed in the MSE version.

\begin{table}[h!]
\caption{{Mean absolute percentage (MAPE) and mean squared (MSE) error levels achieved by the FR3D model trained using the GAN approach on the validation dataset for reconstruction from point measurements.} }
\label{tab:fr3d:gan_results}
\begin{tabular}{@{}llcccc@{}}
\toprule
Var.                   & Input to $\mathcal{D}$ & MAPE    & Min-max MAPE & MSE                   & Min-max MSE           \\ \midrule
$p$  & $\mathcal{L}$   & 11.15\%  & 7.74\%       & $6.99 \times 10^{-3}$ &  $1.05 \times 10^{-3}$ \\ \midrule
$u$  & $\mathcal{L}$   & 10.98\%  & 8.12\%       & $8.21 \times 10^{-3}$ & $1.57 \times 10^{-3}$ \\ \midrule
$v$  & $\mathcal{L}$   & 44.06\%  & 15.08\%      & $1.66 \times 10^{-2}$ & $3.61 \times 10^{-3}$ \\ \midrule
$w$  & $\mathcal{L}$   & 35.16\%  & 7.29\%       & $3.29 \times 10^{-3}$ & $1.32 \times 10^{-3}$ \\ \bottomrule
\end{tabular}
\end{table}


\clearpage
 \bibliographystyle{elsarticle-num} 
 \bibliography{cas-refs}





\end{document}